\documentclass[aps,prd,reprint,superscriptaddress,nofootinbib]{revtex4-2}
\usepackage[T1]{fontenc}
\usepackage[utf8]{inputenc}
\usepackage{amsmath,amssymb,amsthm,mathtools,bm,mathrsfs}
\usepackage{graphicx}
\usepackage{booktabs}
\usepackage{lmodern}
\usepackage{xcolor}
\usepackage[expansion=false]{microtype}
\usepackage{tikz}
\usetikzlibrary{arrows.meta,positioning,calc,fit}
\usepackage[
  colorlinks=true,
  citecolor=blue!65!black,
  linkcolor=blue!55!black,
  urlcolor=blue!65!black
]{hyperref}

\theoremstyle{definition}
\newtheorem{proposition}{Proposition}
\newtheorem{corollary}{Corollary}
\theoremstyle{plain}

\definecolor{DGreen}{RGB}{0, 100, 0}

\newcommand{\be}{\begin{equation}}
\newcommand{\ee}{\end{equation}}

\newcommand{\M}{\mathcal M}
\newcommand{\Mh}{\widehat{\mathcal M}}

\newcommand{\Hh}{\mathcal H}
\newcommand{\scri}{\mathscr I}
\newcommand{\E}{\mathcal E}

\newcommand{\id}{\mathbf 1}
\newcommand{\Tr}{\operatorname{Tr}}
\newcommand{\dd}{\mathrm d}
\newcommand{\ii}{\mathrm i}
\newcommand{\Ad}{\operatorname{Ad}}

\newcommand{\vev}[1]{\left\langle #1\right\rangle}
\newcommand{\ket}[1]{\left|#1\right\rangle}
\newcommand{\bra}[1]{\left\langle#1\right|}

\begin{document}

\title{The Generalized Second Law from Fluctuation Theorems\texorpdfstring{\\}{ }
in a Quantum Reference Frame}

\author{Feng-Li Lin}
\email{fengli.lin@gmail.com}
\affiliation{Department of Physics, National Taiwan Normal University,
Taipei 11677, Taiwan}

\author{Sayid Mondal}
\email{sayid.mondal@gmail.com}
\affiliation{Instituto de Ciencias Exactas y Naturales, Universidad Arturo Prat,
Playa Brava 3256, 1111346, Iquique, Chile}

\date{Revised November 25, 2026}

\begin{abstract}
We prove the generalized second law (GSL) for black-hole exteriors undergoing unitary driving and relaxation on a fixed semifinite algebra. Wall's influential proof [\href{https://doi.org/10.1103/PhysRevD.85.104049}{Phys.\ Rev.\ D \textbf{85}, 104049 (2012)}] assumes semiclassical geometry and restricts fixed parent states to algebras of successive horizon cuts. We instead evolve the state on a fixed dressed exterior algebra. With complete characteristic data, the same semiclassical exterior supports both descriptions, analogous to the Heisenberg and Schr\"odinger pictures through state--operator duality. The asymptotic horizon charge is the momentum of a quantum clock; its conjugate gives the reading. The modular crossed product admits density operators whose entropy with respect to its trace reproduces generalized entropy up to clock contributions. Our approach parallels thermodynamics in regulated field theory and formulates the GSL through a quantum channel fluctuation theorem (FT) for the open exterior. We develop three complementary formulations under specified dynamical assumptions. Separated unitary driving and relaxation combine modular work, the physical-process first law relating horizon absorption to area gain, and relative-entropy contraction through data processing. A time-local description accommodates simultaneous unitary driving and dissipation. A two-time measurement of surprisal, the negative logarithm of the density operator, gives an FT for the complete exterior channel. It identifies the correction for a nonunital channel and gives a GSL bound for unital evolution without resolving intermediate dynamics. Relaxation that preserves every dressed-charge sector and the full faithful dressed reference state is unital. Unitality is sufficient for this FT proof but is not necessary for the GSL along a particular evolution. The bounds establish the GSL in the slow-clock limit when the errors vanish and quantify what is needed at finite clock width.
\end{abstract}

\maketitle
\tableofcontents

\section{Introduction}
\label{sec:introduction}
Black-hole thermodynamics asks how the second law survives when matter and quantum information pass beyond a horizon. The entropy of the exterior fields alone can decrease as matter falls into a black hole, while Hawking radiation allows the horizon area itself to decrease. The generalized second law (GSL) concerns the combined entropy of the geometry and the exterior fields rather than either contribution separately \cite{Bekenstein:1973ur,Hawking:1975vcx}. For a semiclassical black-hole exterior, the generalized entropy is
\begin{equation}
 S_{\rm gen}=\frac{\vev{A}}{4G} + S_{\rm out}.
 \label{eq:Sgen_def}
\end{equation}
Here $A$ is the horizon cross-sectional area and $S_{\rm out}$ is the entropy of the exterior quantum fields, with a common renormalization prescription. We use $\hbar=c=k_B=1$. The GSL asserts that the generalized entropy at the later stage of the specified comparison is no smaller than at the earlier stage. Understanding why this combination obeys a second law is a basic test of the thermodynamic interpretation of black holes and of the interplay between quantum fields and gravitation.

\begin{figure*}[tbh]
\centering
\begin{minipage}[t]{0.48\linewidth}
\centering
\textbf{(a) Wall's cut comparison}\par
{\small Heisenberg-like: fixed parent states, changing observables}\par\medskip
\begin{tikzpicture}[scale=.90,>=Latex,font=\small]
 \coordinate (B) at (0,0);
 \coordinate (ip) at (3,3);
 \coordinate (i0) at (6,0);
 \coordinate (im) at (3,-3);
 \coordinate (C1) at (.65,.65);
 \coordinate (C2) at (1.50,1.50);
 \fill[black!2] (B)--(ip)--(i0)--(im)--cycle;
 \draw[black!45,line width=.8pt] (B)--(ip)--(i0)--(im)--cycle;
 \draw[blue!70!black,line width=2.1pt] (C1)--(ip);
 \draw[red!75!black,line width=1.7pt,densely dashed] (C2)--(ip);
 \draw[teal!70!black,line width=1.6pt] (ip)--(i0);
 \draw[blue!70!black,line width=1.2pt]
   (C1)..controls (2,.75) and (4.3,.35)..(i0);
 \draw[red!75!black,line width=1.2pt,dashed]
   (C2)..controls (2.6,1.55) and (4.5,.55)..(i0);
 \fill[blue!70!black] (C1) circle (2pt);
 \fill[red!75!black] (C2) circle (2pt);
 \node[left=4pt,blue!70!black] at (C1) {$C_1$};
 \node[left=4pt,red!75!black] at (C2) {$C_2$};
 \node[blue!70!black,below] at (3.60,.48) {$\Sigma_{1,\rm out}$};
 \node[red!75!black,above] at (3.75,.98) {$\Sigma_{2,\rm out}$};
 \draw[->,black!65,line width=.8pt] (2.95,.74)--(2.95,1.19);
 \node[font=\scriptsize,anchor=east,black!65] at (2.82,.96) {later};
 \draw[->,black!55,dotted,line width=1pt] (2.65,1.32)--(2.02,1.95);
 \draw[->,black!55,dotted,line width=1pt] (3.88,.82)--(4.50,1.44);
 \node[font=\scriptsize,black!65] at (3.02,2.27) {future data};
 \node[left] at (B) {$\mathcal B$};
 \node[above] at (ip) {$i^+$};
 \node[right] at (i0) {$i^0$};
 \node[below] at (im) {$i^-$};
 \node[rotate=45,above] at (2.35,2.35) {$\Hh_R^+$};
 \node[rotate=-45,above,teal!70!black] at (4.65,1.35) {$\scri_R^+$};
 \node[rotate=-45,below,black!55] at (1.35,-1.35) {$\Hh_R^-$};
 \node[rotate=45,below,black!55] at (4.65,-1.35) {$\scri_R^-$};
 \draw[->,black!55] (5.65,-2.45)--(5.65,-1.60);
 \node[font=\scriptsize,below,black!55] at (5.65,-2.45) {future};
\end{tikzpicture}\par
\footnotesize
$\M(\Sigma_{i,\rm out})\simeq\M_{C_i>}$,\quad
$\M_{C_2>}\subseteq\M_{C_1>}$.
\par\smallskip
Fixed $\Phi,\Psi$; compare $\Phi\circ\iota_{C_i},\Psi\circ\iota_{C_i}$.
\end{minipage}\hfill
\begin{minipage}[t]{0.48\linewidth}
\centering
\textbf{(b) Our fixed-algebra protocol}\par
{\small Schr\"odinger-like: fixed algebra, evolving state}\par\medskip
\begin{tikzpicture}[scale=.90,>=Latex,font=\small]
 \coordinate (B) at (0,0);
 \coordinate (ip) at (3,3);
 \coordinate (i0) at (6,0);
 \coordinate (im) at (3,-3);
 \fill[black!2] (B)--(ip)--(i0)--(im)--cycle;
 \draw[black!45,line width=.8pt] (B)--(ip)--(i0)--(im)--cycle;
 \draw[blue!70!black,line width=2pt] (B)--(im);
 \draw[teal!70!black,line width=2pt] (im)--(i0);
 \draw[->,blue!60!black,line width=.9pt] (1.25,-1.18)--(2.15,-.28);
 \draw[->,teal!70!black,line width=.9pt] (4.75,-1.18)--(3.85,-.28);
 \node[align=center] at (3,.65)
   {fixed $\M(R)$\\[4pt]state evolves on $\Mh(R)$};
 \node[font=\scriptsize,align=center,black!65] at (3,-1.50)
   {incoming boundary algebra};
 \node[left] at (B) {$\mathcal B$};
 \node[above] at (ip) {$i^+$};
 \node[right] at (i0) {$i^0$};
 \node[below] at (im) {$i^-$};
 \node[rotate=45,above,black!55] at (1.35,1.35) {$\Hh_R^+$};
 \node[rotate=-45,above,black!55] at (4.65,1.35) {$\scri_R^+$};
 \node[rotate=-45,below,blue!70!black] at (1.35,-1.35) {$\Hh_R^-$};
 \node[rotate=45,below,teal!70!black] at (4.65,-1.35) {$\scri_R^-$};
 \draw[->,black!55] (5.65,-2.45)--(5.65,-1.60);
 \node[font=\scriptsize,below,black!55] at (5.65,-2.45) {future};
\end{tikzpicture}\par
\footnotesize
$\M(R)\simeq\M(\Hh_R^-)\,\bar\otimes\,\M(\scri_R^-)$\quad(fixed).
\par\smallskip
Fixed $\Mh(R)$; evolve $\rho_0\mapsto\rho_T=\mathcal C_T(\rho_0)$.
\end{minipage}
\caption{Complementary state--observable descriptions of the right exterior of a bifurcate black hole, analogous to the Heisenberg and Schr\"odinger pictures. Angular directions and the interior extensions of the Cauchy surfaces are suppressed. (a) In Wall's Heisenberg-like comparison, $\Sigma_{i,\rm out}$ meets the future horizon at $C_i$. Solid blue and dashed red mark earlier and later slices and their nested horizon tails, with the same future-infinity algebra included in both comparisons. The parent states $\Phi,\Psi$ are fixed, while their restrictions through the inclusions $\iota_{C_i}$ change with the accessible algebra. (b) In our Schr\"odinger-like protocol, the incoming-data algebra and its dressed crossed product $\Mh(R)$ are fixed, while the state evolves under $\mathcal C_T$. The analogy concerns state--observable duality; it does not identify cut restriction with the physical channel combining unitary driving and relaxation. See Section~\ref{sec:state_observable_duality}. Here $\mathcal B$ is the bifurcation surface, and $i^\pm$ and $i^0$ denote timelike and spatial infinities.}
\label{fig:Wall_Cauchy}
\end{figure*}

Formulating a proof is subtler than applying the ordinary second law to a reduced density operator. The exterior-field entropy in Eq.~\eqref{eq:Sgen_def} is ultraviolet divergent in continuum quantum field theory (QFT), and its renormalization must be treated consistently with the gravitational contribution. Moreover, the exterior field algebra is generally type III and has no intrinsic reduced density operator, although relative entropy remains meaningful \cite{Araki:1976zv,Haag:1996,Witten:2018zxz}. A proof must therefore specify the exterior observables being compared, define the combined entropy, and relate its change to a quantity with a definite sign. Even when the complete matter--gravity system evolves unitarily, exterior measurements alone cannot determine its correlations with degrees of freedom beyond the horizon. Preservation of the full system's fine-grained entropy consequently does not by itself determine the exterior generalized-entropy change.

Wall's influential proof addresses these difficulties by combining monotonicity of relative entropy for nested horizon algebras with geometric modular flow and the focusing equation \cite{Wall:2011hj}. Within a semiclassical geometry, it compares fixed parent states restricted to the observables available beyond successive horizon cuts. Subsequent developments include second laws in gravitational effective field theory, crossed-product constructions for horizon cuts, and extensions involving dynamical cuts \cite{HollandsKovacsReall:2022,FaulknerSperanza:2024,Kirklin:2024,Shajiee:2026}.

In this paper, we develop a complementary approach motivated by quantum fluctuation theorems, which connect the statistics of work and entropy production to the ordinary second law \cite{Jarzynski:1997,Crooks:1999,Kurchan:2000,Tasaki:2000,Campisi:2011}. We keep the dressed right-wedge algebra fixed and follow the state through unitary driving and relaxation. Building on the gravitational crossed-product construction of Witten and collaborators \cite{Witten:2022qdt,Chandrasekaran:2022eqq,Chandrasekaran:2022cip,Jensen:2023yxy,KudlerFlam:2023qfl}, we use the asymptotic horizon charge and its conjugate as a quantum clock. The joint field--clock algebra admits density operators, and their trace entropy can be related to the area and exterior-field entropy once the clock contribution and approximation error are accounted for. It makes the GSL argument parallel to the thermodynamic reasoning in regulated QFT and allows a formulation in terms of quantum-channel fluctuation statistics. We develop three complementary settings, beginning with separated unitary driving and relaxation, then allowing simultaneous open-system dynamics, and finally considering a two-time surprisal fluctuation theorem for the complete channel. Each formulation connects the entropy balance to specified properties of the physical evolution. Our type-I matter argument parallels earlier GSL derivations using a fluctuation theorem for horizon transitions~\cite{IsoOkazawaZhang:2010} and stochastic open-system dynamics for an exterior scalar field~\cite{IsoOkazawa:2011}.

The following subsections explain these viewpoints and their relation. Section~\ref{sec:wall_cut_proof} reviews Wall's cut-based proof; Section~\ref{sec:proposal_overview} outlines our fixed-algebra proposal and its three channel formulations; and Section~\ref{sec:state_observable_duality} discusses their common semiclassical exterior and the analogy with Heisenberg--Schr\"odinger state--observable duality, illustrated in Fig.~\ref{fig:Wall_Cauchy}.

\subsection{Wall's cut-indexed proof}
\label{sec:wall_cut_proof}

Wall formulates the GSL without a density operator intrinsic to the continuum exterior algebra
\cite{Wall:2011hj}.  A cut $C$ of the future causal horizon determines an algebra $\M_{C>}$ of future boundary data, including the common asymptotic sector specified below.  If $C_2$ is nowhere to the past of $C_1$, then the later-cut algebra is contained in the earlier-cut algebra, and monotonicity of Araki relative entropy \cite{Araki:1976zv} gives
\begin{equation}
 \M_{C_2>}\subseteq\M_{C_1>},
 \qquad
 D_{C_1}(\Phi\Vert\Psi)\geq D_{C_2}(\Phi\Vert\Psi).
 \label{eq:Wall_DPI}
\end{equation}
Here $D_C$ denotes the relative entropy of the two states restricted to
$\M_{C>}$.  No reduced density operator for the type-III algebra is needed.

Figure~\ref{fig:Wall_Cauchy}(a) relates this boundary description to a Cauchy surface $\Sigma$.  Let $C=\Sigma\cap\Hh_R^+$ and let $\Sigma_{\rm out}$ be its exterior portion, a partial Cauchy surface.  Under Wall's determinism assumption, its data can be evolved to the horizon above $C$ together with future infinity \cite{Wall:2011hj}:
\begin{equation}
 \begin{aligned}
 \M(\Sigma_{\rm out})&:=\M\bigl(D(\Sigma_{\rm out})\bigr)\simeq\M_{C>},\\
 \M_{C>}&:=\M(\Hh^+_{R,C>})\vee\M(\scri_R^+).
 \end{aligned}
 \label{eq:Wall_Cauchy_boundary}
\end{equation}
Here $D$ denotes the domain of dependence and $\vee$ denotes the generated von Neumann algebra; with independent complete boundary sectors, $\vee$ is their von Neumann tensor product.  Thus $C$ labels the boundary algebra: the data do not live on the codimension-two cut alone. Any required timelike-infinity or zero-mode data are included in the asymptotic sector in Eq.~\eqref{eq:Wall_Cauchy_boundary}.

As $\Sigma$ moves upward in the diagram, its cut moves forward along the horizon.  The future-horizon portion shrinks while the asymptotic sector is held fixed, yielding Eq.~\eqref{eq:Wall_DPI}.  In this sense, the accessible algebra evolves.  This is a restriction to nested subalgebras, not merely unitary transport of the same algebra.  A deformation preserving both $C$ and the domain of dependence would not change the algebra.

Wall assumes that the boundary data determine the exterior, that fields on distinct horizon generators supply independent degrees of freedom (ultralocality), that the horizon algebra is locally Lorentz invariant, and that the generator of future-directed null translations has nonnegative spectrum (stability).  These assumptions have been checked in Ref.~\cite{Wall:2011hj} for several free fields and selected two-dimensional conformal theories.

Rapid variation of the quantum energy flux does not by itself imply a large change in the geometry. A short, weak pulse, for example, can vary rapidly while producing only a perturbative change in the horizon area. Wall's analysis does not require the flux to vary slowly: it uses quantum fields on the classical background and computes the leading geometric response semiclassically from the expected stress tensor \cite[Sec.~2.2]{Wall:2011hj}. In the classically stationary case relevant here, the background horizon therefore provides common generators for comparing cuts, even though the area responds to the flux. This description presupposes that the backreaction remains perturbative.

Choose $\Psi$ to be the horizon vacuum times a fixed faithful asymptotic reference state.  For an arbitrary state $\Phi$, write
\begin{equation}
 D_C(\Phi\Vert\Psi)
 =\Delta\vev{K_C}-\Delta S_{\rm out}[C].
 \label{eq:Wall_relent}
\end{equation}
Here $\Delta$ denotes the value in $\Phi$ minus the value in $\Psi$ at the same cut. The modular energy $K_C$ includes the common asymptotic contribution; its horizon part generates boosts about $C$. Its expectation difference is understood with the same subtraction as $\Delta S_{\rm out}[C]$.  The linearized Raychaudhuri equation, with a common future-stationary boundary condition, gives schematically
\begin{equation}
 \Delta\vev{K_C}
 =C_{\rm future}-\frac{\Delta\vev{A[C]}}{4G}.
 \label{eq:Wall_area}
\end{equation}
The constant $C_{\rm future}$ contains the late-time area difference fixed by the boundary condition and the common asymptotic modular-energy difference. Neither depends on which cut is chosen. Equations~\eqref{eq:Wall_relent} and \eqref{eq:Wall_area} imply
\begin{equation}
 D_C(\Phi\Vert\Psi)
 =C_{\rm future}-\Delta S_{\rm gen}[C].
 \label{eq:Wall_identity}
\end{equation}
The reference has cut-independent generalized entropy on this stationary background. Thus the decrease of $D_C$ in Eq.~\eqref{eq:Wall_DPI} gives $S_{\rm gen}[C_2]\geq S_{\rm gen}[C_1]$ for $\Phi$.  The argument combines geometric modular flow \cite{Takesaki:1970,Sorce:2023}, focusing/area balance, and the monotonicity of relative entropy.

Faulkner and Speranza extended this cut-based framework to crossed-product gravitational algebras associated with arbitrary cuts of a Killing horizon \cite{FaulknerSperanza:2024}. Given a reference state with geometric modular flow on the horizon, they show that the trace entropy of the gravitational algebra reproduces the generalized entropy of the cut in a semiclassical limit. The comparison between cuts still follows from monotonicity of relative entropy for the underlying quantum field algebras. Their result thus provides a gravitational algebraic formulation of Wall's cut comparison. In this paper, we develop a complementary approach to the GSL, sketched in the next subsection. We use a crossed product to construct a fixed dressed right-wedge algebra and study how the generalized entropy changes as the state evolves through a physical protocol.

\subsection{A fixed dressed algebra and three formulations of the GSL}
\label{sec:proposal_overview}

Throughout this paper, \emph{exterior}, \emph{right exterior}, and \emph{right wedge} refer to the same region $R$ associated with the chosen right asymptotic region. This region excludes the black-hole interior and the opposite exterior; Fig.~\ref{fig:eternal_horizon} makes the distinction explicit. For the massless radiative sector considered here, we assume that data on the past horizon $\Hh_R^-$ and past null infinity $\scri_R^-$ uniquely determine the exterior solution and exhaust the chosen radiative phase space. These are characteristic data because both initial surfaces are null. Completeness also requires that no additional modes or symplectic flux be omitted at the limiting cross-sections or timelike infinity. With the stated decay and independence of the two boundary sectors, this gives \cite{DimockKay:1987,KudlerFlam:2023qfl}
\begin{equation}
 \M(R)\simeq
 \M(\Hh_R^-)\,\bar\otimes\,\M(\scri_R^-).
 \label{eq:wedge_boundary}
\end{equation}
The bar denotes the von Neumann tensor product for independent complete incoming sectors. Equation~\eqref{eq:wedge_boundary} fixes a presentation of the parent exterior algebra illustrated in Fig.~\ref{fig:Wall_Cauchy}(b). Section~\ref{sec:state_observable_duality} relates it to Wall's cut algebras. The bifurcation surface and the gravitational charges at the asymptotic horizon ends enter the construction of the observables and their entropy.

The quantum clock consists of an asymptotic horizon charge and its conjugate reading \cite{AharonovSusskind:1967,AharonovKaufherr:1984,PageWootters:1983,Popescu:2018,Giacomini:2019,DeVuyst:2024fxc}. The modular crossed product gives a fixed type-$\mathrm{II}_\infty$ dressed algebra $\Mh(R)$ with a faithful normal semifinite trace $\tau$ \cite{Witten:2022qdt,Chandrasekaran:2022eqq,Chandrasekaran:2022cip,Jensen:2023yxy,KudlerFlam:2023qfl}. Semifiniteness means that positive operators can be approximated from below by positive operators of finite trace. Here $\tau(\id)=\infty$: the trace is not a state, but a dressed state can have a density operator normalized to unit trace. Both ordinary type-I quantum mechanics and the type-II dressed algebra admit normalized density operators and the same trace-based formulas for entropy and measurement probabilities. This common framework allows quantum fluctuation theorems to be applied to the dressed exterior. For the preparations described in Section~\ref{sec:crossed_product}, the trace entropy contains generalized entropy, the entropy of the clock charge distribution, and a remainder. For the rescaled packets used below, broadening the charge distribution narrows the distribution of the conjugate reading. The packet entropy itself need not be small; only its change enters the entropy balance. Section~\ref{sec:construction} develops this construction and its entropy interpretation.

\begin{figure*}[t]
\centering
\begin{minipage}[t]{0.48\linewidth}
\centering
\textbf{(a) Unitary driving in the right exterior}\par\medskip
\begin{tikzpicture}[scale=.90,>=Latex,font=\small]
  \coordinate (B) at (0,0);
  \coordinate (ip) at (3,3);
  \coordinate (i0) at (6,0);
  \coordinate (im) at (3,-3);
  \fill[black!2] (B)--(ip)--(i0)--(im)--cycle;
  \draw[black!45,line width=.8pt] (B)--(ip)--(i0)--(im)--cycle;
  \draw[blue!70!black,line width=2pt] (B)--(ip);
  \draw[teal!70!black,line width=2pt] (ip)--(i0);
  \node[left] at (B) {$\mathcal B$};
  \node[above] at (ip) {$i^+$};
  \node[right] at (i0) {$i^0$};
  \node[below] at (im) {$i^-$};
  \node[rotate=45,above,blue!70!black] at (1.35,1.35) {$\Hh_R^+$};
  \node[rotate=-45,above,teal!70!black] at (4.65,1.35) {$\scri_R^+$};
  \node[rotate=-45,below,black!55] at (1.35,-1.35) {$\Hh_R^-$};
  \node[rotate=45,below,black!55] at (4.65,-1.35) {$\scri_R^-$};

  \node[align=center,black!75] at (3,-1.90)
    {reference\\state $\sigma_g$};
  \draw[->,black!55,dotted,line width=.9pt] (3,-1.40)--(3,-1.04);
  \filldraw[fill=red!7,draw=red!75!black,line width=.8pt]
    (3,-.50) ellipse (1.22 and .50);
  \node[red!75!black,align=center] at (3,-.50)
    {unitary drive\\$U$};
  \draw[->,blue!70!black,line width=1.2pt]
    (2.35,-.10)..controls (2.10,.30) and (1.90,.90)..(1.60,1.60);
  \draw[->,teal!70!black,line width=1.2pt]
    (3.65,-.10)..controls (3.90,.30) and (4.10,.90)..(4.40,1.60);
  \node[blue!70!black,align=center] at (1.25,.20)
    {absorbed\\$E_H$};
  \node[teal!70!black,align=center] at (4.75,.20)
    {exterior\\$\Delta E_{\rm out}$};
  \draw[->,black!55,dotted,line width=.9pt] (3,.12)--(3,1.12);
  \node[right,black!65] at (3,.65) {$\E$};
  \node[align=center] at (3,1.68)
    {late exterior\\state $\chi$};
  \draw[->,black!55] (5.65,-2.45)--(5.65,-1.60);
  \node[font=\scriptsize,below,black!55] at (5.65,-2.45) {future};
\end{tikzpicture}\par
\footnotesize
State stages on the same $\Mh(R)$.
\end{minipage}\hfill
\begin{minipage}[t]{0.48\linewidth}
\centering
\textbf{(b) State protocol and energy balance}\par\medskip
\small
\setlength{\abovedisplayskip}{5pt}
\setlength{\belowdisplayskip}{5pt}
\setlength{\abovedisplayshortskip}{3pt}
\setlength{\belowdisplayshortskip}{5pt}
\[
 \sigma_g\xrightarrow{\;U\;}\omega\xrightarrow{\;\E\;}\chi,
 \qquad \E(\sigma_g)=\sigma_g .
\]
\raggedright
\begin{enumerate}
  \setlength{\itemsep}{5pt}
  \item \textit{Reference preparation.} Begin in $\sigma_g$, which is stationary under the undriven relaxation channel.
  \item \textit{Unitary driving.} The localized unitary drive
  produces $\rho_\omega=U\rho_{\sigma_g}U^\dagger$.
  \item \textit{Exterior relaxation.} The effective channel gives
  $\rho_\chi=\E(\rho_\omega)$ after restricting to exterior observables.
\end{enumerate}
\textit{Killing-energy balance:}
\[
 W=\underbrace{E_H}_{\text{horizon}}
       +\underbrace{\Delta E_{\rm out}}_{\text{exterior}} .
\]
\textit{Physical-process area response:}
\[
 \frac{\Delta\vev{A^{(2)}}}{4G}=\beta_HE_H .
\]
\end{minipage}
\caption{Spacetime and state descriptions of the separated protocol used in Section~\ref{sec:FT}. The unitary drive deposits Killing energy in the exterior fields. During subsequent relaxation, part is absorbed through the future horizon and part remains in the exterior energy account, including radiation reaching future null infinity. The physical-process first law gives the area change between the initial and final geometric cross-sections. Section~\ref{subsec:area_endpoint} specifies when it equals the area-charge change during relaxation. The state labels indicate stages on one dressed algebra. Sections~\ref{sec:simultaneous} and \ref{sec:fullcore_FT} do not require a separate relaxation stage following the unitary drive.}
\label{fig:physical_process}
\end{figure*}

We treat the dressed exterior as an open quantum system interacting with degrees of freedom excluded from the reduced description, such as infalling modes after they enter the black-hole interior. Restricting to exterior observables can then yield nonunitary evolution even when the complete matter--gravity dynamics is unitary. We describe this evolution by a quantum channel on the fixed dressed algebra, without assuming a tensor-product decomposition across the horizon. The gravitational clock remains part of the system, and radiation reaching $\scri_R^+$ remains in the exterior energy balance. Figure~\ref{fig:protocol} summarizes the three formulations and the entropy inequality underlying each GSL argument:
\begin{enumerate}
 \item \emph{Separated unitary driving and relaxation} (Section~\ref{sec:FT}). A unitary drive takes the initial dressed reference state $\sigma_g$ to $\omega$, and reference-preserving relaxation $\E$ yields $\chi$:
 \begin{equation}
  \sigma_g\xrightarrow{\;U\;}\omega
   \xrightarrow{\;\E\;}\chi,
   \qquad \E(\sigma_g)=\sigma_g .
  \label{eq:protocol_intro}
 \end{equation}
 Measuring the reference surprisal (the negative logarithm of the reference density operator) before and after the unitary drive gives a fluctuation theorem (FT) that relates the mean modular work of the unitary drive to the relative entropy $D_\tau(\omega\Vert\sigma_g)$ of dressed states, computed with the trace $\tau$ (defined in Section~\ref{subsec:modular_crossed_product}). The data-processing inequality (DPI) bounds the distinguishability left after relaxation. Matching the horizon energy and area response and controlling the clock terms turn this contraction into a conditional GSL. Figure~\ref{fig:physical_process} illustrates this protocol; the cutoff calculation in Section~\ref{sec:regulated_GSL} gives the basic energy balance.
 \item \emph{Simultaneous unitary driving and horizon exchange} (Section~\ref{sec:simultaneous}). A time-local effective dynamics allows unitary driving and relaxation throughout the same interval. There is no distinguished state $\omega$ between unitary driving and relaxation. The integrated dissipative entropy production replaces the relative-entropy contraction of the separated protocol. The horizon response must match the area change assigned to the dissipative part of the dynamics.
 \item \emph{Two-time surprisal FT} (Section~\ref{sec:fullcore_FT}). Here the second measurement uses the surprisal of the actual final state. The mean difference of the two outcomes equals the trace-entropy change under the spectral assumptions of Section~\ref{sec:two_time_experiments}. A general channel produces a nonunital correction in the exponential average, so this relation alone need not prove the GSL. Unitality removes that correction and normalizes the reverse experiment. We then establish sufficient physical conditions for dressed relaxation to be unital, giving a GSL bound with the stated clock errors and without a time-local description of the intermediate evolution.
\end{enumerate}
Table~\ref{tab:gsl_formulations} compares the roles and assumptions of the three formulations. The separated protocol is the main result. The time-local formulation extends its entropy balance to overlapping unitary driving and dissipation, while the two-time formulation asks which property of the complete channel makes the surprisal FT imply the GSL. A sufficient answer is unitality, established under preservation of every dressed-charge sector during relaxation and stationarity of the full faithful reference state in Section~\ref{sec:unital_physical_conditions}. These sufficient conditions are stronger than mean energy--area balance. They constrain how relaxation redistributes the joint field--clock states within each charge sector. Unitality removes the FT correction for all allowed inputs. It is not necessary for the GSL along a particular evolution, and Sections~\ref{sec:FT} and \ref{sec:simultaneous} do not assume it. A microscopic application must establish these channel properties and bound the clock errors for the states it produces. These formulations build on work on fluctuation relations and gravitational algebras \cite{Cirafici:2024a,Cirafici:2024b,Cirafici:2025,Rastegin:2013,Kwon:2019} and complement studies of null-infinity GSLs and open-system black-hole thermodynamics \cite{RignonBret:2026a,RignonBret:2026b}.

\begin{figure*}[t]
\centering
\begin{tikzpicture}[>=Latex,
  panel/.style={draw=black!32,rounded corners=3pt,inner sep=8pt,
  align=center,text width=.285\textwidth,minimum height=43mm,
  font=\small},
  banner/.style={fill=black!6,rounded corners=2pt,inner ysep=5pt,
  text width=.94\textwidth,align=center,font=\small}]
 \node[banner] (common) {\textbf{One fixed dressed algebra} $\Mh(R)$
   \quad\textbar\quad
   \mbox{\thinmuskip=2mu\medmuskip=2mu\thickmuskip=3mu
   $S_\tau(\nu)=S_{\rm gen}(\nu)+S[g_\nu]+r_\epsilon(\nu)$}};
 \node[panel,fill=blue!4,anchor=north] (b) at ([yshift=-9pt]common.south)
  {\textbf{(b) Simultaneous dynamics}\\[-1pt]
   \footnotesize Section~\ref{sec:simultaneous}\\[6pt]
   $\dot\rho_t=-\ii[H_t,\rho_t]+\mathcal L_t(\rho_t)$\\[6pt]
   $\Sigma_{\rm diss}=\int_0^T\dot\Sigma_{\rm diss}\,\dd t\geq0$\\[4pt]
   Integrated dissipative area response};
 \node[panel,fill=teal!4,left=6pt of b] (a)
  {\textbf{(a) Separated protocol}\\[-1pt]
   \footnotesize Section~\ref{sec:FT}\\[6pt]
   $\sigma_g\xrightarrow{\;U\;}\omega\xrightarrow{\;\E\;}\chi$\\[6pt]
   $D_\tau(\omega\Vert\sigma_g)-D_\tau(\chi\Vert\sigma_g)\geq0$\\[4pt]
   Relaxation fixes the reference};
 \node[panel,fill=orange!5,right=6pt of b] (c)
  {\textbf{(c) Two-time surprisal FT}\\[-1pt]
   \footnotesize Section~\ref{sec:fullcore_FT}\\[6pt]
   $\rho_0\xrightarrow{\;\mathcal C_T\;}\rho_T$\\[6pt]
   $\langle e^{-\Sigma_{\rm 2t}}\rangle_F=\Gamma_T$\\[4pt]
   Unital channel: $\Gamma_T=1$\\[4pt]
   $\langle\Sigma_{\rm 2t}\rangle_F=\Delta S_\tau\geq0$};
\end{tikzpicture}
\caption{Entropy inequalities for three formulations on the same gravitationally dressed right-wedge algebra. (a) Reference-preserving relaxation gives a nonnegative relative-entropy loss by data processing. (b) Data processing for each reference-preserving dissipative step gives a nonnegative integrated production. (c) Under the support and convergence assumptions of Section~\ref{sec:two_time_experiments}, the two-time relation contains $\Gamma_T=\tau[\rho_T\mathcal C_T(\id)]$. For a unital channel, $\Gamma_T=1$ and Jensen's inequality gives $\langle\Sigma_{\rm 2t}\rangle_F=\Delta S_\tau\geq0$, the nonnegative quantity used in this formulation. Unitality also normalizes the reverse experiment, which starts in $\rho_T$ and need not return it to $\rho_0$. The entropy dictionary, Eqs.~\eqref{eq:entropy_equals_Sgen} and \eqref{eq:generic_entropy_target}, and the stated gravitational and clock conditions turn these inequalities into the respective GSL bounds.}
\label{fig:protocol}
\end{figure*}

\begin{table*}[t]
\caption{Roles of the three formulations in Fig.~\ref{fig:protocol}. They form a pedagogical progression; their dynamical assumptions are distinct.}
\label{tab:gsl_formulations}
\centering
\small
\begin{tabular}{@{}lll@{}}
\toprule
\parbox[t]{.20\textwidth}{\raggedright \textbf{Formulation}} &
\parbox[t]{.29\textwidth}{\raggedright \textbf{Question addressed}} &
\parbox[t]{.46\textwidth}{\raggedright \textbf{Nonnegative entropy quantity and its role}}\\
\midrule
\parbox[t]{.20\textwidth}{\raggedright Separated protocol (Sec.~\ref{sec:FT})} &
\parbox[t]{.29\textwidth}{\raggedright How do the work of the unitary drive, horizon absorption, and exterior entropy fit together?} &
\parbox[t]{.46\textwidth}{\raggedright Measuring the fixed reference surprisal gives modular-work statistics. DPI makes the relative-entropy loss during reference-preserving relaxation nonnegative.}\\
\addlinespace[3pt]
\parbox[t]{.20\textwidth}{\raggedright Simultaneous dynamics (Sec.~\ref{sec:simultaneous})} &
\parbox[t]{.29\textwidth}{\raggedright What replaces the intermediate state $\omega$ when unitary driving and relaxation overlap?} &
\parbox[t]{.46\textwidth}{\raggedright DPI gives a nonnegative dissipative production rate. The integrated mean balance requires matching the dissipative energy and area changes to the horizon response.}\\
\addlinespace[3pt]
\parbox[t]{.20\textwidth}{\raggedright Two-time surprisal FT (Sec.~\ref{sec:fullcore_FT})} &
\parbox[t]{.29\textwidth}{\raggedright When does the full-channel surprisal FT imply the GSL?} &
\parbox[t]{.46\textwidth}{\raggedright A general channel gives $\Delta S_\tau\geq-\log\Gamma_T$. Unitality gives $\Delta S_\tau=\langle\Sigma_{\rm 2t}\rangle_F\geq0$ and normalizes the reverse experiment; the entropy dictionary then gives a GSL bound with clock errors.}\\
\bottomrule
\end{tabular}
\end{table*}

\subsection{A common exterior algebra and state--observable duality}
\label{sec:state_observable_duality}

Wall's cut comparison and our fixed-algebra protocol use the same semiclassical exterior and its parent field algebra. In Wall's approach, complete outgoing data on $\Hh_R^+\cup\scri_R^+$ are assumed to determine the exterior solution, with no omitted modes or boundary fluxes; a horizon cut selects the future horizon tail together with the same asymptotic sector \cite{Wall:2011hj,FaulknerSperanza:2024}. Our approach uses complete incoming data on $\Hh_R^-\cup\scri_R^-$ to define the fixed parent algebra for the eternal Schwarzschild exterior and its massless radiative sector, as in Eq.~\eqref{eq:wedge_boundary}. Scattering relates this presentation to Wall's outgoing one \cite{DimockKay:1987}. Thus a fixed parent algebra accommodates changing algebras of accessible observables.

Moving a future-horizon cut or the outgoing portion of a slice at $\scri_R^+$ does not redefine the incoming algebra. Within perturbation theory, changes in the future geometry can nevertheless modify scattering, fluxes, and the identification of outgoing observables. Our $E_{\rm out}$ includes radiation reaching $\scri_R^+$; comparisons using finite outgoing cuts must include the intervening flux. Wall keeps the same future-infinity algebra at every horizon cut, so its contribution to the reference modular energy is cut independent. Changing that asymptotic algebra requires reconsidering this cancellation.

\begingroup\clubpenalty=10000
Figure~\ref{fig:Wall_Cauchy} illustrates the Heisenberg--Schr\"odinger analogy. Wall's inclusions change which observables are accessible while keeping the parent states fixed. Our protocol instead specifies a channel and the evolving reduced state on a fixed dressed algebra. These are complementary uses of state--observable duality, without identifying cut restriction with physical relaxation.\par
\endgroup

The Schr\"odinger description connects the GSL to open-system thermalization through unitary driving, energy exchange, and horizon response. For a unital channel on the full field--clock algebra, the two-time surprisal FT implies that the trace entropy cannot decrease. The relation between trace entropy and generalized entropy then gives a GSL bound when the changes in the clock contribution and entropy remainder are controlled. The sufficient unitality condition for this FT proof places a physical constraint on the quantum-channel model of black-hole dynamics. This direct connection to the unital quantum channel is a conceptual advantage of our approach over Wall's.

Section~\ref{sec:construction} reviews the dressed algebra and entropy dictionary. Sections~\ref{sec:FT}--\ref{sec:fullcore_FT} develop the channel formulations, and Section~\ref{sec:conclusion} concludes. Appendices~\ref{app:density}--\ref{app:charge_transport} give supporting derivations and clock-error estimates.

\section{Right-wedge observables and generalized entropy: a review}
\label{sec:construction}

This section collects the ingredients established in earlier work rather than proposing a new construction of horizon algebras or generalized entropy \cite{Witten:2022qdt,Chandrasekaran:2022eqq,Chandrasekaran:2022cip,KudlerFlam:2023qfl,Jensen:2023yxy}. We specify the observables, charges, and entropy to which our fluctuation-theorem proposal applies. Reconstructing the full exterior from its null boundary requires the completeness assumptions of Section~\ref{sec:proposal_overview}. The horizon flux and area relations also require the falloff and perturbative conditions given below.

Figure~\ref{fig:eternal_horizon} fixes the horizon notation using the maximally extended eternal Schwarzschild spacetime. In Kruskal coordinates, $R$ is the region $U<0<V$. We follow the complete-sheet convention of Ref.~\cite{KudlerFlam:2023qfl}: $\Hh^-:=\{V=0\}$ extends through the bifurcation surface $\mathcal B$, with
\[
 \Hh_R^-:=\{V=0,\ U<0\},\qquad
 \Hh_L^-:=\{V=0,\ U>0\}.
\]
Thus $\Hh^-=\Hh_R^-\cup\mathcal B\cup\Hh_L^-$. The superscript $-$ labels this complete null sheet; its $U>0$ branch is the future horizon of the left exterior. Only $\Hh_R^-$ belongs to the past boundary of $R$. The second complete sheet is $\Hh^+=\{U=0\}$, whose right branch $V>0$ is the future horizon of $R$.

For the horizon reference used below, the full modular generator of the right-horizon algebra is represented by the Killing flux over all of $\Hh^-$, whereas the one-sided energy involves only $\Hh_R^-$:
\[
 K_{\Psi,\Hh^-}=\beta_HF_\xi^{\Hh^-},\qquad
 k_\Psi=\beta_HF_\xi^{\Hh_R^-}.
\]
We abbreviate $K_{\Psi,\Hh^-}$ to $K_\Psi$ in the horizon-sector formulae. This distinction concerns the integration domain and the action of the full modular generator on the algebra and its commutant; it does not assert an intrinsic type-III density operator for either branch. The one-sided energy is used through regulated, reference-subtracted expectation values. The independent infinity sector must also be included when the full exterior modular generator is required, as explained below.

\begin{figure*}[t]
\centering
\begin{minipage}[c]{.68\textwidth}
\centering
\begin{tikzpicture}[scale=.78,>=Latex,font=\small]
 \coordinate (B) at (0,0);
 \coordinate (rp) at (3,3); \coordinate (r0) at (6,0);
 \coordinate (rm) at (3,-3);
 \coordinate (lp) at (-3,3); \coordinate (l0) at (-6,0);
 \coordinate (lm) at (-3,-3);
 \fill[teal!28] (B)--(rp)--(r0)--(rm)--cycle;
 \fill[black!12] (B)--(lp)--(l0)--(lm)--cycle;
 \fill[red!16] (B)--(lp)--(rp)--cycle;
 \fill[orange!22] (B)--(lm)--(rm)--cycle;
 \draw[black!50,line width=.8pt] (lp)--(l0)--(lm);
 \draw[black!50,line width=.8pt] (rp)--(r0)--(rm);
 \draw[black!45,line width=.9pt] (lm)--(B)--(rp);
 \draw[blue!75!black,line width=2pt] (rm)--(B)--(lp);
 \draw[teal!80!black,line width=3pt] (rm)--(B);
 \draw[black!75,line width=.8pt] (-3,3)--(-2.8,3.06);
 \foreach \j in {0,...,13}{
  \pgfmathsetmacro{\x}{-2.8+.4*\j}
  \draw[black!75,line width=.8pt] (\x,3.06)--({\x+.2},2.94)--({\x+.4},3.06);
  \draw[black!75,line width=.8pt] (\x,-3.06)--({\x+.2},-2.94)--({\x+.4},-3.06);
 }
 \draw[black!75,line width=.8pt] (2.8,3.06)--(3,3);
 \draw[black!75,line width=.8pt] (-3,-3)--(-2.8,-3.06);
 \draw[black!75,line width=.8pt] (2.8,-3.06)--(3,-3);
 \node[above=4pt] at (0,3.08) {$r=0$ singularity};
 \node[below=4pt] at (0,-3.08) {$r=0$ singularity};
 \node[align=center] at (0,2.20) {black-hole\\interior};
 \node[align=center] at (0,-2.20) {white-hole\\interior};
 \node[align=center,teal!70!black] at (3.48,0) {right exterior $R$\\[3pt]$U<0<V$};
 \node[align=center,black!65] at (-3.48,0) {left exterior\\[3pt]$V<0<U$};
 \fill[black] (B) circle(2.2pt);
 \node[fill=white,inner sep=1.5pt] at (0,-.33) {$\mathcal B$};
 \node[above right=2pt] at (rp) {$i_R^+$};
 \node[right=3pt] at (r0) {$i_R^0$};
 \node[below right=2pt] at (rm) {$i_R^-$};
 \node[above left=2pt] at (lp) {$i_L^+$};
 \node[left=3pt] at (l0) {$i_L^0$};
 \node[below left=2pt] at (lm) {$i_L^-$};
 \node[rotate=-45,above] at (4.65,1.35) {$\scri_R^+$};
 \node[rotate=45,below] at (4.65,-1.35) {$\scri_R^-$};
 \node[rotate=45,above] at (-4.65,1.35) {$\scri_L^+$};
 \node[rotate=-45,below] at (-4.65,-1.35) {$\scri_L^-$};
 \node[rotate=-45,above,blue!75!black] at (-1.60,1.60) {$\Hh_L^-\ (U>0)$};
 \node[rotate=-45,above,teal!80!black] at (1.60,-1.60) {$\Hh_R^-\ (U<0)$};
 \node[rotate=45,below,black!60] at (1.60,1.60) {$\Hh_R^+$};
 \node[rotate=45,below,black!60] at (-1.60,-1.60) {$\Hh_L^+$};
\end{tikzpicture}
\end{minipage}\hfill
\begin{minipage}[c]{.29\textwidth}
\small\raggedright
\textbf{Complete horizon sheet}\par\smallskip
Blue and teal together give
\[
 \Hh^-:\quad V=0,\quad U\in\mathbb R.
\]
The horizon modular generator is
\[
 K_\Psi=\beta_HF_\xi^{\Hh^-}.
\]
\medskip
\textbf{Right past-horizon branch}\par\smallskip
Teal marks $\Hh_R^-$, with $U<0$.
Its one-sided energy is
\[
 k_\Psi=\beta_HF_\xi^{\Hh_R^-}.
\]
The incoming boundary of $R$ is $\Hh_R^-\cup\scri_R^-$.
\end{minipage}
\caption{Penrose diagram of the eternal Schwarzschild spacetime, with angular directions suppressed. The shaded right diamond is $R$; the black-hole interior and left exterior are distinct regions. The complete sheet $\Hh^-\ (V=0)$ runs from $i_R^-$ through $\mathcal B$ to $i_L^+$, and contains both the teal right branch and the blue left branch. The other diagonal is $\Hh^+\ (U=0)$. The $\pm$ labels follow the complete-sheet convention; in particular, $\Hh_L^-$ is future-directed from $\mathcal B$ and is the left exterior's future horizon. The horizon contribution to $K_\Psi$ uses the full $U$ range, while $k_\Psi$ uses only $U<0$. The latter is understood through the regulated, subtracted expectations specified in the text.}
\label{fig:eternal_horizon}
\end{figure*}

The distinction that matters for entropy is whether the algebra admits a trace. A type-I factor has minimal projections and the usual operator trace.  A type-$\mathrm{II}_1$ factor has no minimal projections but has a finite normalized trace.  A type-$\mathrm{II}_\infty$ factor instead has a faithful normal semifinite trace with $\tau(\id)=\infty$.  A type-III factor has no faithful normal semifinite trace \cite{Connes:1973,Haag:1996}.  Local QFT algebras are generally type $\mathrm{III}_1$; the crossed product by the modular action of a faithful normal state is semifinite \cite{Takesaki:1973}.  It provides density operators $\rho_\nu$ for normal states, their surprisals $k_\nu=-\log\rho_\nu$, and the two-time measurement statistics used below.

We first identify the scalar and gravitational fluxes that determine the horizon charges. We then use the right asymptotic charge as the momentum of a quantum clock and construct the dressed exterior algebra relative to that clock.

\subsection{Scalar horizon algebra and flux balance}
\label{sec:scalar}

This subsection reviews the construction in \cite{KudlerFlam:2023qfl} but presents only the ingredients needed by the GSL argument.  The construction is performed on a stationary background with a bifurcate Killing horizon and Killing field $\xi^a$ with surface gravity $\kappa$.  We set $\beta_H=2\pi/\kappa$.

\subsubsection{Bulk observables and characteristic data}

For a real Klein--Gordon field $\phi$ with equation $P\phi=0$, a real compactly supported test function $f$ defines the smeared observable
\begin{equation}
 \phi_f=\int_M f\phi\,\epsilon_{\bar g},
 \qquad f\sim f+Pf_0\,,
 \label{eq:scalar_smear}
\end{equation}
Here $\bar g_{ab}$ is the stationary background metric and $\epsilon_{\bar g}=\sqrt{-\det\bar g}\,\dd^4x$ its spacetime volume form, while $f_0\in C_0^\infty(M)$. All covariant derivatives below are taken with respect to $\bar g$ (or its induced metric on a cross-section); we write $\nabla=\bar\nabla$ for the background Levi-Civita connection. Let $\tilde{G}=\tilde{G}_{\rm ret}-\tilde{G}_{\rm adv}$ be the causal propagator.  The covariant symplectic form on a Cauchy surface $\Sigma$ is
\begin{equation}
 \begin{aligned}
 \Omega_\Sigma(\phi_1,\phi_2)
 &=\int_\Sigma
 \left(\phi_1\nabla^a\phi_2-\phi_2\nabla^a\phi_1\right)
 \dd\Sigma_a,\\
 \phi_f&=\Omega_\Sigma(\tilde{G}f,\phi).
 \end{aligned}
 \label{eq:scalar_symplectic}
\end{equation}
Evaluating this bilinear form on two solutions is called a symplectic pairing; in the second line, it constructs the smeared observable from $\tilde{G}f$ and the field $\phi$. On shell, the symplectic current is conserved. When the past horizon $\Hh_R^-$ and past null infinity $\scri_R^-$ supply complete scattering data and no flux is lost at their ends, deforming $\Sigma$ to this boundary gives
\begin{equation}
 \phi_f
 =\Omega_{\Hh_R^-}(\tilde{G}f,\phi)+\Omega_{\scri_R^-}(\tilde{G}f,\phi).
 \label{eq:scalar_boundary_split}
\end{equation}
This decomposition underlies Eq.~\eqref{eq:wedge_boundary}: the global wedge state is specified on a complete null boundary, while a horizon cut selects only part of that data. For massive fields, bound states, or zero modes, additional data (including data at timelike infinity when required) must be included in the characteristic description. We use the massless radiative setting when writing only $\Hh_R^-\cup\scri_R^-$.

Let $U$ be affine on the complete sheet $\Hh^-$ of Fig.~\ref{fig:eternal_horizon}, with $U<0$ on its right branch and $U=0$ at the bifurcation surface $\mathcal B$. Write $\varphi=\phi|_{\Hh^-}$ and
$\Pi=\partial_U\varphi$. With $\dd\mu$ the background transverse area measure, the horizon part of the symplectic form is
\begin{equation}
 \Omega_{\Hh^-}(\varphi_1,\varphi_2)
 =\int\dd U\,\dd\mu\,
 (\varphi_1 \Pi_2-\varphi_2\Pi_1).
 \label{eq:horizon_symplectic_scalar}
\end{equation}
For a real test function $s$ on $\Hh_R^-$, let $\Pi(s):=\int_{\Hh_R^-}\dd U\,\dd\mu\,s\Pi$. The horizon factor and the complete wedge algebra are distinguished by
\begin{align}
 \M_H&:=\M(\Hh_R^-)
 =\{e^{\ii\Pi(s)}:s\in C_0^\infty(\Hh_R^-,\mathbb R)\}^{\prime\prime},
 \label{eq:scalar_Weyl}\\
\M(R)&\simeq\M_H\,\overline\otimes\,\M_\infty,
 \quad \M_\infty:=\M(\scri_R^-).
 \label{eq:complete_wedge}
\end{align}
The tensor product presupposes independent complete boundary sectors; otherwise one uses the algebra they generate. Suppressing $\M_\infty$ in notation does not remove its dynamics, entropy, or correlations with the horizon sector.

\subsubsection{Killing flow as modular flow}

For the Gaussian affine vacuum, the horizon two-point function is
\begin{equation}
 \vev{\Pi(U,y)\Pi(U',y')}_\Psi
 =-c_\Pi\frac{\delta_\mu(y,y')}{(U-U'-\ii0)^2},
 \label{eq:horizon_2pt}
\end{equation}
Here $y,y'$ label transverse points, $\delta_\mu(y,y')$ is the delta distribution with respect to $\dd\mu$, and $c_\Pi>0$ is fixed by the symplectic normalization. This is a specification of the vacuum correlation function, not just its Hadamard singularity. Under the standard local QFT hypotheses, the horizon algebra is a type-$\mathrm{III}_1$ factor. The Hadamard short-distance singularity alone does not establish thermality for an arbitrary state.

The Killing vector on the past horizon is
\begin{equation}
 \xi=-\kappa U\partial_U.
 \label{eq:Killing_horizon}
\end{equation}
The classical Killing energy that generates this flow on the radiative phase space is
\begin{equation}
 F_{\xi,\phi}^{\Hh^-}
 = -\tfrac12\Omega_{\Hh^-}(\varphi,\mathcal L_\xi\varphi)
 =-\kappa\int_{\Hh^-}\dd U\,\dd\mu \,U \, \Pi^2.
 \label{eq:scalar_flux}
\end{equation}
On the right branch, $U=-e^{-\kappa u}$ turns Killing flow into translation in $u$. The two-point function of $\partial_u\varphi$ obtained from Eq.~\eqref{eq:horizon_2pt} is analytic in the thermal strip; its boundary values at imaginary-time separation $2\pi/\kappa$ exchange the operator order. For the quasifree field vacuum, whose higher correlation functions are determined by this two-point function, this gives the Kubo--Martin--Schwinger (KMS) condition \cite{Kubo:1957,MartinSchwinger:1959,HaagHugenholtzWinnink:1967}.  The
corresponding Bisognano--Wichmann identification is 
\begin{equation}
K_{\Psi,\phi}=\beta_HF_{\xi,\phi}^{\Hh^-}.
 \label{eq:BW_horizon}
\end{equation}
This is the curved-spacetime Killing-horizon analog of boost modular flow \cite{Bisognano:1975ih,Sewell:1982,KayWald:1991}.  The automorphism is outer on the type-III algebra; adjoining its implementing translation is precisely the crossed-product step.

\subsubsection{Raychaudhuri equation and asymptotic horizon charge}

Let $\theta$ denote the expansion of the generators of $\Hh^-$ with respect to $U$. We choose the scalar sector with no independent first-order metric perturbation and impose the first-order stationary horizon boundary condition. Then $\theta^{(1)}=0$; this is a sector and boundary-condition choice, not a consequence of background stationarity alone. The second-order null Raychaudhuri equation reads
\begin{equation}
 \partial_U\theta^{(2)}=-8\pi G\,\Pi^2.
 \label{eq:Raychaudhuri_scalar}
\end{equation}
To fix the second-order convention and avoid a factor-of-two ambiguity, take a family $Y(\lambda)$ and define
\begin{equation}
 \begin{aligned}
 Y(\lambda)&=Y^{(0)}+\lambda Y^{(1)}+\lambda^2Y^{(2)}+O(\lambda^3),\\
 Y^{(2)}&=\tfrac12\left.\frac{\dd^2Y}{\dd\lambda^2}\right|_0
 =\tfrac12\delta^2Y.
 \end{aligned}
 \label{eq:perturbation_convention}
\end{equation}
The symbol $\delta$ denotes variation with respect to $\lambda$; $\lambda$ is the small parameter. We compare radiative preparations with the same background and first-order charges, so their geometric entropy differences begin at quadratic order. Higher perturbative orders are separate from the clock-width corrections considered below.

Define the local horizon charge
\begin{equation}
 q_U={1\over 4G}\int_{S_U} \dd\mu_U\,(1-U\theta). 
 \label{eq:Q_U}
\end{equation}
Here $\dd\mu_U$ is the perturbed area element on $S_U$, with $\partial_U\dd\mu_U=\theta\dd\mu_U$; $\dd\mu$ continues to denote its background value in quadratic flux integrals. Under stationary falloff $U\theta\to0$ as $U\to\pm\infty$, the asymptotic horizon charges 
\begin{equation}
 q_R := \lim_{U\rightarrow - \infty}q_U^{(2)}, \qquad q_L := \lim_{U\rightarrow \infty}q_U^{(2)}
  \label{eq:endpoint_charges}
\end{equation}
equal the corresponding second-order area coefficients divided by $4G$.
These charges are defined at the asymptotic ends of the complete null horizon. The affine coordinate is chosen so that $\mathcal B$ is at $U=0$ and the right past-horizon branch is $-\infty<U<0$. Thus $q_R$ is defined by the limiting cross-sections $S_U$ as $U\to-\infty$, at the $i_R^-$ end in the schematic asymptotically flat diagrams; $q_L$ is defined at the opposite affine end $U\to+\infty$ of the complete horizon. The bifurcation surface is a finite cross-section, not either asymptotic end. The limits require the stated falloff and regularity; they are not additional local observables at an ordinary spacetime point \cite{KudlerFlam:2023qfl}.

Combining Eq.~\eqref{eq:Raychaudhuri_scalar} and Eq.~\eqref{eq:Q_U} gives 
\begin{equation}\label{charge_balance}
 \partial_U q_U^{(2)}= 2\pi \int\dd\mu\,U \, \Pi^2 \;. 
\end{equation}
Integrating it from $U=-\infty$ to $\infty$ gives the full-horizon constraint
\begin{equation}
 q_L=q_R-  K_{\Psi,\phi},
 \label{eq:scalar_balance}
\end{equation}
where we have used Eq.~\eqref{eq:BW_horizon}.
The charge difference thus measures the integrated Killing flux; under stationary falloff it also measures the difference of the limiting area charges, as in the physical-process first law \cite{Bardeen:1973gs,IyerWald:1994,GaoWald:2001}.

\subsection{Metric perturbations on the horizon}
\label{sec:metric}

This subsection reviews the construction in \cite{Klinger:2026,Mondal:2026} of the symplectic form for metric perturbations on the horizon.

\subsubsection{Gauge-invariant observables}

Let $h_{ab}:=\delta g_{ab}=\left.\partial_\lambda g_{ab}(\lambda)\right|_{\lambda=0}$ be a solution of the linearized Einstein equation, with $g_{ab}(0)=\bar g_{ab}$.  A compactly supported symmetric tensor $f^{ab}$ defines the gauge-invariant smeared observable
\begin{equation}
 h_f=\int_M f^{ab}h_{ab}\,\epsilon_{\bar g},
 \qquad \bar\nabla_af^{ab}=0,
 \qquad f\sim f+P_{\bar g}f_0,
 \label{eq:metric_smear}
\end{equation}
where $f_0$ is a compactly supported test tensor and $P_{\bar g}$ is the gauge-fixed linearized Einstein operator on the appropriate quotient. The Lee--Wald symplectic current gives
\begin{equation}
 h_f=\Omega_\Sigma(\tilde{G}_{\bar g}f,h).
 \label{eq:metric_pairing}
\end{equation}
With residual gauge transformations removed and boundary conditions that eliminate the extra surface terms, current conservation gives the same incoming decomposition as for the scalar:
\begin{equation}
 h_f=\Omega_{\Hh_R^-}(\tilde{G}_{\bar g}f,h)
 +\Omega_{\scri_R^-}(\tilde{G}_{\bar g}f,h).
 \label{eq:metric_boundary_split}
\end{equation}
The full construction requires covariant phase-space and Noether-charge technology of Refs.~\cite{LeeWald:1990,IyerWald:1994,WaldZoupas:2000}; here we only need the radiative horizon sector.

\subsubsection{Shear/news algebra and flux}

Choose a horizon-adapted gauge in which $\Hh^-$ and its affine generator $\ell^a=(\partial_U)^a$ remain fixed, with $h_{ab}\ell^b=0$ on the horizon \cite{Hollands:2012sf,Klinger:2026}. Let $e_A^a$ be tangent to its two-dimensional cross-sections and transported along $\ell^a$. Their background metric is $q_{AB}=\bar g_{ab}e_A^ae_B^b$, with inverse $q^{AB}$ and area measure $\dd\mu$; these tensors raise and lower transverse indices. The pullback $h_{AB}=h_{ab}e_A^ae_B^b$ is the perturbation of the cross-section metric. Its trace-free part is the radiative metric datum
\[
 C^\Hh_{AB}:=h_{AB}-\tfrac12q_{AB}q^{CD}h_{CD}.
\]
Since $\partial_Uq_{AB}=0$ on the background Killing horizon, varying the decomposition of the null extrinsic curvature into expansion and shear gives $\delta\theta=\tfrac12\partial_U(q^{AB}h_{AB})$ and
\begin{equation}
 \begin{aligned}
 \delta\sigma_{AB}
 &=\tfrac12\partial_Uh_{AB}-\tfrac12q_{AB}\delta\theta
 =\tfrac12\partial_U C^\Hh_{AB},\\
 N^\Hh_{AB}&:=2\delta\sigma_{AB}=\partial_U C^\Hh_{AB},
 \quad q^{AB}N^\Hh_{AB}=0.
 \end{aligned}
 \label{eq:horizon_news}
\end{equation}
Thus $N^\Hh_{AB}$ is the horizon news, measuring the variation of the radiative metric along the null generators. The linearized Raychaudhuri equation gives $\partial_U\delta\theta=0$ in this sector; the stationary first-order boundary condition sets $\delta\theta=0$ \cite{Hollands:2012sf,Klinger:2026}.

Pulling the Einstein--Hilbert Lee--Wald symplectic current back to $\Hh^-$ in this gauge leaves the antisymmetrized product of $C^\Hh_{AB}$ and its affine derivative. With the orientation used in Eq.~\eqref{eq:horizon_symplectic_scalar}, the radiative part is \cite{LeeWald:1990,Hollands:2012sf}
\begin{equation}
 \begin{aligned}
 \Omega_{\Hh^-}(h_1,h_2)
 &=\frac{1}{32\pi G}\int_{\Hh^-}
 \dd U\,\dd\mu \\[-0.2ex]
 &\quad\times
 \left(C_1^{AB}\partial_UC_{2AB}
 -C_2^{AB}\partial_UC_{1AB}\right),
 \end{aligned}
 \label{eq:graviton_symplectic}
\end{equation}
where $C_{iAB}=C^\Hh_{AB}[h_i]$. Equation~\eqref{eq:graviton_symplectic} assumes radiative falloff or boundary conditions that remove the surface terms; independent zero modes and boundary charges require separate treatment. Weyl operators constructed from the smeared news generate the gravitational horizon factor, as in Eq.~\eqref{eq:scalar_Weyl}; the complete right-wedge algebra also includes any independent data at infinity.  The radiative affine vacuum has the corresponding Gaussian horizon correlations and, under the regularity and Killing-horizon assumptions, provides a geometric KMS modular flow. A shared Hadamard short-distance singularity alone would not establish the KMS property of an arbitrary graviton state.

The gravitational Killing generator obtained from
Eq.~\eqref{eq:graviton_symplectic} is
\begin{equation}
F_{\xi,\rm grav}^{\Hh^-}
 =-\frac{\kappa}{8 \pi G}
 \int_{\Hh^-}\dd U\,\dd\mu\,
 U\,\delta\sigma_{AB}\delta\sigma^{AB}.
 \label{eq:graviton_flux}
\end{equation}
The associated modular generator is
\be
K_{\Psi,{\rm grav}} = \beta_H F_{\xi,\rm grav}^{\Hh^-}.
\ee

The second-order Raychaudhuri equation including both scalar and metric perturbations is
\begin{equation}
\partial_U \theta^{(2)}
 = -8\pi G\,T^{(2)}_{UU}. 
 \label{eq:Raychaudhuri_metric}
\end{equation}
Here the effective quadratic flux includes both the scalar stress and the shear of the linearized metric:
\be
T^{(2)}_{UU} := \Pi^2 + {1\over 8\pi G} \delta\sigma_{AB}\delta\sigma^{AB}.  
\ee
The resulting total Killing flux is
\be
F_\xi^{\Hh^-} := F_{\xi,\phi}^{\Hh^-} + F_{\xi,{\rm grav}}^{\Hh^-} = -\kappa \int_{\Hh^-} \dd U\,\dd\mu \, U \, T^{(2)}_{UU}\;. 
\ee
Consequently, Eq.~\eqref{eq:scalar_balance} holds with the sum of matter and shear fluxes:
\begin{equation}
q_L= q_R-  K_{\Psi} 
 \label{eq:total_balance}
\end{equation}
with 
\be\label{Ktot}
K_\Psi := K_{\Psi,\phi} + K_{\Psi,{\rm grav}} = \beta_H F^{\Hh^-}_\xi. 
\ee
This calculation treats radiative modes on a smooth horizon. It does not fix independent boundary degrees of freedom, the treatment of caustics, or the matching of charges at infinity
\cite{Hollands:2012sf,Hollands:2024vbe,Mondal:2026}. The fuller symmetry and edge-mode construction can add state-dependent entropy terms \cite{Klinger:2026}; these must be included in the entropy and charge balances when those modes are active.

For a faithful product reference on the complete exterior algebra, the full modular generator is $K_{\Psi,\Hh^-}+K_{\Psi,\scri^-}$. The flux formulas just derived concern the horizon factor, for which we write $K_\Psi=K_{\Psi,\Hh^-}$. In the crossed-product formulas below, $\M$ and $K_\Psi$ must always refer to the same chosen field algebra and its full modular generator. The horizon constraint identifies the charge for $\M_H$. For $\M(R)$ we additionally assume that the total gravitational charges implement the modular action of both boundary sectors; the horizon constraint alone does not establish that identification. Independent sectors can be suppressed in entropy differences only when their contributions and correlations do not change.

Two distinctions are important for the GSL application:
\begin{enumerate}
    \item  The full-horizon flux is a well-defined generator acting on both branches of $\Hh^-$ shown in Fig.~\ref{fig:eternal_horizon}; it is not the one-sided energy flux relevant to the right-exterior area balance. The latter is formally
\begin{equation}
F_\xi^{\Hh^-_R}
 =-\kappa\int_{\Hh^-_R} \dd U\,\dd\mu\,U \, T^{(2)}_{UU}\;, 
 \label{eq:one_sided_total_flux}
\end{equation}
which is nonnegative for classical radiative data since $-U>0$ on this branch. However, a sharply one-sided quantum flux need not define a positive self-adjoint operator with finite fluctuations. 

\item The modular operator of $(\M_H,\Psi)$ is well defined in its Hilbert-space representation. A type-III factor, however, has no intrinsic trace density operator for $\Psi$ whose negative logarithm would define the right-branch energy. We therefore reserve the notation
\be\label{kPsi}
k_\Psi := \beta_H F_\xi^{\Hh^-_R}
\ee
for the one-sided energy, used through its regulated, reference-subtracted expectation values. No decomposition into two one-sided modular Hamiltonians defined by intrinsic type-III density operators is asserted \cite{Bisognano:1975ih,Sewell:1982,KayWald:1991}.

To obtain the right-branch area constraint, differentiate the charge in Eq.~\eqref{eq:Q_U}. Using $\partial_U\dd\mu_U=\theta\dd\mu_U$ and $\theta^{(1)}=0$, terms quadratic in the expansion do not contribute at second perturbative order. Thus
\begin{equation}
 \begin{aligned}
 \partial_Uq_U^{(2)}
 &=\frac{1}{4G}\int\dd\mu\,
 [\theta^{(2)}-\theta^{(2)}-U\partial_U\theta^{(2)}]\\
 &=2\pi\int\dd\mu\,U T_{UU}^{(2)},
 \end{aligned}
 \label{eq:right_branch_charge_derivative}
\end{equation}
where Eq.~\eqref{eq:Raychaudhuri_metric} includes both matter and shear flux. Integrating from $U=-\infty$ to $U=0$ and using Eq.~\eqref{kPsi}, with the asymptotic falloff $U\theta^{(2)}\to0$ in Eq.~\eqref{eq:endpoint_charges} eliminating the expansion boundary term at $U=-\infty$, gives
\begin{equation}
 q_{\mathcal B}=q_R-k_\Psi,\qquad
 q_{\mathcal B}:=\frac{A_{\mathcal B}^{(2)}}{4G}.
 \label{eq:right_branch_area_constraint}
\end{equation}
Here $q_{\mathcal B}$ is the bifurcation-area charge, with its second-order superscript suppressed as for $q_R$ and $q_L$. This relation connects the asymptotic horizon charge to the bifurcation area. In the quantum theory, we use this relation through finite, reference-subtracted expectation values with a common regulator and charge convention, not as an equality between independently defined bare operators. It is distinct from the full-horizon relation \eqref{eq:total_balance}, which contains $K_\Psi$ rather than the one-sided energy $k_\Psi$.

\end{enumerate}

\subsection{Tomita--Takesaki theory and quantum reference frame}
\label{sec:QRF}
For a field algebra $\M$ and a cyclic and separating reference vector $\Psi$, the closure of
$S_\Psi a\Psi=a^\dagger\Psi$ has polar decomposition
$S_\Psi=J_\Psi\Delta_\Psi^{1/2}$. The modular automorphisms are \cite{Takesaki:1970,Sorce:2023}
\begin{equation}
\alpha_t^\Psi(a)=\Delta_\Psi^{it}a\Delta_\Psi^{-it},
 \qquad K_\Psi=-\log\Delta_\Psi.
\end{equation}
Geometric identification with Killing flow needs the horizon vacuum/KMS and regularity assumptions; it is not true for an arbitrary state \cite{Bisognano:1975ih,Sewell:1982,KayWald:1991}. The full standard-form generator acts on both the algebra and its commutant. With $K_\Psi=\beta_HF_\xi$, the convention above corresponds to Killing time $-\beta_Ht$ for modular parameter $t$. 
 
For a second vector $\Phi$, the relative Tomita operator is defined initially on the dense domain $\M\ket\Psi$ by $S_{\Phi\Psi}a\Psi=a^\dagger\Phi$, and $\Delta_{\Phi\Psi}=S_{\Phi\Psi}^\dagger S_{\Phi\Psi}$. We also use $\Phi,\Psi$ for the normal states that these normalized vectors induce on $\M$. With this index convention, their Araki relative entropy is \cite{Araki:1976zv,Witten:2018zxz}
\begin{equation}
 D_\M(\Phi\Vert\Psi)
 :=-\bra\Phi\log\Delta_{\Psi\Phi}\ket\Phi.
 \label{eq:Araki_relative_entropy_definition}
\end{equation}
Here $\Delta_{\Psi\Phi}$ is obtained from $S_{\Psi\Phi}a\Phi=a^\dagger\Psi$, with the standard support extension for nonfaithful states. Relative entropy measures distinguishability on the specified algebra and can be infinite; we assume finiteness where entropy balances use it. Its definition requires no type-III density operator. The crossed product adds the modular clock and provides a trace representation of the same notion of relative entropy on the dressed algebra.

Before imposing the horizon charge constraint, the independent charges $q_R$ and $q_L$ serve as clock momenta conjugate to readings $p_R$ and $p_L$, with $[q_I,p_J]=-\ii\delta_{IJ}$. On the right branch, $p_R$ reads relational boost time. This implements the reference-frame idea of Refs.~\cite{AharonovSusskind:1967,AharonovKaufherr:1984,PageWootters:1983,Rovelli:1991clock,Marolf:1995clock,Popescu:2018,Giacomini:2019,DeVuyst:2024fxc}.
In the representation with only the right clock explicit, Eq.~\eqref{eq:total_balance} fixes $q_L$ in terms of $q_R$ and the field generator. The joint modular transformation of the field and translation of the right clock then has generator 
\begin{equation}
C_{\Psi} := -q_L = K_{\Psi} - q_R,
\end{equation}
so that for $\widehat{a} \in \Mh$,
\begin{equation}
    [\widehat{a}, C_{\Psi}] = 0. 
\end{equation}
The invariant relational observables associated with $a\in\M$ are given by Refs.~\cite{Hoehn:2019trinity,DeVuyst:2024fxc} (see also Eq.~(9.16) of Ref.~\cite{Liu:2025krl}):
 \begin{equation}
 O_R^t(a)=e^{\ii(p_R-t) K_\Psi}
 a e^{-\ii(p_R-t) K_\Psi}.
 \label{eq:relational_observable}
\end{equation}
$O_R^t(a)$ gives the field observable $a$ when $p_R$ reads $t$. It commutes with the diagonal constraint in the ideal-clock limit.  This is the Page--Wootters relational mechanism applied to a horizon charge \cite{PageWootters:1983,Giacomini:2019}.

The quantum clock is not merely an auxiliary ultraviolet cutoff.  The operator $q_R$ is a physical asymptotic horizon charge, related to an area variation by Eq.~\eqref{eq:endpoint_charges}, while its conjugate $p_R$ supplies the relational boost time. We denote their eigenstates by $\ket q$ and $\ket p$, respectively, with $q_R\ket q=q\ket q$. Throughout, $q_R$ denotes the operator and $q$ its real spectral value; the wavefunction $g(q)$ is a function of this value. A normalized quantum clock wavepacket
\begin{equation}
 \ket g=\int_{-\infty}^{\infty}\dd q\,g(q)\ket q,
 \qquad \int\dd q\,|g(q)|^2=1,
 \label{eq:clock_packet}
\end{equation}
specifies the preparation of the clock momentum: $|g(q)|^2$ is the probability density for measuring the horizon charge. The conjugate reading $p_R$ has the distribution determined by the Fourier transform of $g$. When the entropy expansion applies in the initial and final states of a protocol, using the same clock wave packet in both states cancels the difference of their explicit contributions $S[g]$. This cancellation does not control the other packet-dependent terms in the dynamical balance; the preparation and the channel must be checked separately.

\subsection{Crossed product, density operator, and generalized entropy}
\label{sec:crossed_product}

The clock of Section~\ref{sec:QRF} provides a quantum reference for the modular evolution of the exterior fields. We replace a field observable at a prescribed coordinate time with a relational observable~\cite{BrownKuchar:1995dust,Rovelli:2002partial,Dittrich:2007complete,FrancoisRavera:2024}, here conditioned on the clock reading as in Eq.~\eqref{eq:relational_observable}. A joint translation of the clock and modular evolution of the fields leaves this relation invariant. The crossed product describes the algebra of such dressed observables, including the asymptotic horizon charge. Thus the quantum reference frame is the physical clock and its preparation; the crossed product is the observable algebra of the exterior relative to that frame \cite{DeVuyst:2024fxc}. Its semifinite trace allows us to assign a density operator to a normal dressed state and compare its entropy along the protocols below.

\subsubsection{The modular crossed product}
\label{subsec:modular_crossed_product}

We abbreviate \emph{quantum reference frame} as QRF and use two unitarily equivalent representations of the dressed exterior. In $\mathrm{QRF}_1$, the field generators are the relational observables $O_R^0(a)$ and the horizon charge is represented by $q_R$. In $\mathrm{QRF}_0$, the field generators are $a\in\M$ and the same charge is represented by $X$. The transformation $U_\Psi=e^{-\ii K_\Psi p_R}$ conjugates observables and acts on state vectors to relate the two representations.

The dressed algebra in $\mathrm{QRF}_1$ is \cite{Liu:2025krl}
\begin{equation}
 \Mh_1=\{e^{\ii K_\Psi p_R}a e^{-\ii K_\Psi p_R},
 e^{\ii s q_R}:a\in\M,s\in\mathbb R\}^{\prime\prime},
 \label{eq:dimensionless_clock}
\end{equation}
whose commutant is
\begin{equation}
 \Mh_1'=\{a',e^{\ii s C_\Psi}:a'\in\M',s\in\mathbb R\}^{\prime\prime}.
 \label{eq:commutant_qrf1}
\end{equation}
The transformed charge in $\mathrm{QRF}_0$ is
\begin{equation}
 X:=U_\Psi q_RU_\Psi^\dagger=q_R+K_\Psi.
 \label{eq:X_definition}
\end{equation}
The dressed algebra and its commutant in $\mathrm{QRF}_0$ are then
\begin{equation}
 \begin{aligned}
 \Mh_0&:=U_\Psi\Mh_1U_\Psi^\dagger\\
 &=\{a,e^{\ii s X}:a\in\M,s\in\mathbb R\}^{\prime\prime},
 \end{aligned}
 \label{eq:crossed_rotated}
\end{equation}
\begin{equation}
 \begin{aligned}
 \Mh_0'&:=U_\Psi\Mh_1'U_\Psi^\dagger\\
 &=\{e^{-\ii K_\Psi p_R}a'e^{\ii K_\Psi p_R},e^{-\ii s q_R}:\\
 &\hspace{3em}a'\in\M',\ s\in\mathbb R\}^{\prime\prime}.
 \end{aligned}
 \label{eq:commutant_qrf0}
\end{equation}
Hereafter we express observables and states in $\mathrm{QRF}_0$ and write $\Mh:=\Mh_0$. This frame change is distinct from the physical unitary driving considered later. The modular flow of $\M$ is inner in the crossed product (it is implemented by unitaries in $\Mh$), which is a type-$\mathrm{II}_\infty$ factor when $\M$ is type $\mathrm{III}_1$ \cite{Takesaki:1973,Witten:2022qdt}.

For the formal preparation $\ket\Psi\otimes\ket{p=0}$, flat in the charge variable $q$, the positive operators representing the weights on $\Mh$ and its commutant are, respectively, $\rho_{\widehat\Psi}$ and $\rho^{\prime}_{\widehat\Psi}$:
\begin{equation}
 \begin{aligned}
 \rho_{\widehat\Psi}&=e^{-X},
 \qquad \rho'_{\widehat\Psi}=e^{-q_R},\\
 \Delta_{\widehat\Psi}
 &=\rho_{\widehat\Psi}(\rho'_{\widehat\Psi})^{-1}
 =e^{-K_\Psi}.
 \end{aligned}
 \label{eq:left_right_densities}
\end{equation}
The flat-clock vector is not normalizable. The algebra admits a faithful normal semifinite trace $\tau$, unique up to a positive scale, with the normalization specified in Appendix~\ref{app:density}. In particular,
\begin{equation}
 \tau(\id)=\int_{-\infty}^{\infty}e^q\dd q=\infty.
 \label{eq:trace_infinite}
\end{equation}
Thus $\tau$ is not a normalized tracial state. A normal state $\nu$ is represented by a positive density operator $\rho_\nu$ with $\tau(\rho_\nu)=1$. Its spectral projections belong to $\Mh$, although the operator itself may be unbounded. For normal states $\mu,\nu$, we use
\begin{align}
 \vev O_\nu=\nu(O)&=\tau(\rho_\nu O),
 \label{eq:state_trace_expectation}\\
 S_\tau(\nu)&=-\tau(\rho_\nu\log\rho_\nu),
 \label{eq:full_trace_entropy}\\
 D_\tau(\mu\Vert\nu)&=\tau[\rho_\mu(\log\rho_\mu-\log\rho_\nu)].
 \label{eq:full_relative_entropy}
\end{align}
The last expression is Araki relative entropy on $\Mh$, written using its trace. We assume the required trace expressions are finite and that the support of $\rho_\mu$ is contained in that of $\rho_\nu$; otherwise relative entropy is defined in the extended sense and is infinite when support inclusion fails. No commutativity is assumed. The same formulas apply in type-I quantum mechanics with $\tau=\Tr$. Throughout, entropy arguments denote states, while $\rho_\nu$ denotes the density operator representing $\nu$.

A normalized packet $g$ defines the normal reference state $\sigma_g$ through the vector $U_\Psi(\ket\Psi\otimes\ket g)$ in $\mathrm{QRF}_0$. Its density operator and surprisal are
\begin{equation}
 \begin{aligned}
 \rho_{\sigma_g}&=|g(X)|^2e^{-X},\qquad \tau(\rho_{\sigma_g})=1,\\
 k_g&:=-\log\rho_{\sigma_g}=X+h_g(X),\\
 h_g(X)&:=-\log|g(X)|^2.
 \end{aligned}
 \label{eq:packet_surprisal}
\end{equation}
The term $h_g(X)$ is the clock contribution to the reference surprisal; it remains fixed by $g$ even when the evolving state's packet changes. For $g\ne0$ almost everywhere, $\sigma_g$ is faithful and KMS for its own modular flow; it need not be KMS for the geometric flow generated by $X$. The factor $|g(X)|^2$ makes the reference density operator normalizable with respect to $\tau$; without this factor, $\tau(e^{-X})=\int_{\mathbb R}\dd q=\infty$.

The dressed bifurcation-area charge and one-sided modular energy in $\mathrm{QRF}_0$ are
\begin{equation}
 \mathsf{q}_{\mathcal B}:=U_\Psi q_{\mathcal B}U_\Psi^\dagger,
 \qquad \mathsf{k}_\Psi:=U_\Psi k_\Psi U_\Psi^\dagger.
 \label{eq:transported_area_energy}
\end{equation}
Conjugating the integrated constraint \eqref{eq:right_branch_area_constraint} gives
\begin{equation}
 \mathsf{q}_{\mathcal B}
 =U_\Psi(q_R-k_\Psi)U_\Psi^\dagger
 =X-\mathsf{k}_\Psi.
 \label{eq:transported_constraint_derivation}
\end{equation}
These observables and the states in which they are evaluated are henceforth expressed in $\mathrm{QRF}_0$. The one-sided energy $\mathsf{k}_\Psi$ is distinct from both the full modular generator $K_\Psi$ and the reference surprisal $k_g$; its finite, reference-subtracted interpretation and identification with exterior Killing energy are discussed in Appendix~\ref{app:charge_transport}.

Table~\ref{tab:qrf_dictionary} summarizes the physical interpretation of the crossed product. The trace belongs to the algebra, whereas the clock packet specifies the preparation.

\begin{table}[t]
\caption{Quantum reference frame interpretation of the modular crossed product.}
\label{tab:qrf_dictionary}
\small
\setlength{\tabcolsep}{3pt}
\begin{tabular}{@{}ll@{}}
\toprule
\parbox[t]{.28\linewidth}{\textbf{Physical ingredient}} &
\parbox[t]{.66\linewidth}{\textbf{Algebraic description}}\\
\midrule
\parbox[t]{.28\linewidth}{Clock} &
\parbox[t]{.66\linewidth}{Horizon charge $q_R$ and conjugate reading $p_R$.}\\[6pt]
\parbox[t]{.28\linewidth}{Dressed observables} &
\parbox[t]{.66\linewidth}{Relational fields and the horizon charge generate $\Mh$.}\\[6pt]
\parbox[t]{.28\linewidth}{Normal state} &
\parbox[t]{.66\linewidth}{$\nu(A)=\tau(\rho_\nu A)$, with $\tau(\rho_\nu)=1$.}\\[6pt]
\parbox[t]{.28\linewidth}{Clock preparation} &
\parbox[t]{.66\linewidth}{$|g|^2$ is the charge distribution; $S[g]$ is its differential entropy.}\\
\bottomrule
\end{tabular}
\end{table}

\subsubsection{Entropy and generalized entropy in quantum reference frame}
\label{subsec:qrf0_entropy}

Consider a normalized state $\ket\Phi\otimes\ket g$ in $\mathrm{QRF}_1$, and calculate its generalized entropy in $\mathrm{QRF}_0$. In the latter frame,
\begin{equation}
 \begin{aligned}
 \ket{\Phi_g^{(0)}}&=U_\Psi(\ket\Phi\otimes\ket g),\\
 \nu(\widehat A)&=\bra{\Phi_g^{(0)}}\widehat A\ket{\Phi_g^{(0)}}
 =\tau(\rho_\nu\widehat A),\qquad \widehat A\in\Mh.
 \end{aligned}
 \label{eq:density_definition}
\end{equation}
The state and density-operator conventions are those of Eqs.~\eqref{eq:state_trace_expectation}--\eqref{eq:full_relative_entropy}. Appendix~\ref{app:density} gives an explicit density operator for natural-cone preparations. The reference $\sigma_g$ uses the same packet and the field state $\Psi$; we take $g\ne0$ almost everywhere so that this reference is faithful.

Since $U_\Psi^\dagger XU_\Psi=q_R$, the distribution of $X$ in this preparation is exactly $|g(x)|^2$:
\begin{equation}
 \vev{f(X)}_\nu=\int\dd x\,|g(x)|^2f(x).
 \label{eq:qrf0_clock_distribution}
\end{equation}
This holds for bounded $f$, and for unbounded $f$ when the integral is absolutely convergent. In particular,
\begin{equation}
 \begin{aligned}
 \vev X_\nu&=\int\dd x\,x|g(x)|^2,\\
 \vev{h_g(X)}_\nu&=S[g]
 :=-\int\dd x\,|g(x)|^2\log|g(x)|^2.
 \end{aligned}
 \label{eq:clock_entropy}
\end{equation}
The differential entropy $S[g]$ describes the clock's charge distribution; it is not the von Neumann entropy of the pure clock vector. Assuming the displayed moments and entropies are finite, Eq.~\eqref{eq:full_relative_entropy} and $\log\rho_{\sigma_g}=-k_g$ give
\[
 \begin{aligned}
 D_\tau(\nu\Vert\sigma_g)
 &=\tau(\rho_\nu\log\rho_\nu)+\tau(\rho_\nu k_g)\\
 &=-S_\tau(\nu)+\vev{k_g}_\nu\\
 &=-S_\tau(\nu)+\vev X_\nu+S[g].
 \end{aligned}
\]
The last line uses $k_g=X+h_g(X)$ and Eq.~\eqref{eq:clock_entropy}; rearranging yields
\begin{equation}
 S_\tau(\nu)=\vev X_\nu+S[g]-D_\tau(\nu\Vert\sigma_g).
 \label{eq:qrf0_exact_entropy}
\end{equation}
This identity is exact for the specified preparation, before any slow-clock approximation, and requires no commutativity of the density operators.

For the slow-clock limit, keep the field state $\Phi$ fixed with finite Araki relative entropy $D_\M(\Phi\Vert\Psi)$ as defined in Eq.~\eqref{eq:Araki_relative_entropy_definition} and broaden the clock packet according to
\begin{equation}
 g_\epsilon(x)=\sqrt\epsilon\,g_0(\epsilon(x-q_0)),
 \qquad \epsilon\to0,
 \label{eq:slow_clock_packet}
\end{equation}
where $g_0$ is a fixed normalized profile and $q_0$ a fixed charge offset. The two relative entropies refer to different observable algebras. $D_\M(\Phi\Vert\Psi)$ measures how well the original field preparations can be distinguished using the exterior field algebra $\M$. By contrast, $D_\tau(\nu_\epsilon\Vert\sigma_{g_\epsilon})$ compares the corresponding dressed preparations on $\Mh$, using its trace $\tau$. Both preparations use the same clock packet, so the clock by itself carries no information about which field state was prepared. For these states,
\begin{equation}
 \begin{aligned}
 D_\tau(\nu_\epsilon\Vert\sigma_{g_\epsilon})
 &=D_\M(\Phi\Vert\Psi)-r_\epsilon(\nu_\epsilon),\\
 0\leq r_\epsilon(\nu_\epsilon)&\xrightarrow{\epsilon\to0^+}0.
 \end{aligned}
 \label{eq:slow_clock_relative_entropy}
\end{equation}
The gap $r_\epsilon$ measures the distinguishability lost by restricting joint field--clock measurements to the dressed algebra. Why does it vanish? Even a measurement of the field alone, in $\mathrm{QRF}_0$, sees only an average over modular-time shifts set by the clock. As the charge packet broadens, these shifts concentrate at zero. Appendix~\ref{app:slow_clock_limit} shows that the field measurements then recover the original field relative entropy in the limit, while no dressed measurement can exceed it. The arrow is a limit of numbers for fixed $\Phi,\Psi$ and normalized packets at every $\epsilon>0$; it asserts neither a rate nor monotonicity in $\epsilon$.

Substituting Eq.~\eqref{eq:slow_clock_relative_entropy} into Eq.~\eqref{eq:qrf0_exact_entropy}, and suppressing the $\epsilon$ labels on $\nu$ and $g$, yields the crossed-product entropy formula \cite{Witten:2022qdt,Chandrasekaran:2022eqq,KudlerFlam:2023qfl,Liu:2025krl}
\begin{equation}
 S_\tau(\nu)=\vev X_\nu-D_\M(\Phi\Vert\Psi)
 +S[g]+r_\epsilon(\nu).
 \label{eq:crossed_entropy}
\end{equation}
The remainder satisfies $r_\epsilon=o(1)$ as $\epsilon\to0$. Here $f(\epsilon)=o(1)$ means $\lim_{\epsilon\to0}f(\epsilon)=0$, without a specified rate; in contrast, $O(1)$ only means boundedness and need not tend to zero.

To identify the geometric contribution, compare $\nu$ with a fixed reference preparation $\sigma_{g_{\rm ref}}$. In this comparison $\Delta$ denotes the prepared value minus the reference value. Introduce a common ultraviolet (UV) regulator $\Lambda$ for the exterior fields, including radiative gravitons if they are quantized. A regulated field state $\mu_\Lambda$ has density operator $\rho_{\mu_\Lambda}$ and entropy $S(\mu_\Lambda)=-\Tr(\rho_{\mu_\Lambda}\log\rho_{\mu_\Lambda})$. We write $D(\mu_\Lambda\Vert\nu_\Lambda)$ for relative entropy with the ordinary trace, as in Eqs.~\eqref{eq:full_trace_entropy} and \eqref{eq:full_relative_entropy} with $\tau=\Tr$.

Apply Eq.~\eqref{eq:Araki_relative_entropy_definition} to the regulated field states $\Phi_\Lambda,\Psi_\Lambda$. In the matrix representation, $\Phi_\Lambda$ is represented by $\sqrt{\rho_{\Phi_\Lambda}}$, with inner product $\Tr(B^\dagger C)$ for matrices $B,C$. The relative modular operator acts on $B$ as $\rho_{\Psi_\Lambda}B\rho_{\Phi_\Lambda}^{-1}$ on the appropriate support. Left and right multiplication commute, so its logarithm acts as $(\log\rho_{\Psi_\Lambda})B-B\log\rho_{\Phi_\Lambda}$. Taking the expectation in $\sqrt{\rho_{\Phi_\Lambda}}$ gives \cite{Witten:2018zxz}
\[
 \begin{aligned}
 &D(\Phi_\Lambda\Vert\Psi_\Lambda)\\
 &\quad=\Tr[\rho_{\Phi_\Lambda}(\log\rho_{\Phi_\Lambda}-\log\rho_{\Psi_\Lambda})]\\
 &\quad=\Tr[(\rho_{\Phi_\Lambda}-\rho_{\Psi_\Lambda})(-\log\rho_{\Psi_\Lambda})]
       -\Delta S_{{\rm out},\Lambda}.
 \end{aligned}
\]
The last line uses $D(\Psi_\Lambda\Vert\Psi_\Lambda)=0$ and $\Delta S_{{\rm out},\Lambda}:=S(\Phi_\Lambda)-S(\Psi_\Lambda)$. For the product reference on $\Hh_R^-\cup\scri_R^-$, the reference modular Hamiltonian $-\log\rho_{\Psi_\Lambda}$ is the sum of its horizon and past-null-infinity terms, each acting on its own factor. We keep the incoming state on $\scri_R^-$ fixed at its reference value, so the infinity term cancels in the expectation difference. This condition does not constrain outgoing radiation on $\scri_R^+$.

The right-horizon reference is KMS at inverse temperature $\beta_H$ for Killing evolution. With a cutoff preserving this evolution, its regulated state $\Psi_{H,\Lambda}$ has modular Hamiltonian
\[
 -\log\rho_{\Psi_{H,\Lambda}}
 = k_{\Psi,\Lambda}+(\log Z_{H,\Lambda})\id ,
\]
where $k_{\Psi,\Lambda}$ is the regulated one-sided Killing energy in modular units [Eq.~\eqref{kPsi}]. The normalization constant cancels on subtraction, leaving $\Delta\vev{k_{\Psi,\Lambda}}$ as the energy term in the cutoff identity. Transporting the states and observable together by $U_\Psi$ preserves this difference (Appendix~\ref{app:charge_transport}).

We now remove the common cutoff, assuming that the regulated relative entropy converges to $D_\M(\Phi\Vert\Psi)$ and that the subtracted energy and entropy have finite limits \cite{CovariantRegulator:2023}. In $\mathrm{QRF}_0$ this gives
\begin{equation}
 D_\M(\Phi\Vert\Psi)
 =\Delta\vev{\mathsf{k}_\Psi}-\Delta S_{\rm out}.
 \label{eq:Araki_entropy_difference}
\end{equation}
Under the fixed-incoming-state assumption only the horizon contributes to the modular-energy difference, but $\Delta S_{\rm out}$ compares $\Phi$ and $\Psi$ on the full exterior, including correlations between the two sectors. This field-theory identity requires no slow-clock approximation.

The transported constraint \eqref{eq:transported_constraint_derivation} and the geometric identification give
\begin{equation}
 \begin{aligned}
 \Delta\vev{\mathsf{q}_{\mathcal B}}
 &=\Delta\vev X-\Delta\vev{\mathsf{k}_\Psi}\\
 &=\frac{\Delta\vev{A_{\mathcal B}^{(2)}}}{4G}.
 \end{aligned}
 \label{eq:area_clock_relation}
\end{equation}
Subtracting Eq.~\eqref{eq:crossed_entropy} for the reference, whose field relative entropy vanishes, now yields
\begin{equation}
 \begin{aligned}
 \Delta S_\tau
 &=\Delta\vev{\mathsf{q}_{\mathcal B}}+\Delta S_{\rm out}
   +\Delta S[g]+\Delta r_\epsilon\\
 &=\Delta S_{\rm gen}+\Delta S[g]+\Delta r_\epsilon,
 \end{aligned}
 \label{eq:entropy_to_area}
\end{equation}
where
\begin{equation}
 S_{\rm gen}(\mathcal B;\nu)
 =\frac{\vev{A_{\mathcal B}}_\nu}{4G}+S_{\rm out}(\nu).
 \label{eq:Sgen_bifurcation}
\end{equation}
For the transported product preparations, $S_{\rm out}(\nu)$ denotes the renormalized exterior-field entropy associated with the input $\Phi$; its finite difference is the one in Eq.~\eqref{eq:Araki_entropy_difference}. It excludes the clock. Absorbing the common state-independent constant into the generalized-entropy convention, with $\tau$ fixed, gives
\begin{equation}
 S_\tau(\nu)=S_{\rm gen}(\nu)+S[g_\nu]+r_\epsilon(\nu).
 \label{eq:entropy_equals_Sgen}
\end{equation}
For the product preparation, $g_\nu=g$. A mixed or correlated output need not have a clock wavefunction. If the packet term uses the measured charge distribution, $|g_\nu(x)|^2\dd x$ denotes the spectral probability of $X$ in that state (when it has a density); no phase or pure-state interpretation is needed. Any other model profile requires an independent justification. One must identify the exterior-field contribution and bound the remainder for the actual output, including field--clock correlations; Eq.~\eqref{eq:slow_clock_relative_entropy} supplies this control only for the stated product preparations. Independently changing asymptotic, edge, or additional frame entropies must be included separately.

For any admissible comparison $\nu_i\to\nu_f$, define $\Delta S[g]:=S[g_{\nu_f}]-S[g_{\nu_i}]$ and $\Delta r_\epsilon:=r_\epsilon(\nu_f)-r_\epsilon(\nu_i)$. Then
\begin{equation}
 \Delta S_{\rm gen}
 =S_\tau(\nu_f)-S_\tau(\nu_i)
 -\Delta S[g]-\Delta r_\epsilon.
 \label{eq:generic_entropy_target}
\end{equation}
We call Eqs.~\eqref{eq:entropy_equals_Sgen} and \eqref{eq:generic_entropy_target} the \emph{entropy dictionary}: they relate dressed trace entropy to generalized entropy, with the packet contribution and remainder displayed separately. The initial and final packets can differ. All comparisons use the same trace normalization and additive entropy convention.

\section{Unitary driving and relaxation: a conditional generalized second law}
\label{sec:FT}
The main result of this section is a conditional GSL for a separated protocol on the semifinite dressed right-wedge algebra constructed in Section~\ref{sec:construction}: a unitary drive is followed by relaxation. Section~\ref{sec:regulated_GSL} uses a UV-regulated QFT as a pedagogical example to show how work, exterior energy, and horizon area enter the argument. Section~\ref{sec:typeI_FT} derives the unitary surprisal fluctuation theorem in semifinite trace notation and identifies its mean with the relative entropy generated by the unitary drive. Section~\ref{sec:FT_to_GSL} combines data processing with the dressed entropy dictionary and relates the relaxation energy balance to the horizon area response, keeping the clock contributions explicit. Section~\ref{sec:GSL_proof} states the assumptions for the conditional GSL and estimates the corrections for a slow Gaussian clock.

\subsection{The separated GSL argument with a UV cutoff}
\label{sec:regulated_GSL}

We use the UV cutoff $\Lambda$ and the state notation of Section~\ref{subsec:qrf0_entropy}. Here the matter states $\sigma_\Lambda$, $\omega_\Lambda$, and $\chi_\Lambda$ describe the exterior before the unitary drive, after the unitary drive, and after relaxation, respectively. With any additional infrared regulator needed for a finite partition function, we take the initial state to have the Gibbs density operator
\begin{equation}
 \begin{aligned}
 \rho_{\sigma_\Lambda}&=Z_\Lambda^{-1}
 e^{-\beta_H H_{{\rm out},\Lambda}},\\
 Z_\Lambda&:=\Tr e^{-\beta_H H_{{\rm out},\Lambda}},
 \end{aligned}
 \label{eq:regulated_Gibbs}
\end{equation}
where $H_{{\rm out},\Lambda}$ is the regulated exterior Killing Hamiltonian. A unitary drive prepares $\omega_\Lambda$ with density operator $\rho_{\omega_\Lambda}=U_\Lambda\rho_{\sigma_\Lambda}U_\Lambda^\dagger$, and subsequent relaxation gives $\rho_{\chi_\Lambda}=\E_\Lambda(\rho_{\omega_\Lambda})$. We assume that $\E_\Lambda$ is completely positive and trace-preserving, with $\E_\Lambda(\rho_{\sigma_\Lambda})=\rho_{\sigma_\Lambda}$. All energy changes use the same $H_{{\rm out},\Lambda}$. For the full protocol the change in the fine-grained entropy of the regulated exterior fields is $\Delta S_{{\rm out},\Lambda}=S(\chi_\Lambda)-S(\sigma_\Lambda)$.

\begin{enumerate}
 \item \emph{Work and energy balance.} With negligible heat transfer during the unitary drive, its mean supplied work is $W=\Tr[(\rho_{\omega_\Lambda}-\rho_{\sigma_\Lambda})H_{{\rm out},\Lambda}]$. The unitary drive preserves entropy, $S(\omega_\Lambda)=S(\sigma_\Lambda)$. Substituting the Gibbs logarithm in the relative entropy gives
 \begin{equation}
  \beta_H W
  =D(\omega_\Lambda\Vert\sigma_\Lambda)
  =\beta_H(E_H+\Delta E_{\rm out}).
  \label{eq:regulated_work}
 \end{equation}
 The first equality is the mean surprisal-work identity derived in Section~\ref{sec:typeI_FT}. The second uses Killing-energy balance: no further work is supplied during relaxation, $E_H$ is the energy absorbed by the horizon, and $\Delta E_{\rm out}=\Tr[(\rho_{\chi_\Lambda}-\rho_{\sigma_\Lambda})H_{{\rm out},\Lambda}]$ is the change in the total exterior Killing energy, including radiation at future null infinity. There is no uncounted flux to infinity.
 \item \emph{Horizon response.} At second perturbative order, the physical-process first law, with the appropriate horizon boundary conditions and common background normalization, gives
 \begin{equation}
  \frac{\Delta A}{4G}=\beta_H E_H .
  \label{eq:regulated_area}
 \end{equation}
 This step imports the gravitational response; the QFT cutoff does not imply it.
 \item \emph{Relative entropy left in the exterior.} The final state need not be thermal. Apply the cutoff identity preceding Eq.~\eqref{eq:Araki_entropy_difference} to the pair $(\chi_\Lambda,\sigma_\Lambda)$. By Eq.~\eqref{eq:regulated_Gibbs}, $-\log\rho_{\sigma_\Lambda}=\beta_H H_{{\rm out},\Lambda}+\log Z_\Lambda$; the constant drops out of the expectation difference, giving
 \begin{equation}
  D(\chi_\Lambda\Vert\sigma_\Lambda)
  =\beta_H\Delta E_{\rm out}-\Delta S_{{\rm out},\Lambda}.
  \label{eq:regulated_final_relative}
 \end{equation}
 \item \emph{Generalized entropy and DPI.} Subtract Eq.~\eqref{eq:regulated_final_relative} from Eq.~\eqref{eq:regulated_work} and use Eq.~\eqref{eq:regulated_area}. Then
 \begin{equation}
  \begin{aligned}
  \Delta S_{{\rm gen},\Lambda}
   &:=\frac{\Delta A}{4G}+\Delta S_{{\rm out},\Lambda}\\
   &=D(\omega_\Lambda\Vert\sigma_\Lambda)
    -D(\chi_\Lambda\Vert\sigma_\Lambda)\geq0 .
  \end{aligned}
  \label{eq:regulated_GSL}
 \end{equation}
 The inequality follows from DPI applied to $\E_\Lambda$, since $\E_\Lambda(\rho_{\sigma_\Lambda})=\rho_{\sigma_\Lambda}$.
\end{enumerate}

This proof shows how the exterior energy term cancels, leaving the horizon area gain plus the exterior entropy change. It assumes a regulated Gibbs initial state, relaxation that preserves it in the absence of driving, and that the energy lost by the exterior is the flux responsible for the area gain. It does not construct the continuum exterior density operator, prove that a physical channel has the required fixed point, or justify removing the cutoff. In the dressed formulation, the reference surprisal also contains the clock packet, and the trace entropy already includes the area term. The following subsections use the entropy dictionary of Section~\ref{sec:construction} and explicitly include the clock corrections; Eq.~\eqref{eq:regulated_GSL} cannot simply be applied to a type-III algebra.

\subsection{Surprisal fluctuation theorem}
\label{sec:typeI_FT}
Fluctuation theorems compare the statistics of a forward process with those of a specified reverse process, yielding average inequalities of second-law type. Familiar examples concern work statistics \cite{Jarzynski:1997,Crooks:1999,Kurchan:2000,Tasaki:2000,Campisi:2011}. Repeated measurements of the entropy observable $-\log\rho_\sigma$ are also well established in entropic fluctuation theory and its modular formulation \cite{Jaksic:2011entropic,Benoist:2024two}; related work explicitly discusses surprisal entropy-production statistics \cite{Upadhyaya:2024surprisal}. We refer to the fixed-reference version as the \emph{surprisal fluctuation theorem}. Its trace-based derivation applies both to ordinary type-I quantum mechanics and to our type-II dressed algebra, using the usual operator trace in the former and the semifinite trace $\tau$ in the latter. In both cases, normalization and trace cyclicity yield the same FT identities. For the dressed algebra, the required operator domains, finite trace expressions, and spectral limits must be controlled, and the entropy dictionary relates the resulting mean entropy balance to generalized entropy including the clock corrections.

Using the trace conventions of Eqs.~\eqref{eq:state_trace_expectation}--\eqref{eq:full_relative_entropy}, let $\sigma$ be a faithful reference state with $\tau(\rho_\sigma)=1$. The surprisal operator and its expectation are
\begin{equation}
 k_\sigma:=-\log\rho_\sigma,\qquad
 S_\tau(\sigma)=\tau(\rho_\sigma k_\sigma).
 \label{eq:surprisal_operator_pedagogical}
\end{equation}
This is the quantum counterpart of the information $-\log p$ associated with an outcome of probability $p$. It is also the modular Hamiltonian defined by the density operator, distinct from the full generator $-\log\Delta_\sigma$. Write
\begin{equation}
 k_\sigma=\sum_nk_nP_n,\qquad
 \rho_\sigma=\sum_ne^{-k_n}P_n.
 \label{eq:surprisal_spectrum_pedagogical}
\end{equation}
The outcome probability is $e^{-k_n}\tau(P_n)$: $e^{-k_n}$ is an eigenvalue of the density operator, while $\tau(P_n)$ is the trace of the entire eigenspace. The sums describe discrete spectra; for continuous spectra we assume that the joint spectral measure and the indicated moments exist. All trace expressions, entropies, and exponential moments used below are assumed well defined.

A unitary drive $U$ prepares $\rho_\omega=U\rho_\sigma U^\dagger$. Measure the same reference observable $k_\sigma$ immediately before and after the unitary drive. The two outcomes $n,m$ define the dimensionless surprisal change, or modular work, $w=k_m-k_n$. Write $P_F(n,m)$ for the probability of the ordered outcomes (initial, final) and $\mathcal P_F(w)$ for the distribution of their difference:
\begin{align}
 P_F(n,m)&=\tau(P_mUP_n\rho_\sigma P_nU^\dagger),\\
 \mathcal P_F(w)&=\sum_{mn}P_F(n,m)\delta(w-k_m+k_n).
 \label{eq:surprisal_distribution_pedagogical}
\end{align}
Thus $w$ fluctuates between repeated realizations of the two-measurement experiment even though $U$ is fixed. As in quantum work statistics, it is defined by two measurements rather than by a single work observable \cite{Talkner:2007sf}.

Since $[k_\sigma,\rho_\sigma]=0$, the first measurement leaves the unconditioned initial state unchanged. The characteristic function consequently takes the full-counting-statistics form
\begin{equation}
 \begin{aligned}
 G(u)&:=\vev{e^{\ii uw}}_F\\
 &=\tau(e^{\ii uk_\sigma}Ue^{-\ii uk_\sigma}\rho_\sigma U^\dagger)\\
 &=\tau(\rho_\sigma^{-\ii u}U\rho_\sigma^{1+\ii u}U^\dagger).
 \end{aligned}
 \label{eq:char_function}
\end{equation}
For the reverse experiment, prepare $\sigma$ again and replace $U$ by $U^\dagger$. Its joint law is $P_R(m,n)=\tau(P_nU^\dagger P_m\rho_\sigma P_mU)$. Using Eq.~\eqref{eq:surprisal_spectrum_pedagogical} and trace cyclicity gives $P_F(n,m)=e^{k_m-k_n}P_R(m,n)$. The reversed outcome pair has change $-w$, so on common support
\begin{equation}
 \mathcal P_F(w)=e^w\mathcal P_R(-w),\qquad
 \vev{e^{-w}}_F=\sum_{mn}P_R(m,n)=1.
 \label{eq:IFT_work}
\end{equation}
The integral identity also follows by setting $u=\ii$ in Eq.~\eqref{eq:char_function}. This specifies an inverse-unitary reverse protocol; identifying it with a physical time reversal can require reversing the external controls as well.

Jensen's inequality implies $\vev w_F\geq0$. To understand this average, note that unitary conjugation preserves the spectrum of the density operator and hence
\begin{equation}
 S_\tau(\omega)=S_\tau(\sigma).
 \label{unitary_SvN}
\end{equation}
Taking the first moment and using this entropy equality yields
\begin{equation}
 \begin{aligned}
 \vev w_F&=-\ii G'(0)
 =\vev{k_\sigma}_\omega-\vev{k_\sigma}_\sigma\\
 &=\tau[\rho_\omega(\log\rho_\omega-\log\rho_\sigma)]\\
 &=D_\tau(\omega\Vert\sigma)\geq0.
 \end{aligned}
 \label{w_ave_relative}
\end{equation}
The nonnegative mean measures the distinguishability of the driven state from the reference, while the fine-grained entropy is unchanged. For the regulated Gibbs density operator in Eq.~\eqref{eq:regulated_Gibbs}, $k_{\sigma_\Lambda}=\beta_H H_{{\rm out},\Lambda}+\log Z_\Lambda$. Measuring the same $H_{{\rm out},\Lambda}$ before and after the unitary drive therefore gives $\vev{w}_F=\beta_H W$, where $W=\Tr[(\rho_{\omega_\Lambda}-\rho_{\sigma_\Lambda})H_{{\rm out},\Lambda}]$ is the mean supplied work when heat transfer during the unitary drive is negligible, as in Eq.~\eqref{eq:regulated_work}. Here work is measured relative to the same Killing Hamiltonian at both times; nonnegative mean work then expresses thermal passivity \cite{PuszWoronowicz:1978}. The general surprisal identity requires only the same reference surprisal in both measurements, with no Gibbs interpretation or restriction on the driving unitary.

For the dressed preparation in Eq.~\eqref{eq:packet_surprisal}, the same formulas apply with $k_\sigma=k_g=X+h_g(X)$. Both measurements use the same $k_g$, even if the clock distribution changes. Section~\ref{sec:fullcore_FT} instead measures the initial and final states' own surprisals. The following subsections explain how this modular work is related to physical energy transfer and generalized entropy, including the horizon response and controlled slow-clock corrections.

\subsection{The full dressed entropy balance}
\label{sec:FT_to_GSL}
We now connect the surprisal FT to generalized entropy. We first describe the exterior as an open system and explain when its evolution can be separated into a short unitary drive followed by relaxation. We then use DPI to derive an entropy balance on the full crossed product, including the clock-packet contributions and the entropy remainder. Finally, we use the horizon energy/area response to identify the geometric term in this balance and explain under which additional assumptions modular work reduces to physical work.

\subsubsection{Subsystem, states, and two different entropies}
We use the fixed dressed algebra $\Mh(R)$, trace $\tau$, and definitions in Eqs.~\eqref{eq:state_trace_expectation}--\eqref{eq:full_relative_entropy}. The trace entropy $S_\tau$ is that of the full dressed state. When the entropy dictionary \eqref{eq:entropy_equals_Sgen} applies, it contains the area contribution as well as the packet entropy and the remainder. By contrast, $\Delta S_{\rm out}$ is the reference-subtracted exterior field entropy evaluated with a common regulator. State symbols label the entropy arguments; density operators are used to express the channel evolution.

Physically, the reduced exterior description need not include all degrees of freedom participating in scattering. In a regulated system, let $R$ include the exterior fields and their gravitational clock, and let $B$ collect the omitted degrees of freedom described in Section~\ref{sec:proposal_overview}, such as infalling modes whose post-crossing observables lie inside the black hole. We trace over $B$ after the interaction. An initially uncorrelated preparation and a joint unitary scattering operator $V$ give
\[
 \begin{aligned}
 \rho_{RB}^{\rm fin}&=V(\rho_R\otimes\eta_B)V^\dagger,\\
 \mathcal C_{\rm BH}(\rho_R)&=\operatorname{Tr}_B\rho_{RB}^{\rm fin}.
 \end{aligned}
\]
The environment state $\eta_B$ is fixed independently of the input $\rho_R$. A physical black-hole application must specify $V$ from the perturbative scattering dynamics. It can include overlapping unitary driving and relaxation; their separation is not a property of a generic channel. Exterior measurements determine the reduced state but cannot in general reconstruct correlations with $B$ after scattering.

This construction illustrates a unitary dilation: an open evolution is represented by coupling to a larger system and then restricting the predictions to observables of the dressed exterior. The algebraic counterpart is Stinespring's representation of the Heisenberg channel as a compression of a representation on a larger Hilbert space. That larger space is a mathematical realization and need not be a physical black-hole interior. In continuum QFT, we therefore formulate the channel directly on $\Mh(R)$, without assuming a physical inside--outside tensor product or an intrinsic partial trace across the horizon.

We begin with the separated protocol because it isolates the two ingredients of the argument: the unitary drive produces the modular work measured in Section~\ref{sec:typeI_FT}, and relaxation reduces distinguishability from the reference state. We first study this separated protocol as an approximation to the exterior evolution when dissipative effects during the interval of unitary driving are negligible. A duration of unitary driving short compared with the relaxation timescale motivates this approximation. The geometric comparison used for the horizon response and the relation between modular work and physical work are specified below. We write
\begin{equation}
 \begin{aligned}
 \mathcal C_{\rm BH}&=\E\circ\Ad_U,\\
 \rho_\omega&=U\rho_{\sigma_g}U^\dagger,\qquad
 \rho_\chi=\E(\rho_\omega).
 \end{aligned}
 \label{eq:BH_channel_stages}
\end{equation}
Here $\Ad_U(\rho):=U\rho U^\dagger$ denotes the unitary channel acting on density operators in the Schr\"odinger picture. The sequence of dressed density operators is
\[
 \rho_{\sigma_g}
 \xrightarrow{\;\mathrm{Ad}_U\;}
 \rho_\omega
 \xrightarrow{\;\E\;}
 \rho_\chi.
\]
The two-time experiment of Section~\ref{sec:typeI_FT} is performed on separate copies of the initial preparation. In the protocol above there is no intervening measurement: relaxation acts on $U\rho_{\sigma_g}U^\dagger$, whose coherences between different $k_g$ eigenspaces could be lost in a measurement. The clock distribution may change in either stage, and the unitary drive or relaxation may create correlations between fields and clock. In the regulated picture, the relaxation map itself can be represented as $\E(\rho)=\operatorname{Tr}_B[V_{\rm rel}(\rho\otimes\eta_B)V_{\rm rel}^\dagger]$.

Assume that the Schr\"odinger-picture map $\E$ is completely positive and trace-preserving, with a normal Heisenberg adjoint, and that
\begin{equation}
 \E(\rho_{\sigma_g})=\rho_{\sigma_g}.
 \label{stationary_c}
\end{equation}
This says that relaxation leaves the chosen dressed reference preparation unchanged. It is an additional assumption about the physical channel, not a consequence of the KMS condition alone. The full channel, including unitary driving and relaxation, need not fix $\sigma_g$. Overlapping unitary driving and relaxation are treated in Section~\ref{sec:simultaneous}.

Unless otherwise indicated, $\Delta$ compares $\chi$ with $\sigma_g$, $\Delta_{\rm rel}$ compares $\chi$ with $\omega$, and $\Delta_{\rm drv}$ compares $\omega$ with $\sigma_g$. All entropies, expectation values, and differences below satisfy the necessary support and domain conditions.

\subsubsection{Exact contraction and clock correction}
Define the loss of relative entropy with respect to the fixed preparation $\sigma_g$ during relaxation:
\begin{equation}
 \Sigma_D:=D_\tau(\omega\Vert\sigma_g)
 -D_\tau(\chi\Vert\sigma_g)\geq0.
 \label{eq:channel_mean_fixed}
\end{equation}
The inequality follows from DPI and Eq.~\eqref{stationary_c}. The first two steps in the derivation of Eq.~\eqref{eq:qrf0_exact_entropy} hold for any admissible state, independently of its clock preparation, and give
\begin{equation}
 D_\tau(\nu\Vert\sigma_g)=\vev{k_g}_\nu-S_\tau(\nu).
 \label{eq:relative_surprisal_identity}
\end{equation}
The reference subtraction is implicit here: $\vev{k_g}_{\sigma_g}=S_\tau(\sigma_g)$. Appendix~\ref{app:reference_subtraction} spells out the equivalent identity after subtracting the reference.
Subtract this identity for $\omega$ and $\chi$. Together with $S_\tau(\omega)=S_\tau(\sigma_g)$, it yields
\begin{equation}
 \Delta S_\tau=\Sigma_D+\Delta_{\rm rel}\vev{k_g}.
 \label{eq:exact_core_balance}
\end{equation}
To obtain generalized entropy, apply the common difference identity \eqref{eq:generic_entropy_target}, derived from Eq.~\eqref{eq:entropy_equals_Sgen}, to the initial state $\sigma_g$ and final state $\chi$. The corresponding packet and remainder differences are $\Delta S[g]=S[g_\chi]-S[g]$ and $\Delta r_\epsilon=r_\epsilon(\chi)-r_\epsilon(\sigma_g)$. These compare the whole protocol, whereas $\Delta_{\rm rel}$ compares $\chi$ with the state $\omega$ prepared by the unitary drive.

Since $k_g=X+h_g(X)$, combining Eqs.~\eqref{eq:exact_core_balance} and \eqref{eq:generic_entropy_target} gives the generalized-entropy balance with clock corrections
\begin{equation}
 \begin{aligned}
 \Delta S_{\rm gen}={}&\Sigma_D+\Delta_{\rm rel}\vev X
 +\Delta_{\rm rel}\vev{h_g(X)}\\
 &-\Delta S[g]-\Delta r_\epsilon.
 \end{aligned}
 \label{eq:channel_Sgen_correction}
\end{equation}
The two relaxation terms use the same operators $X$ and $h_g(X)$ in $\omega$ and $\chi$; $h_g$ is determined by the initial packet even if the final packet differs. The mean-charge contribution vanishes for the geometric comparison below. Given the entropy dictionary \eqref{eq:generic_entropy_target} for the actual initial and final states, the balance includes the full remainder difference $\Delta r_\epsilon$; its size requires an independent estimate.

\subsubsection{Area response and horizon boundary conditions}
\label{subsec:area_endpoint}
In $\mathrm{QRF}_0$, the horizon constraint already derived in Eq.~\eqref{eq:transported_constraint_derivation} expresses the dressed asymptotic charge as
\begin{equation}
 X=\mathsf{q}_{\mathcal B}+\mathsf{k}_\Psi.
 \label{eq:rotated_area_constraint}
\end{equation}
Here $\mathsf{q}_{\mathcal B}$ and $\mathsf{k}_\Psi$ are the bifurcation-area charge and one-sided modular energy defined in Eq.~\eqref{eq:transported_area_energy}; $\mathsf{k}_\Psi$ is distinct from the full modular generator $K_\Psi$. For the complete exterior, this identity requires the total-charge matching described in Section~\ref{sec:metric}, including every active asymptotic contribution. We suppress the common reference subtraction and, where necessary, interpret charge identities through finite expectation-value differences (Appendix~\ref{app:charge_transport}).

Write the one-sided energy in physical units and denote its expectation by
\begin{equation}
 \mathsf{k}_\Psi=\beta_H H_{\rm out},\qquad
 E_{\rm out}(\nu):=\vev{H_{\rm out}}_\nu.
 \label{eq:exterior_energy_definition}
\end{equation}
Identifying $H_{\rm out}$ with the physical exterior Killing generator requires the charge-and-scattering correspondence discussed in Appendix~\ref{app:charge_transport}. Assuming that this correspondence holds and that all additional boundary charges remain unchanged during relaxation, Eq.~\eqref{eq:rotated_area_constraint} implies
\begin{equation}
 \Delta_{\rm rel}\vev X
 =\Delta_{\rm rel}\vev{\mathsf{q}_{\mathcal B}}
 +\beta_H\Delta_{\rm rel}E_{\rm out}.
 \label{eq:combined_charge_balance}
\end{equation}

We evaluate the horizon response from a regular bifurcation cross-section $C_i$ to a cross-section $C_f$ in the stationary future limit. On $\Hh_R^+$, let $V$ be affine, with $V_i=0$ and $\xi=\kappa V\partial_V$; it is distinct from the past-horizon parameter $U$ defining the incoming algebra. At the second perturbative order used in Section~\ref{sec:scalar}, $\theta^{(1)}=0$, and
\[
 \begin{aligned}
 \partial_V\theta^{(2)}&=-8\pi G\,T^{(2)}_{VV},\\
 E_H&=\kappa\int_{V_i}^{V_f}\dd V\int\dd\mu_0\,V T^{(2)}_{VV},
 \end{aligned}
\]
where $\dd\mu_0$ is the background area element. Here $T^{(2)}_{VV}$ is the effective focusing source: the matter null stress plus the gravitational shear term $\sigma^{(1)}_{AB}\sigma^{(1)AB}/(8\pi G)$, evaluated with the common reference subtraction in quantum expectation values. Thus $E_H$ is the net absorbed Killing energy, consistently with Eq.~\eqref{eq:regulated_area}. Since $\partial_V(\dd\mu)^{(2)}=\theta^{(2)}\dd\mu_0$, the area difference is $\Delta A^{(2)}:=A^{(2)}[C_f]-A^{(2)}[C_i]=\int_{V_i}^{V_f}\dd V\int\dd\mu_0\,\theta^{(2)}$.

Using $\beta_H\kappa=2\pi$ and integrating by parts gives \cite{GaoWald:2001,Mishra:2018}
\begin{equation}
 \begin{aligned}
 \beta_HE_H
 &=-\frac{1}{4G}\int\dd\mu_0\int_{V_i}^{V_f}\dd V\,
 V\partial_V\theta^{(2)}\\
 &=\frac{\Delta A^{(2)}}{4G}
 -\frac{1}{4G}\int\dd\mu_0[V\theta^{(2)}]_{V_i}^{V_f}.
 \end{aligned}
 \label{eq:future_horizon_area_balance}
\end{equation}
Regularity at $V_i=0$ gives $\int_{C_i}V\theta^{(2)}\dd\mu_0=0$, and the stationary future limit is taken with $\int_{C_f}V\theta^{(2)}\dd\mu_0\to0$. Both boundary integrals therefore vanish. To represent this geometric process by the relaxation $\omega\to\chi$, we match the dressed area-charge difference to the area change between these cross-sections:
\begin{equation}
 \Delta_{\rm rel}\vev{\mathsf{q}_{\mathcal B}}
 =\frac{\Delta\vev{A^{(2)}}}{4G},
 \label{eq:relaxation_area_identification}
\end{equation}
with a common reference subtraction. This matching of channel states to geometric data is distinct from the representation change $U_\Psi$. Combining Eqs.~\eqref{eq:future_horizon_area_balance} and \eqref{eq:relaxation_area_identification} yields
\begin{equation}
 \Delta_{\rm rel}\vev{\mathsf{q}_{\mathcal B}}=\beta_HE_H.
 \label{eq:FT_GSL_area_response}
\end{equation}
The result is understood through semiclassical expectation values at second perturbative order, with regular horizon generators throughout the comparison.
For relaxation without external work or unaccounted outgoing flux, conservation of Killing energy gives
\begin{equation}
 \Delta_{\rm rel}E_{\rm out}=-E_H.
 \label{eq:relaxation_energy_conservation}
\end{equation}
Here $\Delta_{\rm rel}E_{\rm out}=E_{\rm out}(\chi)-E_{\rm out}(\omega)$ measures the change during relaxation. It is distinct from $\Delta E_{\rm out}=E_{\rm out}(\chi)-E_{\rm out}(\sigma_g)$, the full-protocol difference used in its regulated form in Eq.~\eqref{eq:regulated_work}; the two satisfy $\Delta E_{\rm out}=\Delta_{\rm drv}E_{\rm out}+\Delta_{\rm rel}E_{\rm out}$. Radiation escaping to infinity must remain within the exterior energy account or be included explicitly as an additional flux. Substituting Eqs.~\eqref{eq:FT_GSL_area_response} and \eqref{eq:relaxation_energy_conservation} into Eq.~\eqref{eq:combined_charge_balance} therefore gives
\begin{equation}
 \Delta_{\rm rel}\vev X=\beta_HE_H-\beta_HE_H=0.
 \label{eq:relaxation_charge_conservation}
\end{equation}
Thus relaxation preserves the mean combined area--field charge for the states $\omega$ and $\chi$ to this order. This is weaker than conservation of the whole distribution of $X$ for every input state. For other horizon cross-sections, Eq.~\eqref{eq:future_horizon_area_balance} retains the expansion boundary contribution; any residual area--energy mismatch contributes to $\Delta_{\rm rel}\vev X$.

\subsubsection{Relative entropies in terms of work, energy, matter entropy, and clock correction}
We now express the relative entropies after unitary driving and after relaxation in physical terms, identifying the work supplied by the unitary drive and the exterior energy--entropy contribution that remains after relaxation, with the clock corrections included explicitly.

For the unitary drive, $S_\tau(\omega)=S_\tau(\sigma_g)$ implies $D_\tau(\omega\Vert\sigma_g)=\Delta_{\rm drv}\vev{k_g}$ (see Appendix~\ref{app:reference_subtraction}). Using $k_g=X+h_g(X)$ and $X=\mathsf{q}_{\mathcal B}+\mathsf{k}_\Psi$, and matching $\Delta_{\rm drv}\vev{\mathsf{k}_\Psi}=\beta_H W$ with $W:=\Delta_{\rm drv}E_{\rm out}$, gives
\[
 D_\tau(\omega\Vert\sigma_g)
 =\Delta_{\rm drv}\vev{\mathsf{q}_{\mathcal B}}+\beta_H W
 +\Delta_{\rm drv}\vev{h_g(X)}.
\]
Here $W$ is the supplied physical work when heat transfer during the unitary drive is negligible. If the area-charge change caused by the unitary drive vanishes, this reduces to
\begin{equation}
 D_\tau(\omega\Vert\sigma_g)=\beta_HW
 +\Delta_{\rm drv}\vev{h_g(X)}.
 \label{eq:physical_modular_work}
\end{equation}
Thus the mean modular work created by the unitary drive is the physical Killing work in thermal units, plus the explicit clock-packet correction. The condition $\Delta_{\rm drv}\vev{\mathsf{q}_{\mathcal B}}=0$ is an additional assumption, not a consequence of a short unitary drive.

For the final state, the reference-subtracted relative-entropy identity and the entropy dictionary give
\[
 \begin{aligned}
 D_\tau(\chi\Vert\sigma_g)
 ={}&\Delta\vev X-\Delta S_{\rm gen}\\
 &+\Delta\vev{h_g(X)}-\Delta S[g]-\Delta r_\epsilon.
 \end{aligned}
\]
When the same area charge and exterior Killing energy enter both the horizon constraint and the entropy dictionary, $\Delta\vev X=\Delta\vev{\mathsf{q}_{\mathcal B}}+\beta_H\Delta E_{\rm out}$ and $\Delta S_{\rm gen}=\Delta\vev{\mathsf{q}_{\mathcal B}}+\Delta S_{\rm out}$. The area contributions therefore cancel directly, yielding
\begin{equation}
 \begin{aligned}
 D_\tau(\chi\Vert\sigma_g)={}&\beta_H\Delta E_{\rm out}
 -\Delta S_{\rm out}\\
 &+\Delta\vev{h_g(X)}-\Delta S[g]-\Delta r_\epsilon.
 \end{aligned}
 \label{eq:corrected_late_relative}
\end{equation}
The final relative entropy is then expressed in terms of exterior Killing energy and exterior field entropy (matter entropy in the scalar model), together with the clock-packet terms and entropy remainder. The appearance of $S_{\rm out}$ results from this area cancellation; the entropy of the dressed state remains $S_\tau$. This derivation for the final state does not require the area change during the unitary drive to vanish.

\subsection{Conditional GSL and Gaussian clock bounds}
\label{sec:GSL_proof}
We consider the separated physical process of Section~\ref{subsec:area_endpoint}, for which the area response and Killing-energy conservation give $\Delta_{\rm rel}\vev X=0$ by Eq.~\eqref{eq:relaxation_charge_conservation} at second perturbative order. All GSL bounds in this section are at that perturbative order; the errors estimated below are clock errors. We collect the algebraic and state assumptions used below:
\begin{enumerate}
 \item[(A1)] A fixed dressed algebra $\Mh(R)$ and a fixed faithful normal semifinite trace $\tau$ describe all stages of the process. The initial state is the normalized faithful preparation $\sigma_g$ with surprisal $k_g=X+h_g(X)$.
 \item[(A2)] The separated protocol is $\rho_\omega=U\rho_{\sigma_g}U^\dagger$ followed by $\rho_\chi=\E(\rho_\omega)$, where $\E$ is completely positive and trace preserving and fixes $\rho_{\sigma_g}$, as in Eqs.~\eqref{eq:BH_channel_stages}--\eqref{stationary_c}. If this protocol approximates the physical evolution, we must also control its errors.
 \item[(A3)] The required relative entropies, entropy differences, and reference-subtracted charge expectation values are finite and satisfy the support and domain conditions. The reference and final states obey the entropy dictionary \eqref{eq:entropy_equals_Sgen}, with the same trace and reference convention, identified clock distributions $|g_\nu|^2$ in that dictionary, no independent entropy change in any additional sector, and independently controlled entropy remainders in both states.
\end{enumerate}

\begin{proposition}[Clock-controlled generalized-entropy bound]
\label{prop:GSL}
For this separated physical process, under assumptions (A1)--(A3) and using the charge balance \eqref{eq:relaxation_charge_conservation}, the generalized-entropy change obeys
\begin{equation}
 |\Delta S_{\rm gen}-\Sigma_D|\leq\eta,
 \qquad \Delta S_{\rm gen}\geq-\eta,
 \label{eq:GSL_theorem}
\end{equation}
where $\Sigma_D=D_\tau(\omega\Vert\sigma_g)-D_\tau(\chi\Vert\sigma_g)\geq0$ and $\eta:=\eta_h+\eta_{\rm pkt}+\eta_r$, provided the nonnegative error bounds satisfy
\begin{equation}
 |\Delta_{\rm rel}\vev{h_g(X)}|\leq\eta_h,
 \label{eq:GSL_charge_error_bounds}
\end{equation}
\begin{equation}
 |\Delta S[g]|\leq\eta_{\rm pkt},\qquad
 |\Delta r_\epsilon|\leq\eta_r.
 \label{eq:GSL_frame_error_bounds}
\end{equation}
\end{proposition}
The packet and remainder bounds have different statuses. The quantity $\eta_{\rm pkt}$ bounds the explicit difference $|S[g_\chi]-S[g]|$. In contrast, $\eta_r$ must come from an independent estimate of the entropy-expansion remainder. For example, if the admissible initial and final states, parameterized by $\epsilon$, satisfy $|r_\epsilon(\nu)|\leq b_\nu(\epsilon)$ for $\nu=\chi,\sigma_g$, then
\[
 |\Delta r_\epsilon|\leq b_\chi(\epsilon)+b_{\sigma_g}(\epsilon),
 \qquad \eta_r:=b_\chi+b_{\sigma_g}.
\]
Appendix~\ref{app:slow_clock_limit} proves convergence for fixed field states with the specified product preparations, but gives no universal rate or bound for a general channel output. In the exact reference preparation of Section~\ref{sec:crossed_product}, the reference entropy equals its packet expression, so $r_\epsilon(\sigma_g)=0$ in that convention. The Gaussian moment calculation below bounds a different term and does not supply $b_\chi$.
\begin{proof}[\normalfont\bfseries Proof]
Using the charge balance \eqref{eq:relaxation_charge_conservation}, Eq.~\eqref{eq:channel_Sgen_correction} can be reduced to
\[
 \Delta S_{\rm gen}=\Sigma_D
 +\Delta_{\rm rel}\vev{h_g(X)}
 -\Delta S[g]-\Delta r_\epsilon.
\]
DPI gives $\Sigma_D\geq0$. The triangle inequality with Eqs.~\eqref{eq:GSL_charge_error_bounds} and \eqref{eq:GSL_frame_error_bounds} then yields Eq.~\eqref{eq:GSL_theorem}.
\end{proof}

\begin{corollary}[Asymptotic GSL with a finite-packet error bound]
\label{cor:GSL_limit}
Under the conditions of Proposition~\ref{prop:GSL}, suppose the independently established clock-error bounds obey
\[
 \eta(\epsilon):=\eta_h(\epsilon)+\eta_{\rm pkt}(\epsilon)
 +\eta_r(\epsilon)\longrightarrow0.
\]
Then the finite-packet entropy change satisfies
\[
 |\Delta S_{\rm gen}(\epsilon)-\Sigma_D(\epsilon)|\leq\eta(\epsilon),
 \qquad \Delta S_{\rm gen}(\epsilon)\geq-\eta(\epsilon),
\]
and consequently
\[
 \liminf_{\epsilon\to0}\Delta S_{\rm gen}(\epsilon)\geq0.
\]
Thus the GSL holds asymptotically in the controlled slow-clock limit. This conclusion does not assert nonnegative entropy change for every finite $\epsilon$.
\end{corollary}
 The following discussion distinguishes the finite-width GSL criterion from the slow-clock limit, then identifies the physical corrections and evaluates them for a Gaussian reference. We close by explaining how a broad clock packet can register a finite charge transfer during the unitary drive.

\subsubsection{Finite-width and asymptotic GSL criteria}
At a fixed packet width, Eq.~\eqref{eq:GSL_theorem} gives the stronger lower bound $\Delta S_{\rm gen}\geq\Sigma_D-\eta$. Thus the finite-width GSL follows if $\Sigma_D\geq\eta$, or if a sharper estimate shows that the signed correction $\Delta S_{\rm gen}-\Sigma_D$ is at least $-\Sigma_D$. In particular, a nonnegative correction suffices even at saturation. When $\Sigma_D$ is smaller than the error bound and the correction has unknown sign, the estimate does not decide the sign of $\Delta S_{\rm gen}$. In the slow-clock limit, vanishing errors give $\liminf_{\epsilon\to0}\Delta S_{\rm gen}\geq0$; if in addition $\liminf_{\epsilon\to0}\Sigma_D(\epsilon)>0$, they give a positive change for all sufficiently small finite $\epsilon$. A vanishing error by itself gives no such finite-width guarantee near saturation. All limits are taken along families of normal states at finite $\epsilon$, with the duration and perturbative accuracy of the protocol under control. Sending $\epsilon$ to zero does not remove higher-order gravitational corrections or errors in the channel approximation.

\subsubsection{Physical sources of the corrections}
With the mean-charge term eliminated by Eq.~\eqref{eq:relaxation_charge_conservation}, three clock corrections remain. The term $\Delta_{\rm rel}\vev{h_g(X)}$ measures how relaxation changes the average of the fixed function $h_g$ of charge. Equality of the two mean charges does not determine this average for nonlinear $h_g$. The packet-entropy and remainder terms compare the initial and final states of the full protocol. Table~\ref{tab:gsl_error_budget} summarizes sufficient bounds on these three terms. They are useful when the signs are unknown; cancellations can give sharper estimates. Errors in approximating the physical evolution by the separated protocol, and higher-order gravitational corrections, require separate control.

\begin{table*}[t]
\caption{Clock corrections to $\Delta S_{\rm gen}-\Sigma_D$ after imposing the charge balance \eqref{eq:relaxation_charge_conservation}, and estimates for a broad Gaussian reference packet. Here $V_\nu$ is the variance of the dressed charge $X$, while $v_\nu$ is the variance of the packet distribution $|g_\nu|^2$.}
\label{tab:gsl_error_budget}
\centering
\small
\begin{tabular}{@{}lll@{}}
\toprule
\parbox[t]{.25\textwidth}{\raggedright\textbf{Correction}} &
\parbox[t]{.35\textwidth}{\raggedright\textbf{Conditions for vanishing as $\epsilon\to0$}\textsuperscript{a}} &
\parbox[t]{.34\textwidth}{\raggedright\textbf{Gaussian estimate}}\\
\midrule
\parbox[t]{.25\textwidth}{\raggedright Clock term in reference surprisal\\ $\Delta_{\rm rel}\vev{h_g(X)}$} &
\parbox[t]{.35\textwidth}{\raggedright The expectation change from $\omega$ to $\chi$ is $o(1)$.} &
\parbox[t]{.34\textwidth}{\raggedright $O(\epsilon^2)$ if $V_\chi-V_\omega=O(1)$; the means agree by the charge balance.}\\
\addlinespace[3pt]
\parbox[t]{.25\textwidth}{\raggedright Packet entropy\\ $-\Delta S[g]$} &
\parbox[t]{.35\textwidth}{\raggedright $|S[g_\chi]-S[g]|=o(1)$.} &
\parbox[t]{.34\textwidth}{\raggedright Zero for a shift in mean charge with unchanged packet shape; $O(\epsilon^2)$ for Gaussian initial and final clock packets with $v_\chi-v_{\sigma_g}=O(1)$.}\\
\addlinespace[3pt]
\parbox[t]{.25\textwidth}{\raggedright Entropy remainder\\ $-\Delta r_\epsilon$} &
\parbox[t]{.35\textwidth}{\raggedright Show independently that $|\Delta r_\epsilon|=o(1)$.} &
\parbox[t]{.34\textwidth}{\raggedright No rate follows from Gaussianity.}\\
\bottomrule
\end{tabular}\par\smallskip
\parbox{.97\textwidth}{\footnotesize\raggedright
\textsuperscript{a}\enspace As $\epsilon\to0$, $f(\epsilon)=o(1)$ means $f(\epsilon)\to0$, with no specified rate. By contrast, $f(\epsilon)=O(1)$ means that $|f(\epsilon)|$ stays bounded for sufficiently small $\epsilon$; it may approach a nonzero constant.}
\end{table*}

\subsubsection{Error bounds for a broad Gaussian reference}
Appendix~\ref{app:gaussian} derives the moment and packet-entropy estimates. The broad Gaussian reference packet, with $0<\epsilon\ll1$, is
\begin{equation}
 |g_\epsilon(x)|^2=\frac{\epsilon}{\sqrt{2\pi}}
 e^{-\epsilon^2(x-x_0)^2/2}.
 \label{eq:gaussian_packet}
\end{equation}
Write $h_\epsilon:=h_{g_\epsilon}$ for the clock term in the fixed reference surprisal. It is a constant plus $\epsilon^2(X-x_0)^2/2$. Consequently,
\begin{equation}
 \Delta_{\rm rel}\vev{h_\epsilon(X)}
 =\frac{\epsilon^2}{2}\Delta_{\rm rel}\vev{(X-x_0)^2}.
 \label{eq:gaussian_potential_exact}
\end{equation}
Write $\mu_\nu=\vev X_\nu$ and $V_\nu=\vev{(X-\mu_\nu)^2}_\nu$, so that $\vev{(X-x_0)^2}_\nu=V_\nu+(\mu_\nu-x_0)^2$. Equation~\eqref{eq:relaxation_charge_conservation} gives $\mu_\chi=\mu_\omega$ at second perturbative order, so the squared-mean terms cancel. Substituting Eq.~\eqref{eq:gaussian_potential_exact} into the entropy balance, with $\Delta_{\rm rel}\vev X=0$, gives (see Appendix~\ref{app:gaussian})
\begin{equation}
 \Delta S_{\rm gen}=\Sigma_D
 +\frac{\epsilon^2}{2}(V_\chi-V_\omega)
 -\Delta S[g]-\Delta r_\epsilon.
 \label{eq:gaussian_balance}
\end{equation}
Suppose $|V_\chi-V_\omega|\leq M$ uniformly as $\epsilon\to0$ for the protocols considered. The triangle inequality in Eq.~\eqref{eq:gaussian_balance} then gives the following bound, with $\eta_{\rm pkt}$ bounding $|\Delta S[g]|=|S[g_\chi]-S[g]|$ and $\eta_r$ bounding $|\Delta r_\epsilon|$ as in Eq.~\eqref{eq:GSL_frame_error_bounds}:
\begin{equation}
 |\Delta S_{\rm gen}-\Sigma_D|
 \leq\tfrac12M\epsilon^2+\eta_{\rm pkt}(\epsilon)+\eta_r(\epsilon).
 \label{eq:gaussian_bound}
\end{equation}
Shifting the packet in charge without changing its shape, $g_\chi(q)=g(q-d)$ with constant $d$, changes its mean by $d$ but leaves $S[g]$ unchanged; hence $\Delta S[g]=0$. For Gaussian packets, a bounded variance change instead gives $\Delta S[g]=O(\epsilon^2)$ (Appendix~\ref{app:gaussian}). If $|g_\nu|^2$ is the actual spectral distribution of $X$, its variance $v_\nu$ equals $V_\nu$; a packet assigned in an approximate preparation model needs this identification justified. Even when they agree state by state, the packet correction compares $v_\chi$ with $v_{\sigma_g}$, while the relaxation term compares $V_\chi$ with $V_\omega$. With both variance changes bounded and an independent estimate $|\Delta r_\epsilon|=O(\epsilon^p)$, $p>0$, the total error is $O(\epsilon^{\min\{2,p\}})$. Under the same variance bounds, $\Delta r_\epsilon=o(1)$ without a rate gives only an $o(1)$ total error. Thus the GSL holds asymptotically; at finite width, it is guaranteed when $\Sigma_D$ is at least the error bound in Eq.~\eqref{eq:gaussian_bound}.

\subsubsection{Mean clock charge during the unitary drive}
The unitary drive can change the exterior field state and the mean clock charge while preserving the entropy of the full dressed state. This observation is useful for interpreting the work in Eq.~\eqref{eq:physical_modular_work}: a broad packet can register a finite charge transfer with no change of packet entropy. The GSL bound itself does not require the following example.

Suppose the state $\omega$ prepared by the unitary drive, like the reference $\sigma_g$, admits the packet entropy expansion \eqref{eq:crossed_entropy}. This is an extra assumption beyond (A3). Subtracting the reference expansion, with matter state $\Psi$, from that of $\omega$, with matter state $\phi_\omega$, and using unitary entropy preservation gives
\begin{equation}
 \begin{aligned}
 0=\Delta_{\rm drv} S_\tau
 ={}&-D_\M(\phi_\omega\Vert\Psi)+\Delta_{\rm drv}\vev X\\
 &+\Delta_{\rm drv} S[g]+\Delta_{\rm drv} r_\epsilon .
 \end{aligned}
 \label{eq:drive_clock_response}
\end{equation}
Here $D_\M$ is the relative entropy on the matter algebra. All state expectations are in $\mathrm{QRF}_0$, where the physical clock charge is $X$. For the transported product preparations, its mean equals the packet mean of $q_R$ in $\mathrm{QRF}_1$, by Eq.~\eqref{eq:qrf0_clock_distribution}. If the packet entropy is unchanged and the remainder difference tends to zero, the shift in mean charge balances the matter relative entropy up to that remainder.

To see how this can happen, let $p_\epsilon(q)=|g_\epsilon(q)|^2$ be the Gaussian in Eq.~\eqref{eq:gaussian_packet}, with mean $x_0$ and standard deviation $\sigma_q=\epsilon^{-1}$. Suppose the unitary drive translates the packet by a fixed charge $d$, so that $g_{\epsilon,d}(q)=g_\epsilon(q-d)$. Changing variables to $u=q-d$ gives
\[
\begin{aligned}
 \vev{q_R}_{g_{\epsilon,d}}
 &=\int\dd u\,(u+d)p_\epsilon(u)=x_0+d,\\
 S[g_{\epsilon,d}]
 &=-\int\dd u\,p_\epsilon(u)\log p_\epsilon(u)
 =S[g_\epsilon].
\end{aligned}
\]
The mean charge therefore changes by $d$ while the packet keeps its shape and entropy. In the slow-clock limit, the displacement in units of the width is $|d|/\sigma_q=\epsilon|d|\to0$, even though the change in mean charge remains $d$. For the transported product preparations above, if this is the charge distribution after the unitary drive and $\Delta_{\rm drv}r_\epsilon=o(1)$, Eq.~\eqref{eq:drive_clock_response} requires $d=D_\M(\phi_\omega\Vert\Psi)+o(1)$. This example illustrates a possible clock response. It gives no estimate of the dressed charge variance during relaxation or of the initial- and final-state corrections in Eq.~\eqref{eq:gaussian_bound}.

\section{GSL from open-system dynamics on the dressed algebra}
\label{sec:simultaneous}
The exterior fields and their gravitational clock constitute the system described by the dressed algebra $\Mh(R)$. We treat this system as open when it interacts with additional degrees of freedom whose observables are excluded from the reduced description. Such interactions can exchange energy and generate correlations, so the dressed exterior density operator need not evolve unitarily even when the complete matter--gravity evolution is unitary. The gravitational clock remains part of the system; identifying the energy exchange with horizon absorption must therefore be accompanied by the corresponding area response. Section~\ref{sec:FT} describes this reduced evolution through relaxation after a unitary drive. Here we allow unitary driving and horizon interactions to act simultaneously. A controlled time-local description then replaces the two-stage relative-entropy loss by integrated dissipative entropy production, yielding a GSL bound when the dissipative energy and area changes satisfy the required horizon balance and the clock corrections are controlled.

\subsection{Simultaneous unitary driving and relaxation}
Let $t\in[0,T]$ denote the time labeling the applied protocol, distinct from the operator $p_R$ that measures the gravitational clock reading. Write $\nu_t$ for the dressed state and $\rho_t:=\rho_{\nu_t}$ for its density operator, with $\nu_0=\sigma_g$. We assume that the instantaneous change in $\rho_t$ depends only on $t$ and the current density operator:
\begin{equation}
 \dot\rho_t=-\ii[H_t,\rho_t]+\mathcal L_t(\rho_t),\qquad
 \mathcal L_t(\rho_{\sigma_g})=0.
 \label{eq:driven_generator}
\end{equation}
At each fixed $t$, the dissipator $\mathcal L_t$ is assumed to generate a completely positive semigroup that preserves the trace $\tau$ and fixes the same reference state $\sigma_g$. The full evolution need not fix $\sigma_g$. This Markovian assumption permits simultaneous unitary driving and dissipation, but excludes memory effects not captured by $\rho_t$. We assume that all traces, time integrals, and derivatives below exist; for unbounded observables the formulas may be established with a common regulator before taking a controlled limit. Related time-dependent open-system FTs are discussed in Refs.~\cite{ChetriteMallick:2012,DeChiaraImparato:2022,Manzano:2018}.

In Eq.~\eqref{eq:driven_generator}, both terms act on the exterior fields and their gravitational clock. The Hamiltonian $H_t$ generates the unitary driving and includes interaction-induced Hamiltonian corrections; its commutator preserves $S_\tau$. The dissipator describes the effect of interactions with additional degrees of freedom $B$ whose observables are excluded. Writing $V_{t,\delta t}$ for the joint unitary over the interval, the regulated model of Section~\ref{sec:FT_to_GSL} gives
\[
 \begin{aligned}
 \rho_{t+\delta t}
 &=\Tr_B\!\left[V_{t,\delta t}\rho_{RB}(t)V_{t,\delta t}^\dagger\right],\\
 \rho_t&=\Tr_B\rho_{RB}(t).
 \end{aligned}
\]
Two joint states can have the same $\rho_t$ and different field--environment correlations. In general, access to $\rho_t$ alone is insufficient to undo the joint unitary. The sector $B$ may include modes that have crossed the future horizon; no tensor-product decomposition across the horizon is assumed in the continuum. Identifying the dissipative energy loss with horizon Killing flux requires the matching described below.

The time-local approximation uses intervals $\tau_B\ll\delta t\ll\tau_{\rm sys}$, where $\tau_B$ is the memory time of the interaction and $\tau_{\rm sys}$ is the shortest timescale for appreciable changes in the unitary drive or exterior state. In the regulated model, replacing $\rho_{RB}(t)$ by $\rho_t\otimes\eta_{B,t}$, with a prescribed density operator $\eta_{B,t}$, must leave the reduced prediction unchanged to first order in $\delta t$. Radiation that returns to interact with the exterior can invalidate this approximation. Freezing the generators over the interval gives
\[
 \begin{aligned}
 &e^{\delta t\mathcal L_t}
 \!\left(e^{-\ii H_t\delta t}\rho_t e^{\ii H_t\delta t}\right)\\
 &\quad=\rho_t+\delta t\bigl(-\ii[H_t,\rho_t]
       +\mathcal L_t(\rho_t)\bigr)+o(\delta t).
 \end{aligned}
\]
Applying the dissipative map first and the unitary map second gives the same coefficient of $\delta t$; the two orders can differ at higher order. This expansion explains the sum in Eq.~\eqref{eq:driven_generator}. Whether it describes a given black-hole evolution, and whether its dissipator fixes the dressed reference, are separate physical questions.

\subsection{Entropy production and the generalized horizon response}
To derive the production rate, freeze $t$ and apply only the auxiliary dissipative channel $\E_{t,s}=e^{s\mathcal L_t}$ for $s\geq0$. This does not separate the physical unitary driving and relaxation intervals. On states, $\E_{t,s}(\nu)$ denotes the state with density operator $\E_{t,s}(\rho_\nu)$. Since $\E_{t,s}(\sigma_g)=\sigma_g$, DPI gives
\[
 D_\tau(\E_{t,s}(\nu_t)\Vert\sigma_g)
 \leq D_\tau(\nu_t\Vert\sigma_g).
\]
For a variation of $\nu$ with $\tau(\delta\rho_\nu)=0$, the trace derivative is
\[
 \delta D_\tau(\nu\Vert\sigma_g)
 =\tau[\delta\rho_\nu(\log\rho_\nu-\log\rho_{\sigma_g})].
\]
This is a trace-functional derivative and does not assume that the density operators commute. Differentiating the DPI inequality at $s=0^+$ therefore yields the Spohn-type rate \cite{Spohn:1978}
\begin{equation}
 \begin{aligned}
 \dot\Sigma_{\rm diss}(t)
 &:=-\left.\frac{\dd}{\dd s}
 D_\tau(\E_{t,s}(\nu_t)\Vert\sigma_g)\right|_{s=0^+}\\
 &=-\tau[\mathcal L_t(\rho_t)
 (\log\rho_t-\log\rho_{\sigma_g})]\geq0.
 \end{aligned}
 \label{eq:Spohn_rate}
\end{equation}
This derivative measures how rapidly dissipation alone would reduce the distinguishability of the current state from $\sigma_g$. It is generally different from the derivative along the actual driven evolution. Its nonnegativity follows from DPI, without a measurement protocol or a reverse experiment.

Next, trace preservation gives $\tau(\dot\rho_t)=0$, and hence
\[
 \frac{\dd S_\tau}{\dd t}=-\tau(\dot\rho_t\log\rho_t).
\]
The Hamiltonian contribution vanishes by trace cyclicity and $[\rho_t,\log\rho_t]=0$:
\[
 \ii\tau([H_t,\rho_t]\log\rho_t)
 =\ii\tau(H_t[\rho_t,\log\rho_t])=0.
\]
Thus $\dot S_\tau=-\tau[\mathcal L_t(\rho_t)\log\rho_t]$. Since $\log\rho_{\sigma_g}=-k_g$, Eq.~\eqref{eq:Spohn_rate} reads $\dot\Sigma_{\rm diss}=\dot S_\tau-\tau[\mathcal L_t(\rho_t)k_g]$. Rearranging gives
\begin{equation}
 \frac{\dd S_\tau}{\dd t}
 =\dot\Sigma_{\rm diss}(t)+\tau[\mathcal L_t(\rho_t)k_g].
 \label{eq:Spohn_balance}
\end{equation}
The second term is the rate at which dissipation changes the mean reference surprisal $\langle k_g\rangle$. It can have either sign: reference preservation alone need not make the dressed trace entropy increase.

Integrate Eq.~\eqref{eq:Spohn_balance}, substitute $k_g=X+h_g(X)$, and apply the common entropy-difference identity \eqref{eq:generic_entropy_target}. It follows that
\begin{equation}
 \begin{aligned}
 \Delta S_{\rm gen}={}&\Sigma_{\rm diss}+R_X+R_g\\
 &-\Delta S[g]-\Delta r_\epsilon,\\
 \Sigma_{\rm diss}&:=\int_0^T\dd t\,\dot\Sigma_{\rm diss}(t)\geq0,\\
 R_X&:=\int_0^T\dd t\,\tau[\mathcal L_t(\rho_t)X],\\
 R_g&:=\int_0^T\dd t\,\tau[\mathcal L_t(\rho_t)h_g(X)].
 \end{aligned}
 \label{eq:simultaneous_balance}
\end{equation}
Here $\Delta$ means final minus initial. The two clock terms have different origins: $R_g$ measures the dissipative change of the fixed reference observable $h_g(X)$, whereas $\Delta S[g]=S[g_{\nu_T}]-S[g]$ compares the clock distributions entering the entropy dictionary at the two ends of the protocol. The same states determine $\Delta r_\epsilon$. Equation~\eqref{eq:simultaneous_balance} requires that dictionary for the actual output state.

For a fixed observable $O$ in the Schr\"odinger picture, write $\Delta_{\rm diss}\vev O:=\int_0^T\dd t\,\tau[\mathcal L_t(\rho_t)O]$. This is the part of the mean change assigned to the dissipator in Eq.~\eqref{eq:driven_generator}; it is generally not the total change between the two states. For the exterior Killing Hamiltonian, define the net exterior energy lost through that term by
\begin{equation}
 E_H:=-\int_0^T\dd t\,
 \tau[\mathcal L_t(\rho_t)H_{\rm out}].
 \label{eq:simultaneous_heat_definition}
\end{equation}
With this sign, $E_H>0$ denotes energy leaving the exterior. To identify it with absorption at the future horizon, one must account separately for radiation reaching future null infinity and show that the remaining energy transfer equals the horizon Killing flux. Using the matched charge relation $X=\mathsf{q}_{\mathcal B}+\beta_HH_{\rm out}$ gives
\begin{equation}
 R_X=\Delta_{\rm diss}\vev{\mathsf{q}_{\mathcal B}}-\beta_HE_H.
 \label{eq:simultaneous_area_matching}
\end{equation}
If the horizon response, with the appropriate horizon boundary terms, establishes $\Delta_{\rm diss}\vev{\mathsf{q}_{\mathcal B}}=\beta_HE_H$, then $R_X=0$. This is a condition on the dissipative area contribution, not automatically on the total area change. Indeed,
\[
 \Delta\vev{\mathsf{q}_{\mathcal B}}
 =\int_0^T\dd t\,\tau[-\ii[H_t,\rho_t]\mathsf{q}_{\mathcal B}]
 +\Delta_{\rm diss}\vev{\mathsf{q}_{\mathcal B}}.
\]
The physical-process first law in Section~\ref{subsec:area_endpoint} constrains the total area change, whereas Eq.~\eqref{eq:simultaneous_area_matching} involves only its dissipative part. Once the flux is identified with $E_H$, any expansion boundary term in the integrated first law contributes to $R_X$, and the Hamiltonian contribution in the display above must be subtracted. Thus the first law alone does not identify the dissipative term of a proposed generator with the horizon response.

\subsection{Conditional GSL and the role of the unitary drive}
For the Gaussian reference, trace preservation removes the constant in $h_g$, giving
\begin{equation}
 R_g=\frac{\epsilon^2}{2}\int_0^T\dd t\,
 \tau[\mathcal L_t(\rho_t)(X-x_0)^2].
 \label{eq:simultaneous_gaussian}
\end{equation}
The factor $\epsilon^2$ multiplies the dissipative change of a second moment that can itself grow as the packet broadens. The Gaussian shape alone therefore gives no small bound for $R_g$. Using $\Sigma_{\rm diss}\geq0$ in Eq.~\eqref{eq:simultaneous_balance} gives
\begin{equation}
 \Delta S_{\rm gen}\geq
 -|R_X|-|R_g|-|\Delta S[g]|-|\Delta r_\epsilon|.
 \label{eq:simultaneous_GSL_bound}
\end{equation}
Under the matched horizon response, $R_X=0$. If $R_g$, $\Delta S[g]$, and $\Delta r_\epsilon$ tend to zero along the chosen family of protocols, then $\Delta S_{\rm gen}=\Sigma_{\rm diss}+o(1)$ and $\liminf_{\epsilon\to0}\Delta S_{\rm gen}\geq0$, in the sense of Corollary~\ref{cor:GSL_limit}. At finite width, a sufficient condition for $\Delta S_{\rm gen}\geq 0$ is that $\Sigma_{\rm diss}$ be at least as large as the sum of the absolute errors in Eq.~\eqref{eq:simultaneous_GSL_bound}. This is a bound on the mean entropy change; no trajectory FT has been assumed.

If the unitary drive and relaxation have disjoint time supports and the reference is fixed by relaxation, the integrated production reduces to $D_\tau(\omega\Vert\sigma_g)-D_\tau(\chi\Vert\sigma_g)$. On the relaxation interval, the rate is the negative time derivative of the reference-relative entropy, and $R_X$ and $R_g$ become the corresponding differences between the state $\omega$ prepared by the unitary drive and the final state $\chi$. This recovers the balance of Section~\ref{sec:FT} within the time-local model.

For completeness, the role of modular work follows from the same derivatives. Define the dimensionless mean modular work as the contribution of the Hamiltonian term to the change in $\langle k_g\rangle$:
\begin{equation}
 \begin{aligned}
 \dot W_{\rm mod}(t)&:=\tau[-\ii[H_t,\rho_t]k_g],\\
 W_{\rm mod}&:=\int_0^T\dd t\,\dot W_{\rm mod}(t).
 \end{aligned}
 \label{eq:simultaneous_modular_work}
\end{equation}
In the derivative of $D_\tau(\nu_t\Vert\sigma_g)$, the Hamiltonian term containing $\log\rho_t$ vanishes as above, while its reference-logarithm term gives $\dot W_{\rm mod}$. The dissipative part is $-\dot\Sigma_{\rm diss}$ by Eq.~\eqref{eq:Spohn_rate}. Hence
\[
 \frac{\dd}{\dd t}D_\tau(\nu_t\Vert\sigma_g)
 =\dot W_{\rm mod}(t)-\dot\Sigma_{\rm diss}(t).
\]
Integration yields
\begin{equation}
 \begin{aligned}
 \Sigma_{\rm diss}
 ={}&W_{\rm mod}-D_\tau(\nu_T\Vert\sigma_g)\\
 &+D_\tau(\nu_0\Vert\sigma_g).
 \end{aligned}
 \label{eq:continuous_modular_work}
\end{equation}
For $\nu_0=\sigma_g$, the input modular work is therefore the sum of $\Sigma_{\rm diss}$ and the final distinguishability $D_\tau(\nu_T\Vert\sigma_g)$. This separates the work dissipated in the sense of relative entropy from the departure from equilibrium still present at $T$; $W_{\rm mod}$ measures a change of reference surprisal and is not identified with the physical Killing energy $E_H$. 
 
$\Sigma_{\rm diss}$ is a number determined by the model dissipator and the evolving density operators; DPI proves it is nonnegative. It is not itself a stochastic variable endowed with forward and reverse measurement statistics, so no additional forward average $\vev{\Sigma_{\rm diss}}_F$ is intended. A fluctuation theorem resolving the intermediate evolution would require a specified measurement scheme and normalized reverse dynamics, together with a derivation identifying the mean logarithmic probability ratio with $\Sigma_{\rm diss}$. The mean entropy balance alone does not supply such a construction.

Moreover, Eq.~\eqref{eq:simultaneous_balance} gives $\Delta S_\tau=\Sigma_{\rm diss}+R_X+R_g$. When the two-time measurement construction of Section~\ref{sec:fullcore_FT} also applies, its mean identity \eqref{eq:fullcore_FT_mean} therefore implies
\[
 \Sigma_{\rm diss}
 =\vev{\Sigma_{\rm 2t}}_F-R_X-R_g.
\]
Thus positivity of the mean two-time surprisal change does not by itself establish positivity of $\Sigma_{\rm diss}$ without further control of the exchange terms. The integrated production and the mean two-time change coincide when $R_X+R_g=0$. In particular, the charge-preserving unitary-jump model of Eq.~\eqref{eq:charge_preserving_model} preserves both $X$ and $h_g(X)$ under its dissipative evolution, giving $R_X=R_g=0$. This equality of means does not identify distributions for intermediate trajectories.

These distinctions motivate Section~\ref{sec:fullcore_FT}. The mean two-time surprisal change always gives $\Delta S_\tau$ under the stated measurement assumptions, but its exponential average contains a channel-dependent correction in general. We first derive that relation and its generalized-entropy bound, then show how unitality removes the correction and how charge-conserving, reference-preserving dressed relaxation can establish unitality. This construction needs no time-local description of the intermediate evolution.

\section{GSL from the black-hole channel fluctuation theorem}
\label{sec:generic_channel}
\label{sec:fullcore_FT}
We now formulate the GSL through a two-time measurement of surprisal for the reduced black-hole channel from the initial preparation to the final measurement. Unitary driving and horizon interactions may overlap; neither a separated protocol nor a time-local generator is required. Section~\ref{sec:two_time_experiments} derives the fluctuation relation, identifies the correction caused by nonunitality, and applies it to generalized entropy. Section~\ref{sec:unital_physical_conditions} then asks which physical restrictions on dressed black-hole dynamics establish unitality.

\subsection{FT of unital quantum channels and GSL}
\label{sec:two_time_experiments}

\subsubsection{The open-system channel and its reverse}
The system consists of the exterior fields and their gravitational clock, described on the fixed algebra $\Mh(R)$ with trace $\tau$. It is open when degrees of freedom $B$ interact with the exterior but are not measured, as in Section~\ref{sec:simultaneous}. In a regulated model, the exterior and $B$ are prepared independently, evolve under a joint unitary $V_T$, and the observables of $B$ are then discarded. For a fixed density operator $\eta_B$ of $B$, the resulting channel is
\begin{equation}
 \mathcal C_T(\rho)
 =\Tr_B\!\left[V_T(\rho\otimes\eta_B)V_T^\dagger\right].
 \label{eq:generic_BH_dilation}
\end{equation}
The gravitational clock remains part of the system. This unitary representation illustrates the reduction and does not assume an exact inside--outside tensor factorization in continuum gravity.

For the models with Kraus operators in $\Mh(R)$, the forward evolution in Fig.~\ref{fig:two_time_protocol}(a) takes the form
\begin{equation}
 \begin{aligned}
 \rho_T&=\mathcal C_T(\rho_0),\qquad \tau(\rho_0)=1,\\
 \mathcal C_T(\rho)&=\sum_\alpha M_\alpha\rho M_\alpha^\dagger.
 \end{aligned}
 \label{eq:full_channel_kraus}
\end{equation}
We write $\nu_0,\nu_T$ for the corresponding states, with $\rho_0=\rho_{\nu_0}$ and $\rho_T=\rho_{\nu_T}$. In the regulated model, $\eta_B=\sum_b p_b\ket b\bra b$ gives $M_{ab}=\sqrt{p_b}\bra aV_T\ket b$. The channel averages over the initial states $\ket b$ of $B$ with probabilities $p_b$ and sums over its unobserved final states $\ket a$. The exterior observer measures only the fields and their gravitational clock. We use the same trace notation for type-I quantum mechanics and the type-II dressed algebra.\footnote{A density operator need not be bounded in a semifinite algebra. We assume the Schr\"odinger map on operators of finite trace norm admits a compatible normal action on bounded operators, with the two actions agreeing on their common domain; in particular $\mathcal C_T(\id)$ is bounded. In the explicit models, the Kraus sums or integrals converge in these respective senses. Not every type-II channel admits Kraus operators in the algebra itself; the trace-adjoint argument only needs the two compatible actions.}

For the reverse comparison in Fig.~\ref{fig:two_time_protocol}(b), prepare $\rho_T$, interchange the initial and final measurements, and replace each $M_\alpha$ by $M_\alpha^\dagger$. The resulting adjoint Kraus map is
\begin{equation}
 \mathcal C_T^{\rm R}(\rho)
 =\sum_\alpha M_\alpha^\dagger\rho M_\alpha.
 \label{eq:tracial_reverse_channel}
\end{equation}
In the reverse comparison the adjoints act on density operators in the Schr\"odinger picture. The reverse map is completely positive, but is a quantum channel on arbitrary reverse preparations only when it preserves the trace. This choice compares transition weights; it need not coincide with reversing the microscopic interaction $V_T$, or return $\rho_T$ to $\rho_0$ \cite{Albash:2013,Rastegin:2013}.

\subsubsection{Trace preservation and unitality}
\label{sec:unital_dressed_channel}
The two normalization conditions have distinct Kraus expressions,
\begin{equation}
 \begin{aligned}
 \sum_\alpha M_\alpha^\dagger M_\alpha&=\id
 &&\text{(forward trace preservation)},\\
 \sum_\alpha M_\alpha M_\alpha^\dagger&=\id
 &&\text{(forward unitality)}.
 \end{aligned}
 \label{eq:unital_kraus_conditions}
\end{equation}
Forward trace preservation makes $\mathcal C_T^{\rm R}$ unital. Conversely, trace cyclicity gives
\begin{equation}
 \tau[\mathcal C_T^{\rm R}(\rho)]
 =\tau[\rho\mathcal C_T(\id)],
 \label{eq:trace_adjoint}
\end{equation}
so the reverse map preserves the trace for all density operators precisely when $\mathcal C_T(\id)=\id$. Thus unitality makes both sequences in Fig.~\ref{fig:two_time_protocol} physical quantum-channel experiments. Without it, the dashed sequence assigns reverse weights but need not define a normalized experiment.

Physically, a unital channel preserves uniform statistical weights. In a finite-dimensional regulator, $\rho_{\rm unif}=\id/d$ is unchanged. A Gibbs state is close to uniform when its temperature exceeds the full energy spread by a large factor ($k_{\rm B}=1$). Trace-distance contraction then keeps the output close to the same uniform state. This is the precise sense in which an ultrahigh-temperature preparation is difficult to disturb. In the type-$\mathrm{II}_\infty$ algebra, $\tau(\id)=\infty$, so there is no normalized identity state. On a sector with projection $P$ of finite trace, the corresponding uniform density operator is $P/\tau(P)$ and is preserved if $\mathcal C_T(P)=P$. On the full algebra, $\mathcal C_T(\id)=\id$ remains the exact algebraic condition.

\begin{figure*}[t]
\centering
\begin{tikzpicture}[>=Latex,font=\small,
  measurement/.style={draw=blue!60!black,fill=blue!3,
    rounded corners=3pt,line width=.8pt,align=center,
    text width=3.35cm,minimum height=1.65cm,inner sep=5pt},
  channel/.style={draw=blue!60!black,fill=blue!7,
    rounded corners=3pt,line width=.9pt,align=center,
    text width=4.05cm,minimum height=1.65cm,inner sep=5pt},
  reverse measurement/.style={measurement,draw=teal!65!black,fill=teal!3},
  reverse channel/.style={channel,draw=teal!65!black,fill=teal!7,dashed},
  forward arrow/.style={->,blue!65!black,line width=1pt},
  reverse arrow/.style={->,teal!65!black,line width=1pt,dashed}]
 \node[align=center,fill=black!5,rounded corners=2pt,
   text width=16.0cm,inner ysep=5pt] at (8.15,2.0)
   {Fixed dressed algebra $\Mh(R)$ and trace $\tau$
    \quad\mbox{$\rho_T=\mathcal C_T(\rho_0)$}};
 \node[anchor=west,blue!65!black] at (0,1.28)
   {\textbf{(a) Forward experiment}};
 \node[align=center] (finput) at (.10,0) {$\rho_0$};
 \node[measurement] (fmeasure) at (3.0,0)
   {Measure $k_0=-\log\rho_0$\\[4pt]
    Outcome $i$\\[3pt]
    Conditional state: $P_i/\tau(P_i)$};
 \node[channel] (fchannel) at (8.15,0)
   {Forward channel $\mathcal C_T$\\[8pt]
    $\rho\longmapsto\sum_\alpha
     M_\alpha\rho M_\alpha^\dagger$};
 \node[measurement] (ffinal) at (13.3,0)
   {Measure $k_T=-\log\rho_T$\\[4pt]
    Outcome $j$\\[3pt]
    $\Sigma_{\rm 2t}(i,j)=\log(r_i/s_j)$};
 \draw[forward arrow] (finput.east)--(fmeasure.west);
 \draw[forward arrow] (fmeasure.east)--(fchannel.west);
 \draw[forward arrow] (fchannel.east)--(ffinal.west);
 \node[align=center] at (8.15,-1.25)
   {$P_F(i,j)=r_i\tau[Q_j\mathcal C_T(P_i)]$
    \qquad $\sum_{ij}P_F(i,j)=1$};
 \node[anchor=west,teal!65!black] at (0,-2.0)
   {\textbf{(b) Reverse comparison}\quad (read from right to left)};
 \node[align=center] (rinput) at (16.2,-3.3) {$\rho_T$};
 \node[reverse measurement] (rmeasure) at (13.3,-3.3)
   {Measure $k_T=-\log\rho_T$\\[4pt]
    Outcome $j$\\[3pt]
    Conditional state: $Q_j/\tau(Q_j)$};
 \node[reverse channel] (rchannel) at (8.15,-3.3)
   {Adjoint Kraus map $\mathcal C_T^{\rm R}$\\[8pt]
    $\rho\longmapsto\sum_\alpha
     M_\alpha^\dagger\rho M_\alpha$};
 \node[reverse measurement] (rfinal) at (3.0,-3.3)
   {Measure $k_0=-\log\rho_0$\\[4pt]
    Outcome $i$\\[3pt]
    $-\Sigma_{\rm 2t}(i,j)=\log(s_j/r_i)$};
 \draw[reverse arrow] (rinput.west)--(rmeasure.east);
 \draw[reverse arrow] (rmeasure.west)--(rchannel.east);
 \draw[reverse arrow] (rchannel.west)--(rfinal.east);
 \node[align=center] at (8.15,-4.55)
   {$\widetilde P_R(j,i)=s_j\tau[P_i\mathcal C_T^{\rm R}(Q_j)]$
    \qquad $\sum_{ij}\widetilde P_R(j,i)=\Gamma_T$};
 \node[align=center,fill=black!5,rounded corners=2pt,
   text width=16.0cm,inner ysep=6pt] at (8.15,-5.45)
   {\mbox{$\displaystyle\left\langle e^{-\Sigma_{\rm 2t}}\right\rangle_F
      =\Gamma_T=\tau[\rho_T\mathcal C_T(\id)]$}\\[4pt]
    Unitality: \mbox{$\mathcal C_T(\id)=\id
      \Longrightarrow\Gamma_T=1\ \text{and}\ \widetilde P_R=P_R$}
    (normalized reverse experiment)};
\end{tikzpicture}
\caption{Two projective measurements of surprisal on the fixed dressed algebra, shown for discrete spectra and faithful density operators. The maps act on density operators in the Schr\"odinger picture. The conditional density operators are normalized by the projection weights $\tau(P_i)$ and $\tau(Q_j)$. The final forward observable $k_T$ is determined by the unconditioned output $\rho_T$ and is fixed across runs. Forward trace preservation normalizes (a); forward unitality also normalizes the adjoint Kraus experiment (b), shown with dashed arrows to distinguish it from an inverse evolution. The reverse comparison starts in $\rho_T$ and need not return it to $\rho_0$.}
\label{fig:two_time_protocol}
\end{figure*}

\subsubsection{Two-time surprisal statistics and the fluctuation relation}
For clarity, first take faithful density operators with discrete spectra and finite-trace spectral projections,
\begin{equation}
 \rho_0=\sum_i r_iP_i,\qquad
 \rho_T=\sum_j s_jQ_j,\qquad r_i,s_j>0.
 \label{eq:surprisal_spectra}
\end{equation}
In Fig.~\ref{fig:two_time_protocol}(a), measure $k_0=-\log\rho_0$, apply $\mathcal C_T$, and measure $k_T=-\log\rho_T$. These are the surprisals of the actual initial and final states; Section~\ref{sec:typeI_FT} instead measures the same fixed reference surprisal twice. The final observable is fixed by the unconditioned output, rather than by the initial outcome. Outcome $i$ occurs with probability $r_i\tau(P_i)$ and prepares $P_i/\tau(P_i)$. The two-time probability and surprisal change are therefore
\begin{equation}
 \begin{aligned}
 P_F(i,j)&=r_i\sum_\alpha
    \tau(Q_jM_\alpha P_iM_\alpha^\dagger),\\
 \Sigma_{\rm 2t}(i,j)&=\log(r_i/s_j).
 \end{aligned}
 \label{eq:endpoint_sigma}
\end{equation}
Here $\Sigma_{\rm 2t}$ is the final measured surprisal $-\log s_j$ minus the initial value $-\log r_i$. The Kraus label is summed over, not measured. The factors $\tau(P_i)$ and $\tau(Q_j)$ determine the measurement probabilities and normalize the post-measurement density operators. With the ordinary type-I trace, they are the eigenspace dimensions and equal one for rank-one projectors. They must be kept explicitly for general spectral projections, including those of the type-II dressed algebra. Forward trace preservation and $\rho_T=\mathcal C_T(\rho_0)$ give
\begin{equation}
 \sum_jP_F(i,j)=r_i\tau(P_i),\qquad
 \sum_iP_F(i,j)=s_j\tau(Q_j).
 \label{eq:bistochastic_marginals}
\end{equation}
For finite entropy expectations the two marginals give, without assuming unitality,
\begin{equation}
 \begin{aligned}
 \vev{\Sigma_{\rm 2t}}_F
 &=\sum_i r_i\tau(P_i)\log r_i
   -\sum_j s_j\tau(Q_j)\log s_j\\
 &=S_\tau(\nu_T)-S_\tau(\nu_0).
 \end{aligned}
 \label{eq:fullcore_FT_mean}
\end{equation}

The reverse sequence in Fig.~\ref{fig:two_time_protocol}(b) similarly assigns
\begin{equation}
 \begin{aligned}
 \widetilde P_R(j,i)&=s_j\sum_\alpha
    \tau(P_iM_\alpha^\dagger Q_jM_\alpha),\\
 P_F(i,j)&=e^{\Sigma_{\rm 2t}(i,j)}\widetilde P_R(j,i).
 \end{aligned}
 \label{eq:fullcore_forward_reverse}
\end{equation}
The reverse weights $\widetilde P_R(j,i)$ need not sum to one, since the adjoint Kraus map need not preserve the trace. The second line follows from trace cyclicity. To sum over the initial outcomes, completeness, $\sum_iP_i=\id$, and normality of the map give
\[
 \sum_i\mathcal C_T(P_i)=\mathcal C_T(\id)
 =\sum_\alpha M_\alpha M_\alpha^\dagger.
\]
Unitality requires this operator to equal $\id$, whereas trace preservation fixes only $\sum_\alpha M_\alpha^\dagger M_\alpha=\id$. The difference $\mathcal C_T(\id)-\id$ therefore measures nonunitality. Multiplying the second line of Eq.~\eqref{eq:fullcore_forward_reverse} by $e^{-\Sigma_{\rm 2t}}$, summing, and using $\sum_j s_jQ_j=\rho_T$ yields
\begin{equation}
 \begin{aligned}
 \vev{e^{-\Sigma_{\rm 2t}}}_F
 &=\sum_{ij}\widetilde P_R(j,i)\\
 &=\sum_{ij}s_j\tau[Q_j\mathcal C_T(P_i)]\\
 &=\sum_j s_j\tau[Q_j\mathcal C_T(\id)]\\
 &=\tau[\rho_T\mathcal C_T(\id)]=:\Gamma_T,\\
 \Gamma_T&=1+\tau[\rho_T(\mathcal C_T(\id)-\id)].
 \end{aligned}
 \label{eq:nonunital_correction}
\end{equation}
This is the surprisal form of the channel fluctuation relation and its nonunital correction \cite{Albash:2013,RasteginZyczkowski:2014}. For a unital channel, $\widetilde P_R=P_R$ is normalized and
\begin{equation}
 \vev{e^{-\Sigma_{\rm 2t}}}_F=1,
 \qquad \vev{\Sigma_{\rm 2t}}_F=\Delta S_\tau\geq0.
 \label{eq:fullcore_IFT}
\end{equation}
For a general channel, Jensen's inequality instead gives
\begin{equation}
 \Delta S_\tau=\vev{\Sigma_{\rm 2t}}_F\geq-\log\Gamma_T.
 \label{eq:nonunital_entropy_bound}
\end{equation}
A nonunital channel can have $\Gamma_T=1$ for a particular preparation, but this does not make its adjoint map trace preserving for all reverse inputs. Normalizing the formal weights by $\Gamma_T$ likewise does not produce a physical reverse channel.

For continuous spectra the discrete formulas are understood as limits: the measured probabilities must converge without loss of total probability, and both the surprisal mean and its exponential average must converge. These requirements do not follow from normalization alone. Faithfulness means that the support projection of each density operator is $\id$; it does not require a bounded inverse. If $P_{\rm in}=\operatorname{supp}\rho_0\ne\id$, only this subspace is populated in the first measurement. The exponential average is then $\tau[\rho_T\mathcal C_T(P_{\rm in})]$, which can differ from one even for a unital channel. We use the faithful case below.

\subsubsection{From the fluctuation relation to the GSL}
\label{sec:two_time_GSL}
Take $\rho_0=\rho_{\sigma_g}$ and apply the entropy dictionary \eqref{eq:generic_entropy_target} to the actual initial and final dressed states. Equation~\eqref{eq:fullcore_FT_mean} gives
\begin{equation}
 \Delta S_{\rm gen}
 =\vev{\Sigma_{\rm 2t}}_F-\Delta S[g]-\Delta r_\epsilon,
 \label{eq:two_time_entropy_dictionary}
\end{equation}
and hence
\begin{equation}
 \Delta S_{\rm gen}\geq-\log\Gamma_T
 -\Delta S[g]-\Delta r_\epsilon.
 \label{eq:nonunital_GSL_bound}
\end{equation}
When $\Gamma_T>1$, nonunitality permits a negative lower bound even if the clock errors vanish. For that preparation the exponential-average bound alone cannot establish the GSL; the actual mean entropy change can still be nonnegative. Nonunitality can also give $\Gamma_T\leq1$ for a particular process. The term $-\log\Gamma_T$ belongs to the bound from the exponential average; it does not modify the exact mean balances of Sections~\ref{sec:FT} and \ref{sec:simultaneous}.

Unitality removes this correction for every allowed input and yields
\begin{equation}
 \Delta S_{\rm gen}\geq-|\Delta S[g]|-|\Delta r_\epsilon|.
 \label{eq:fullcore_FT_GSL}
\end{equation}
Comparing Eqs.~\eqref{eq:GSL_theorem}, \eqref{eq:simultaneous_balance}, and \eqref{eq:fullcore_FT_GSL} clarifies the role of the channel dynamics. In Eq.~\eqref{eq:GSL_theorem}, the mean-charge term has already been eliminated by Eq.~\eqref{eq:relaxation_charge_conservation}, but $\eta_h$ still bounds $\Delta_{\rm rel}\vev{h_g(X)}$. Equation~\eqref{eq:simultaneous_balance} contains the dissipative contributions $R_X$ and $R_g$ explicitly. In contrast, the two-time FT gives the total dressed entropy change directly from the initial and final density operators. Equation~\eqref{eq:fullcore_FT_GSL} therefore requires no separate balance for $X$ or $h_g(X)$, nor a separated protocol or time-local generator. Unitality removes the channel correction for every allowed input; converting the trace-entropy bound into a GSL bound still requires the entropy dictionary for the two states actually measured.

For a unital channel at finite clock width, the GSL is guaranteed if $\Delta S[g]+\Delta r_\epsilon\leq0$, or more generally if $\Delta S_\tau\geq\Delta S[g]+\Delta r_\epsilon$. If both corrections tend to zero as $\epsilon\to0$, then $\liminf_{\epsilon\to0}\Delta S_{\rm gen}\geq0$. These are statements at the gravitational perturbative order of the entropy dictionary. The remaining question is thus physical: when does dressed black-hole evolution satisfy the unitality condition used by this FT?

\subsection{Unital constraint on black-hole dynamics as a quantum channel}
\label{sec:unital_physical_conditions}

\subsubsection{Charge conservation and the stationary reference}
Charge conservation, reference preservation, and unitality are distinct requirements. We identify a relaxation model consistent with the area--matter balance and explain how preserving the full dressed reference supplies a sufficient condition for unitality in this model.

Undriven relaxation means evolution with the external unitary drive switched off; energy transfer to the horizon and the resulting area response remain included. The physical-process first law and the exterior Killing-energy balance of Section~\ref{subsec:area_endpoint} give, at the perturbative order considered,
\begin{equation}
 \begin{aligned}
 \Delta_{\rm rel}\vev X
 &=\Delta_{\rm rel}\vev{\mathsf{q}_{\mathcal B}}
   +\Delta_{\rm rel}\vev{\mathsf{k}_\Psi}\\
 &=\beta_HE_H-\beta_HE_H=0.
 \end{aligned}
 \label{eq:channel_first_law_charge}
\end{equation}
In the fixed frame $X$ is the bifurcation-area charge plus the one-sided modular energy. A loss of exterior Killing energy can therefore be balanced by an area increase without changing the mean of $X$.

Let $\E$ denote this relaxation channel. As a sufficient model of charge conservation, we impose commutation with $X$ on all its Kraus operators,
\begin{equation}
 \begin{aligned}
 \E(\rho)&=\sum_\alpha M_\alpha\rho M_\alpha^\dagger,\\
 [M_\alpha,X]&=0\qquad\text{for all }\alpha.
 \end{aligned}
 \label{eq:relaxation_kraus_physical}
\end{equation}
Thus no Kraus operator connects states with different total charges, although the area and matter contributions can change separately.\footnote{A Kraus label need not identify a unique microscopic event. Conditioning on an outcome can reweight charge sectors already present in the input; the condition does not assert conservation of the conditional mean charge for an arbitrary input state.} With trace preservation, $\sum_\alpha M_\alpha^\dagger M_\alpha=\id$, the trace dual $\E^*$, acting here on observables rather than on reverse preparations, gives
\begin{equation}
 \begin{aligned}
 \E^*(X)
 &=\sum_\alpha M_\alpha^\dagger XM_\alpha\\
 &=X\sum_\alpha M_\alpha^\dagger M_\alpha=X,\\
 \tau[\E(\rho)X]&=\tau(\rho X).
 \end{aligned}
 \label{eq:microscopic_charge_conservation}
\end{equation}
This model conserves the mean charge for every input with a finite charge average.\footnote{For unbounded $X$, commutation is understood through its spectral projections. We restrict to states with finite absolute first moment, $\tau(\rho|X|)<\infty$, which is preserved by the channel in Eq.~\eqref{eq:relaxation_kraus_physical}.} The first law establishes the mean balance for the physical process under consideration; it does not by itself imply the stronger Kraus commutation condition. Conversely, unitality alone does not enforce charge conservation: a unitary channel is unital but can change $\vev X$ when its unitary does not commute with $X$.

Reference preservation is a separate requirement on the relaxation channel. For the full faithful dressed reference, including its clock dependence, it reads
\begin{equation}
 \begin{aligned}
 \rho_{\sigma_g}&=F_g(X),\qquad F_g(x)=|g(x)|^2e^{-x},\\
 \E(\rho_{\sigma_g})&=\rho_{\sigma_g}.
 \end{aligned}
 \label{eq:charge_reference_conditions}
\end{equation}
Charge conservation alone can still favor particular states within a degenerate charge sector: resetting that sector to one of its states leaves the charge unchanged but is nonunital. Reference preservation alone is also insufficient. In a $d$-dimensional regulator, the reset channel $\E(\rho)=\operatorname{Tr}(\rho)\rho_{\sigma_g}$ fixes the reference but gives $\E(\id)=d\rho_{\sigma_g}$, which equals $\id$ only for a maximally mixed reference. It can also change the mean charge of other inputs.

Thus neither charge conservation alone nor reference preservation alone guarantees unitality. For the channel considered here, commutation of all Kraus operators with $X$, together with preservation of the faithful reference density operator $F_g(X)$, is sufficient to establish unitality. The next subsection derives this implication and applies it to separated and simultaneous unitary driving and relaxation.

\subsubsection{Separated and time-local evolution}
Combining $[M_\alpha,X]=0$ for all $\alpha$ with $\E(F_g(X))=F_g(X)$ gives
\begin{equation}
 \begin{aligned}
 \E(F_g(X))
 &=\sum_\alpha M_\alpha F_g(X)M_\alpha^\dagger\\
 &=F_g(X)\E(\id)=F_g(X),
 \end{aligned}
 \label{eq:reference_implies_unital_step}
\end{equation}
so faithfulness implies $\E(\id)=\id$. Cancellation is justified on spectral regions where $F_g(X)$ is bounded and bounded away from zero, followed by a limit to full support; no globally bounded inverse is required. In a finite regulator, the reference assigns equal weights within each fixed-charge eigenspace. Conserving the charge and preserving those weights prevents relaxation from favoring particular dressed states in that eigenspace. This does not require equality of every forward and backward transition rate.

The separated protocol of Section~\ref{sec:FT}, with relaxation realized in this class, is consequently unital:
\begin{equation}
 (\E\circ\Ad_U)(\id)=\E(U\id U^\dagger)=\id.
 \label{eq:separated_unital_channel}
\end{equation}
The unitary drive need not conserve $X$ or fix $\sigma_g$. Section~\ref{sec:FT} assumes reference preservation, $\E(\rho_{\sigma_g})=\rho_{\sigma_g}$, but its physical-process charge balance, $\tau[(\rho_\chi-\rho_\omega)X]=0$, concerns the actual relaxation $\omega\to\chi$ at the stated perturbative order. It does not require charge conservation for every input or impose $[M_\alpha,X]=0$ on all Kraus operators. Reference preservation and this balance for the specified evolution therefore do not establish unitality. Section~\ref{sec:FT} can consequently include nonunital relaxation maps: DPI supplies the nonnegative relative-entropy loss, while the gravitational balance and clock-error bounds yield the stated GSL conclusion.

For the simultaneous dynamics of Section~\ref{sec:simultaneous}, keep its notation $\mathcal L_t$ for the dissipator. A sufficient condition is
\begin{equation}
 \begin{aligned}
 \mathcal G_t(\rho)&=-\ii[H_t,\rho]+\mathcal L_t(\rho),\\
 \mathcal L_t(\id)&=0
 \quad\Longrightarrow\quad\mathcal G_t(\id)=0.
 \end{aligned}
 \label{eq:driven_unital_generator}
\end{equation}
The identity is then a stationary solution of the evolution equation, so the well-defined time-ordered propagator satisfies $\mathcal C_T(\id)=\id$. This condition is distinct from $\mathcal L_t(\rho_{\sigma_g})=0$ used in Section~\ref{sec:simultaneous}. For example, if the maps $e^{s\mathcal L_t}$ at fixed $t$ have charge-preserving Kraus operators and fix $F_g(X)$, they preserve $\id$ for all $s\geq0$, so $\mathcal L_t(\id)=0$. Both earlier protocols therefore admit realizations satisfying the unitality condition of the two-time FT.

For the GSL along a particular physical evolution, unitality is sufficient but not necessary, subject to the stated gravitational and clock-control assumptions. It guarantees the normalized two-time FT, whereas a nonunital channel can still satisfy the GSL for the state considered. Nonunitality alone therefore does not imply a GSL violation.

\subsubsection{Low-energy interpretation and a concrete model}
Uniform statistical weights motivate the unital description when the interactions do not preferentially select particular accessible states. For example, the regulated channel \eqref{eq:generic_BH_dilation} is unital if the environment is maximally mixed. A finite thermal environment approximates this case when its energy spread is small compared with its temperature. In the dressed black-hole model, uniformity instead concerns complete dressed states within a fixed-charge sector. Even weak relaxation can favor lower-energy states; small exchanged energy relative to the black-hole mass does not imply unitality. The conservation and reference-preservation requirements above specify what the reduced dynamics must satisfy.

A concrete realization is the charge-preserving random-unitary model
\begin{equation}
 \begin{aligned}
 \mathcal L_t(\rho)&=\gamma_t(V_t\rho V_t^\dagger-\rho),
 \qquad \gamma_t\geq0,\\
 V_t^\dagger V_t&=V_tV_t^\dagger=\id,
 \qquad [V_t,X]=0.
 \end{aligned}
 \label{eq:charge_preserving_model}
\end{equation}
Over a short interval $\dd t$, a unitary jump $V_t$ occurs with probability $\gamma_t\dd t$ to first order. If $V_t$ also changes $\mathsf{k}_\Psi$, conservation of $X$ requires the opposite change in $\mathsf{q}_{\mathcal B}$; otherwise this interaction produces no area--energy exchange. The corresponding dissipative evolution fixes the full reference density operator and preserves expectations of bounded functions of $X$ (and of $X$ and $h_g(X)$ when their moments exist). Its generator satisfies $\mathcal L_t(\id)=0$. It therefore has $R_X=R_g=0$ in Section~\ref{sec:simultaneous}.

Adding arbitrary unitary driving preserves unitality by Eq.~\eqref{eq:driven_unital_generator}. Equivalently, the complete channel is an average over unitary evolutions $W_\eta$ for the possible interaction histories,
\begin{equation}
 \mathcal C_T(\rho)=\int\dd\mathbb P(\eta)\,
 W_\eta\rho W_\eta^\dagger,
 \qquad \int\dd\mathbb P(\eta)=1,
 \label{eq:unitary_history_channel}
\end{equation}
with a probability measure independent of the input. This is a continuous Kraus representation; discrete alternatives give $M_\alpha=\sqrt{p_\alpha}W_\alpha$. Each unitary preserves the identity, so their average does too, even when unitary driving and interactions overlap. This model exhibits the channel property needed by the FT. Identifying its transitions with black-hole scattering and its energy transfer with horizon flux remains a physical requirement; the first law by itself does not determine those transition probabilities.

\section{Conclusion}
\label{sec:conclusion}

We have developed a fixed-algebra proof of the GSL that connects the change of generalized entropy to quantum fluctuation theorems and open-system thermodynamics. An asymptotic horizon charge and its conjugate reading supply a physical quantum clock, and the modular crossed product provides a semifinite trace and density operators representing normal states of the dressed exterior. The same trace-based identities then apply to ordinary type-I quantum mechanics and the type-II dressed algebra. This makes the continuum argument parallel to the regulated thermodynamic proof, while the entropy dictionary separates the generalized entropy (area plus exterior-field entropy) from the entropy of the clock charge distribution and the remainder. For the GSL it is the changes in these quantities that matter.

Our approach is complementary to Wall's influential semiclassical proof. Wall compares restrictions of fixed parent states to the algebras of successive horizon cuts; we evolve states on a fixed dressed exterior algebra. When the incoming and outgoing boundary data each determine the exterior field, both descriptions use the same parent field algebra on the semiclassical background. The resulting state--observable duality is analogous to the Schr\"odinger and Heisenberg pictures, although the unitary driving and relaxation protocol differs from restriction to a later cut. The fixed-algebra formulation makes work, energy exchange, and entropy-production statistics explicit ingredients of the GSL.

We present three complementary channel formulations to explain these ingredients and extend the analysis beyond a separated unitary drive and relaxation. Section~\ref{sec:FT} first considers a unitary drive followed by a reference-preserving relaxation channel. Measuring the same reference surprisal on both sides of the unitary drive gives a mean modular work equal to the relative entropy of the driven state with respect to the reference. DPI makes the subsequent loss of this relative entropy nonnegative. Energy conservation and the physical-process first law convert this entropy balance into a GSL bound with explicit clock corrections. This formulation most directly exposes the physical mechanism and its analogy with the regulated proof. Its limitation is the separation of unitary driving from relaxation, which requires justification when used to approximate a physical process.

Section~\ref{sec:simultaneous} allows unitary driving and dissipation to act simultaneously. A time-local generator separates their contributions to the mean entropy balance, and DPI gives a nonnegative dissipative production rate. This describes the interplay between the unitary drive and horizon response without introducing an intermediate state between unitary driving and relaxation. It requires a controlled Markovian approximation and a dissipative generator that preserves the reference state. To infer the GSL, the dissipative energy loss must agree with the horizon flux, and the associated change in area must agree with the physical-process first law. It establishes a mean entropy inequality; a trajectory fluctuation theorem requires additional measurement and reverse-process statistics.

Section~\ref{sec:fullcore_FT} instead uses only the complete channel between the initial and final states. With faithful preparations and the stated spectral limits, the mean two-time surprisal change equals the dressed trace-entropy change, and the exponential average is $\Gamma_T=\tau[\rho_T\mathcal C_T(\id)]$. Unitality sets $\Gamma_T=1$, normalizes the adjoint Kraus reverse experiment, and yields the GSL bound through the entropy dictionary. This formulation needs neither temporal separation nor a time-local generator. Its additional requirement is a physical justification of unitality, or control of the nonunital correction for the process considered. Commutation of the relaxation Kraus operators with the complete dressed charge, together with preservation of the full faithful dressed reference, provides sufficient conditions for unitality. The first law supplies the energy--area interpretation, while these conditions constrain the reduced transition dynamics. Unitality is sufficient for this FT argument, but is not necessary for the GSL along a particular physical evolution. Neither reference preservation alone nor the first-law mean balance alone establishes it.

The three formulations therefore offer different ways to establish the entropy inequality, with explicit dynamical assumptions and quantitative clock errors. Vanishing errors give an asymptotic GSL in the controlled slow-clock limit; at finite clock width, positivity follows when entropy production dominates the corrections or their combined sign is favorable. To apply these bounds to a particular black-hole perturbation, one must derive its reduced exterior channel, match its energy exchange to the horizon flux where that identification is used, and bound the clock terms for the states it actually produces. The two-time formulation also requires convergence of the measured surprisal statistics. Such a construction would connect the fluctuation statistics to concrete perturbations and test the sufficient conditions identified here. More broadly, this framework provides a route from the mean generalized-entropy inequality to a statistical description of black-hole irreversibility, using the common language of quantum channels, work, and entropy production.

\begin{acknowledgments}
We thank Simon Lin for a helpful discussion.
FLL is supported by Taiwan's National Science and Technology Council (NSTC) through Grant Nos.~115-2112-M-003-009-MY3 and 112-2112-M-003-006-MY3. SM is supported by the ANID FONDECYT Postdoctorado grant number 3240055. We acknowledge the use of GPT-6 Astra in deriving some results and preparing the manuscript.
\end{acknowledgments}

\appendix
\section{Trace formula, density operator, and slow-clock limit}
\label{app:density}
\subsection{Trace and density operator}
In the $\mathrm{QRF}_0$ tensor-product realization, $q_R\ket q=q\ket q$ defines the multiplication operator on the clock factor. The physical horizon charge, represented by $q_R$ in $\mathrm{QRF}_1$, is represented by $X=q_R+K_\Psi$ in $\mathrm{QRF}_0$ [Eq.~\eqref{eq:X_definition}]. A weakly dense set of suitably smeared analytic elements of the modular crossed product has the form
\begin{equation}
 \widehat A=\int_{\mathbb R}\dd s\,A(s)e^{\ii s(K_\Psi+q_R)},\qquad A(s)\in\M.
\end{equation}
With the normalization in Section~\ref{sec:crossed_product}, the canonical trace is
\begin{equation}
 \tau(\widehat A)=\int_{\mathbb R}\dd q\,e^q
 \int_{\mathbb R}\dd s\,e^{\ii qs}\bra\Psi A(s)\ket\Psi,
 \label{eq:explicit_core_trace}
\end{equation}
for such smeared elements with $\tau(|\widehat A|)<\infty$. The $s$ integration is performed before the $q$ integration; the formula does not assert absolute convergence of a double integral for arbitrary elements of the crossed product. For the $\mathrm{QRF}_1$ preparation $\ket{\widehat\Phi}=\ket\Phi\otimes\ket g$, the corresponding $\mathrm{QRF}_0$ state $\nu$ of Eq.~\eqref{eq:density_definition} satisfies
\begin{equation}
 \nu(\widehat A)
 =\bra{\widehat\Phi}U_\Psi^\dagger\widehat A U_\Psi\ket{\widehat\Phi}.
 \label{eq:vector_pairing_appendix}
\end{equation}
Choose the standard positive vector representing $\Phi$ (the natural-cone representative, analogous to $\sqrt{\rho_\Phi}$ for matrices). The relative-modular density operator formula \cite{KudlerFlam:2023qfl,Chandrasekaran:2022cip,Liu:2025krl} is
\begin{equation}
 \rho_\nu
 =g(X)e^{-q_R/2}\Delta_{\Phi\Psi}e^{-q_R/2}g(X)^*.
 \label{eq:positive_density_appendix}
\end{equation}
The products of unbounded operators are defined by their closed positive quadratic form. This density operator reproduces the vector-state expectations, $\tau(\rho_\nu\widehat A)=\nu(\widehat A)$. It contains $\Delta_{\Phi\Psi}$, whereas the Araki relative entropy in Eq.~\eqref{eq:Araki_relative_entropy_definition} uses the reversed indices $\Delta_{\Psi\Phi}$. Equation~\eqref{eq:positive_density_appendix} applies to the specified product preparation and its transported state; applying the entropy dictionary to a general channel output requires assumption (A3) of Section~\ref{sec:GSL_proof}.

The clock translation $\bra qU_\Psi\ket g=g(q+K_\Psi)$ explains the shifted packet. Since the clock operator $q_R$ and the field relative modular operator act on separate factors in this realization, the symmetric middle factor can also be written as $e^{-q_R}\Delta_{\Phi\Psi}$. It does not follow that $g(X)$ commutes with that operator. The trace and vector evaluations refer to the same observable in $\mathrm{QRF}_0$.

For $\Phi=\Psi$, $\Delta_{\Psi\Psi}=e^{-K_\Psi}$, so Eq.~\eqref{eq:positive_density_appendix} reduces to $\rho_{\sigma_g}=|g(X)|^2e^{-X}$. The trace check is explicit:
\begin{align}
 \tau(f(X))&=\int\dd x\,e^x f(x),\\
 \tau(\rho_{\sigma_g})&=\int\dd q\,|g(q)|^2=1,\\
 \tau(e^{-X})&=\int_{\mathbb R}\dd x=\infty,\\
 S_\tau(\sigma_g)&=\int\dd q\,|g(q)|^2
 [q-\log|g(q)|^2].
\end{align}
The first identity holds for positive $f$ as an extended trace, or for functions satisfying $\int\dd x\,e^x|f(x)|<\infty$; the entropy identity assumes finite charge and logarithmic moments. Rescaling the trace to $c\tau$, $c>0$, replaces $\rho_\nu$ by $\rho_\nu/c$ and shifts $S_\tau(\nu)$ by $\log c$. Normalized expectations, relative entropy, and entropy differences are unchanged.

\subsection{Relative-entropy limit for a slow clock}
\label{app:slow_clock_limit}
Imagine preparing the exterior field in either $\Phi$ or $\Psi$ and attaching the same normalized clock packet $g_\epsilon$ in both cases. Before dressing, the joint algebra is $\M\,\overline\otimes\,\mathcal B(\mathcal H_C)$, where $\mathcal H_C=L^2(\mathbb R)$ is the clock Hilbert space. The two product states on this algebra have relative entropy $D_\M(\Phi\Vert\Psi)$: they differ only on the exterior field algebra $\M$. Transporting both states and this algebra by $U_\Psi$ preserves the relative entropy. Equation~\eqref{eq:slow_clock_relative_entropy} asks how much of this distinction survives when only the dressed exterior observables are available. We keep $\Phi,\Psi$ fixed, assume $D_\M(\Phi\Vert\Psi)<\infty$, and use the normalized packets of Eq.~\eqref{eq:slow_clock_packet} with the support and moment conditions of Section~\ref{subsec:qrf0_entropy}.

First consider what an observer measuring only a field observable $A\in\M$ would see in $\mathrm{QRF}_0$. At a given value $p$ of the clock reading, the frame change $U_\Psi$ acts on the field as $e^{-\ii pK_\Psi}$. If the reading is not measured, one averages over it. Thus the restriction to the exterior field algebra, $\Phi_\epsilon:=\nu_\epsilon|_\M$, satisfies
\begin{equation}
 \Phi_\epsilon(A)=\int\dd p\,|\widetilde g_\epsilon(p)|^2
 \bra\Phi e^{\ii pK_\Psi} A e^{-\ii pK_\Psi}\ket\Phi,
 \label{eq:slow_clock_modular_average}
\end{equation}
where $\widetilde g_\epsilon(p)=\langle p|g_\epsilon\rangle$. The reference remains exactly $\Psi$ because $K_\Psi\ket\Psi=0$. For the other preparation, the size of the shifts is set by the Fourier transform of the packet:
\[
 |\widetilde g_\epsilon(p)|^2
 =\epsilon^{-1}|\widetilde g_0(p/\epsilon)|^2.
\]
For these rescaled packets the distribution of the conjugate reading $p$ therefore concentrates at zero as $\epsilon\to0^+$. Put $p=\epsilon u$. For each fixed $u$, $e^{-\ii\epsilon uK_\Psi}\ket\Phi$ tends to $\ket\Phi$. For any field observable of norm at most one, the change in its expectation is bounded by $2\|(e^{-\ii\epsilon uK_\Psi}-\id)\ket\Phi\|$. This bound tends to zero for each $u$ and never exceeds four. Its average with the fixed probability distribution $|\widetilde g_0(u)|^2$ therefore tends to zero, independently of the observable. In symbols,
\begin{equation}
 \sup_{A\in\M,\,\|A\|\leq1}
 |\Phi_\epsilon(A)-\Phi(A)|\;\longrightarrow\;0 .
 \label{eq:slow_clock_norm_limit}
\end{equation}

In $\mathrm{QRF}_0$ the field algebra $\M\otimes\id$ lies inside the dressed algebra $\Mh_0$, which in turn lies inside the transported joint algebra $U_\Psi[\M\,\overline\otimes\,\mathcal B(\mathcal H_C)]U_\Psi^\dagger$. Relative entropy cannot increase upon restriction to a smaller algebra. Applying this twice gives
\begin{equation}
 \begin{aligned}
 D_\M(\Phi_\epsilon\Vert\Psi)
 &\leq D_\tau(\nu_\epsilon\Vert\sigma_{g_\epsilon})\\
 &\leq D_\M(\Phi\Vert\Psi).
 \end{aligned}
 \label{eq:slow_clock_sandwich}
\end{equation}
Norm convergence of states does not, by itself, imply continuity of relative entropy. What suffices here is its lower semicontinuity \cite{Araki:1976zv,Araki:1977relative}: by Eq.~\eqref{eq:slow_clock_norm_limit}, $\liminf_{\epsilon\to0^+}D_\M(\Phi_\epsilon\Vert\Psi)\geq D_\M(\Phi\Vert\Psi)$. Yet the left side of Eq.~\eqref{eq:slow_clock_sandwich} is bounded above by this very value for every $\epsilon$. The two bounds must meet. In particular,
\begin{equation}
 0\leq r_\epsilon(\nu_\epsilon)
 \leq D_\M(\Phi\Vert\Psi)-D_\M(\Phi_\epsilon\Vert\Psi)
 \longrightarrow0.
 \label{eq:slow_clock_remainder_bound}
\end{equation}
This argument controls the loss for the specified product preparations without assigning a convergence rate or introducing a state at $\epsilon=0$. For a channel output with clock--field correlations, the average in Eq.~\eqref{eq:slow_clock_modular_average} need not describe its restriction to $\M$, and its entropy remainder requires a separate estimate.

\section{Gaussian clock corrections and error bounds}
\label{app:gaussian}
We use the separated physical process of Proposition~\ref{prop:GSL} and the Gaussian reference in Eq.~\eqref{eq:gaussian_packet}. Let $\mu_\nu=\vev{X}_\nu$ and $V_\nu=\vev{(X-\mu_\nu)^2}_\nu$. At the second perturbative order used for the area response, Eq.~\eqref{eq:relaxation_charge_conservation} gives $\mu_\chi=\mu_\omega$. All bounds in this appendix are at that order; they estimate finite clock width, not higher gravitational variations. Since
\[
 \vev{(X-x_0)^2}_\nu=V_\nu+(\mu_\nu-x_0)^2,
\]
the squared-mean terms cancel during relaxation, and
\begin{equation}
 \Delta_{\rm rel}\vev{h_\epsilon(X)}
 =\frac{\epsilon^2}{2}(V_\chi-V_\omega).
 \label{eq:gaussian_matched_moments}
\end{equation}
Here $h_\epsilon$ is the fixed function $h_{g_\epsilon}=-\log|g_\epsilon|^2$, evaluated in two different states. Substitution in the entropy balance gives Eq.~\eqref{eq:gaussian_balance}; $\Delta S[g]$ and $\Delta r_\epsilon$ compare $\chi$ with $\sigma_g$ over the full protocol. The uniform bound $|V_\chi-V_\omega|\leq M$ and the separate packet-entropy and remainder bounds then give Eq.~\eqref{eq:gaussian_bound}. A shift $g(q)\mapsto g(q-d)$ of the clock packet, with its shape unchanged, preserves $S[g]$ by a change of integration variable. If an independent estimate establishes $|r_\epsilon(\nu)|\leq C_r\epsilon^p$ in both $\sigma_g$ and $\chi$, with $p>0$ and $C_r$ independent of $\epsilon$, then $|\Delta r_\epsilon|\leq2C_r\epsilon^p$. For the exact reference preparation, $r_\epsilon(\sigma_g)=0$, so only the final-state bound is needed. When packet-entropy differences cancel, the resulting error bound is $O(\epsilon^{\min\{2,p\}})$; this is an upper bound, not a determination of the leading error. Merely knowing that the initial variance is $\epsilon^{-2}$ does not bound its change.

If the actual initial and final states have Gaussian clock packets $g_\nu$ with variances $v_\sigma\equiv v_{\sigma_g}=\epsilon^{-2}$ and $v_\chi=\epsilon^{-2}+b_\epsilon$, then their differential entropies $S[g_\nu]=\tfrac12\log(2\pi e\,v_\nu)$ give
\begin{equation}
 \Delta S[g]=\frac12\log(1+\epsilon^2b_\epsilon).
 \label{eq:gaussian_frame_entropy}
\end{equation}
With $|V_\chi-V_\omega|\leq M$, $|b_\epsilon|\leq B$, and $\epsilon^2B<1$, the inequality $|\log(1+z)|\leq|z|/(1-|z|)$ for $|z|<1$ gives
\begin{equation}
 \begin{aligned}
 |\Delta S_{\rm gen}-\Sigma_D|
 \leq{}&\frac{M\epsilon^2}{2}
 +\frac{B\epsilon^2}{2(1-B\epsilon^2)}+\eta_r.
 \end{aligned}
\end{equation}
Here $\eta_r$ bounds $|\Delta r_\epsilon|$ as in Eq.~\eqref{eq:GSL_frame_error_bounds}. As in the main text, additional entropy sectors must either have no changing contribution or be estimated separately. If $|g_\nu|^2$ is the actual distribution of the physical charge, then $v_\nu=V_\nu$: $q_R$ in $\mathrm{QRF}_1$ and $X$ in $\mathrm{QRF}_0$ represent the same measurement. This is automatic for the transported product preparations. A model packet assigned to a channel output cannot be substituted for its measured charge distribution without justification. Even when $v_\nu=V_\nu$ state by state, $v_\chi-v_\sigma$ compares the full protocol and $V_\chi-V_\omega$ only relaxation. A Gaussian charge distribution alone does not bound clock--field correlations or $\Delta r_\epsilon$. The limit concerns differences along a family of normal states at finite $\epsilon$; it does not assert that a normalizable flat-clock state exists at $\epsilon=0$.

A stronger cancellation requires equality of the two variance changes, not merely bounds on them. If $V_\chi-V_\omega=v_\chi-v_\sigma=b_\epsilon$, with $|b_\epsilon|\leq B$, then
\begin{equation}
 \begin{aligned}
 \Delta S_{\rm gen}-\Sigma_D
 &=\tfrac12[z_\epsilon-\log(1+z_\epsilon)]-\Delta r_\epsilon,\\
 z_\epsilon&=\epsilon^2b_\epsilon.
 \end{aligned}
\end{equation}
The explicit clock term is nonnegative for $z_\epsilon>-1$ and equals $b_\epsilon^2\epsilon^4/4+O(\epsilon^6)$ uniformly for bounded $b_\epsilon$. The fourth-order cancellation concerns only the two explicit Gaussian terms; $\Delta r_\epsilon$ still enters the total error. In the charge-preserving channel class of Eq.~\eqref{eq:charge_preserving_model}, the relaxation term $\Delta_{\rm rel}\vev{h_g(X)}$ instead vanishes exactly; packet-entropy differences over the full protocol and the entropy remainder, which can also reflect clock--field correlations, still require separate estimates.

\section{Reference subtraction and state dependence after unitary driving}
\label{app:reference_subtraction}
This appendix explains two steps used in Section~\ref{sec:FT_to_GSL}: why Eq.~\eqref{eq:relative_surprisal_identity} needs no explicit reference subtraction, and how the state prepared by the unitary drive enters the modular-work identity preceding Eq.~\eqref{eq:physical_modular_work}. Throughout, $\tau$ is the fixed trace on the dressed algebra, $\rho_{\sigma_g}$ is the faithful normalized reference density operator, and $k_g=-\log\rho_{\sigma_g}$. We first assume the individual trace expressions and entropies are finite. If only subtracted quantities are finite, the identities require a common regulator and a justified limit; they must not be interpreted as subtraction of two infinite entropies.

\subsection{The implicit reference subtraction}
For any admissible normal state $\nu$, Eq.~\eqref{eq:full_relative_entropy} gives
\begin{equation}
 \begin{aligned}
 D_\tau(\nu\Vert\sigma_g)
 &=\tau[\rho_\nu(\log\rho_\nu-\log\rho_{\sigma_g})]\\
 &=\vev{k_g}_\nu-S_\tau(\nu).
 \end{aligned}
 \label{eq:appendix_reference_identity}
\end{equation}
The first term is the expectation of the reference surprisal in state $\nu$, not the entropy of $\nu$. The reference state therefore already enters through $k_g$. In particular,
\begin{equation}
 \vev{k_g}_{\sigma_g}
 =-\tau(\rho_{\sigma_g}\log\rho_{\sigma_g})
 =S_\tau(\sigma_g).
 \label{eq:appendix_reference_zero}
\end{equation}
Neither side of this equality is generally zero. Their difference is zero, so subtracting it from Eq.~\eqref{eq:appendix_reference_identity} changes nothing:
\begin{equation}
 \begin{aligned}
 D_\tau(\nu\Vert\sigma_g)
 ={}&\bigl(\vev{k_g}_\nu-\vev{k_g}_{\sigma_g}\bigr)\\
 &-\bigl(S_\tau(\nu)-S_\tau(\sigma_g)\bigr).
 \end{aligned}
 \label{eq:appendix_reference_differences}
\end{equation}
Thus Eq.~\eqref{eq:relative_surprisal_identity} and the explicitly reference-subtracted formula are identical. The absence of subtraction symbols in the former does not mean that the reference contribution has been neglected.

\subsection{Unitary driving and the information about the state}
For the separated protocol, $U$ is the physical driving unitary in the fixed dressed algebra, not the representation-changing unitary $U_\Psi$. The unitary drive produces
\begin{equation}
 \rho_\omega=U\rho_{\sigma_g}U^\dagger,\qquad
 \log\rho_\omega=U(\log\rho_{\sigma_g})U^\dagger.
 \label{eq:appendix_unitary_density}
\end{equation}
The second equality follows from functional calculus. Trace invariance then gives
\begin{equation}
 \begin{aligned}
 S_\tau(\omega)
 &=-\tau[U\rho_{\sigma_g}(\log\rho_{\sigma_g})U^\dagger]\\
 &=S_\tau(\sigma_g).
 \end{aligned}
 \label{eq:appendix_unitary_entropy}
\end{equation}
Define the change of a fixed observable under the unitary drive by
\begin{equation}
 \Delta_{\rm drv}\vev O
 :=\vev O_\omega-\vev O_{\sigma_g}
 =\tau[(\rho_\omega-\rho_{\sigma_g})O].
 \label{eq:appendix_drive_difference}
\end{equation}
Using Eq.~\eqref{eq:appendix_reference_differences} and invariance of $S_\tau$ under the unitary drive yields
\begin{equation}
 \begin{aligned}
 D_\tau(\omega\Vert\sigma_g)
 &=\Delta_{\rm drv}\vev{k_g}-\Delta_{\rm drv}S_\tau\\
 &=\Delta_{\rm drv}\vev{k_g}.
 \end{aligned}
 \label{eq:appendix_drive_relative}
\end{equation}
For the normalized reference packet, $k_g=X+h_g(X)$, so
\begin{equation}
 \begin{aligned}
 D_\tau(\omega\Vert\sigma_g)
 &=\Delta_{\rm drv}\vev X
   +\Delta_{\rm drv}\vev{h_g(X)}\\
 &=\tau[(\rho_\omega-\rho_{\sigma_g})X]\\
 &\quad+\tau[(\rho_\omega-\rho_{\sigma_g})h_g(X)].
 \end{aligned}
 \label{eq:appendix_drive_explicit}
\end{equation}
The dependence on $\omega$ is inside both expectation-value differences. The reference observable is fixed; its expectation changes because the state changes. Equivalently, the dependence on the applied unitary can be displayed as
\begin{equation}
 D_\tau(\omega\Vert\sigma_g)
 =\tau[\rho_{\sigma_g}(U^\dagger k_gU-k_g)].
 \label{eq:appendix_drive_heisenberg}
\end{equation}

The reduction to a single reference-surprisal difference relies on obtaining $\omega$ by unitarily conjugating $\sigma_g$. A general output state has the additional entropy-difference term in Eq.~\eqref{eq:appendix_reference_differences}. The unitary drive preserves the fine-grained trace entropy $S_\tau$; the area, exterior-field and clock terms need not be separately constant. The modular-work identity is exact. Identifying physical work additionally requires energy balance, while identifying generalized entropy requires the entropy dictionary.

\subsection{Area response and the physical-work interpretation}
The exact identity \eqref{eq:appendix_drive_explicit} still contains the change of the combined charge $X$. To express it as physical work, use the charge constraint $X=\mathsf{q}_{\mathcal B}+\mathsf{k}_\Psi$ in the same fixed representation:
\begin{equation}
 \Delta_{\rm drv}\vev X
 =\Delta_{\rm drv}\vev{\mathsf{q}_{\mathcal B}}
 +\Delta_{\rm drv}\vev{\mathsf{k}_\Psi}.
 \label{eq:appendix_drive_charge}
\end{equation}
The additional physical input is the matching
\begin{equation}
 \Delta_{\rm drv}\vev{\mathsf{k}_\Psi}
 =\beta_H\Delta_{\rm drv}E_{\rm out}
 =\beta_H W.
 \label{eq:appendix_drive_matching}
\end{equation}
Here $\mathsf{k}_\Psi=\beta_H H_{\rm out}$ and $E_{\rm out}(\nu)=\vev{H_{\rm out}}_\nu$, as defined in Eq.~\eqref{eq:exterior_energy_definition}. Thus $H_{\rm out}$ is the Killing Hamiltonian and $E_{\rm out}$ its expected energy. With energy exchange across the boundaries negligible during the unitary drive, $W=\Delta_{\rm drv}E_{\rm out}$ is the mean work supplied to the exterior. Equations~\eqref{eq:appendix_drive_explicit}--\eqref{eq:appendix_drive_matching} yield the full expression
\begin{equation}
 \begin{aligned}
 D_\tau(\omega\Vert\sigma_g)
 ={}&\Delta_{\rm drv}\vev{\mathsf{q}_{\mathcal B}}+\beta_H W\\
 &+\Delta_{\rm drv}\vev{h_g(X)}.
 \end{aligned}
 \label{eq:appendix_full_physical_work}
\end{equation}
This expression includes the area-charge change caused by the unitary drive. To obtain Eq.~\eqref{eq:physical_modular_work}, impose the further condition
\begin{equation}
 \Delta_{\rm drv}\vev{\mathsf{q}_{\mathcal B}}
 =\vev{\mathsf{q}_{\mathcal B}}_\omega-\vev{\mathsf{q}_{\mathcal B}}_{\sigma_g}=0.
 \label{eq:appendix_zero_drive_area}
\end{equation}
This sets the \emph{change} of the expected area charge to zero, not its absolute expectation. If its reference value has separately been chosen to vanish, the condition can be written $\vev{\mathsf{q}_{\mathcal B}}_\omega=0$ in that convention; no such choice is required. Neither unitary preservation of the full entropy nor a short duration of unitary driving alone establishes Eq.~\eqref{eq:appendix_zero_drive_area}. Without it, the area term in Eq.~\eqref{eq:appendix_full_physical_work} must remain. The relaxation response uses the bifurcation-to-stationary comparison specified in Section~\ref{subsec:area_endpoint}.

\section{Transported horizon charges and physical generator matching}
\label{app:charge_transport}

\subsection{Regulated expectation values and representation covariance}
The classical integrated relation \eqref{eq:right_branch_area_constraint} follows from the right-branch flux balance. In the quantum theory, a sharp one-sided flux can have divergent fluctuations even though the full modular generator exists \cite{KudlerFlam:2023qfl}. The following argument therefore starts at a common regulator, denoted by $\Lambda$, with compatible charge, area, and flux expectation values satisfying the constraint to second perturbative order. Cutoff boundary terms must be included until they cancel or vanish in the limit. In the following formulas $U_\Psi$ denotes the common change of representation at the chosen regulator. Define
\begin{equation}
 \begin{aligned}
 X_\Lambda&=U_\Psi q_{R,\Lambda}U_\Psi^\dagger,\\
 \mathsf{q}_{\mathcal B,\Lambda}&=U_\Psi q_{\mathcal B,\Lambda}U_\Psi^\dagger,\\
 \mathsf{k}_{\Psi,\Lambda}&=U_\Psi k_{\Psi,\Lambda}U_\Psi^\dagger.
 \end{aligned}
 \label{eq:regulated_charge_transport}
\end{equation}
For every admissible vector $\psi$, with $\psi^U=U_\Psi\psi$, unitary covariance of the regulated quadratic forms gives
\begin{equation}
 \langle\psi^U,\mathsf{k}_{\Psi,\Lambda}\psi^U\rangle
 =\langle\psi,k_{\Psi,\Lambda}\psi\rangle,
 \label{eq:transport_pairing_covariance}
\end{equation}
and similarly for the right-branch area and charge. For a regulated state described by a density operator, this is the cyclicity of the trace. To distinguish the frames temporarily, write $\tau_{\rm orig}$ for the trace on $\Mh_1$ in $\mathrm{QRF}_1$ and $\tau$ for the trace on $\Mh_0$ in $\mathrm{QRF}_0$ used in the main text. Their relation is $\tau(U_\Psi A U_\Psi^\dagger)=\tau_{\rm orig}(A)$. This simultaneous transformation of the algebra, trace, and state requires no assumption that $U_\Psi$ belongs to the original crossed product and describes no physical unitary driving process.

Let $\nu_0$ be the common reference, and let $\Delta_\nu\langle O\rangle=\langle O\rangle_\nu-\langle O\rangle_{\nu_0}$. The reference is transported together with the state: for $O^U=U_\Psi O U_\Psi^\dagger$, the transformed difference is $\Delta_{\nu^U}\langle O^U\rangle=\langle O^U\rangle_{\nu^U}-\langle O^U\rangle_{\nu_0^U}$, with $\nu_0^U=\sigma_g$ for the reference preparation of the main text. For an observable $O$, the finite reference-subtracted expectation value means $\Delta_\nu\langle O\rangle=\lim_{\Lambda\to\infty}[\langle O_\Lambda\rangle_\nu-\langle O_\Lambda\rangle_{\nu_0}]$, when this limit exists; the individual unregulated expectations need not exist. If the three subtracted limits exist, conjugating the regulated constraint and taking the same limit yields
\begin{equation}
 \Delta_{\nu^U}\langle\mathsf{q}_{\mathcal B}\rangle
 =\Delta_{\nu^U}\langle X\rangle
 -\Delta_{\nu^U}\langle \mathsf{k}_\Psi\rangle.
 \label{eq:finite_transport_constraint}
\end{equation}
Thus, when separate continuum operators do not exist, $\mathsf{q}_{\mathcal B}=X-\mathsf{k}_\Psi$ denotes the displayed equality of finite, reference-subtracted expectation values. Unitary conjugation alone establishes neither these limits nor affiliation of $\mathsf{k}_\Psi$ with the type-II algebra. The full dressed density operator and its surprisal remain distinct, well-defined objects under the normality and domain assumptions of the main text.

\subsection{What must be checked for a physical right-wedge generator}
At a common regulator, suppose $k_{\Psi,\Lambda}$ generates the right-branch energy flow $\alpha_{s,\Lambda}(A_\Lambda)=e^{\ii s k_{\Psi,\Lambda}}A_\Lambda e^{-\ii s k_{\Psi,\Lambda}}$. Its transported action satisfies
\begin{equation}
 \begin{aligned}
 \alpha_{s,\Lambda}^U(U_\Psi A_\Lambda U_\Psi^\dagger)
 &:=U_\Psi\alpha_{s,\Lambda}(A_\Lambda)U_\Psi^\dagger\\
 &=e^{\ii s \mathsf{k}_{\Psi,\Lambda}}(U_\Psi A_\Lambda U_\Psi^\dagger)e^{-\ii s \mathsf{k}_{\Psi,\Lambda}}.
 \end{aligned}
 \label{eq:transported_energy_flow}
\end{equation}
Here $s$ is the dimensionless parameter of forward Killing evolution ($t_{\rm Killing}=\beta_H s$ under the energy identification). The convention $\alpha_t^\Psi(A)=e^{-\ii tK_\Psi}A e^{\ii tK_\Psi}$ instead assigns the opposite sign to the modular parameter $t$. Conjugation transports the one-sided energy flow; it does not identify its generator with the full $K_\Psi$. If another operator is proposed to generate that flow, agreement of the automorphisms alone does not fix its energy expectation values: implementers can differ by a commuting contribution. The common gravitational charge constraint, reference convention, and sector must also fix that freedom. In the continuum this comparison must be made using the admissible finite, reference-subtracted expectation values, without assuming a bare one-sided operator exists.

Finally, representing the incoming right-branch energy is different from identifying its change with outgoing exterior Killing energy. The GSL application separately assumes
\begin{equation}
 \Delta_{\rm rel}\langle \mathsf{k}_\Psi\rangle
 =\beta_H\Delta_{\rm rel}E_{\rm out},
 \label{eq:transport_scattering_matching}
\end{equation}
with the radiation reaching future null infinity included in the exterior energy account. Other boundary charges must be held fixed or included explicitly in the balance. Neither Eq.~\eqref{eq:transported_energy_flow} nor the identity $X=\mathsf{q}_{\mathcal B}+\mathsf{k}_\Psi$ proves this scattering relation or the future-horizon area response. The focusing calculation in Section~\ref{subsec:area_endpoint} establishes the geometric area response between a regular bifurcation cross-section and the stationary future limit. Its identification with the dressed area-charge change, together with the scattering correspondence above, gives Eq.~\eqref{eq:relaxation_charge_conservation}. Any mismatch remains explicit in $\Delta_{\rm rel}\langle X\rangle$ in the entropy balance.

\bibliographystyle{apsrev4-2}
\bibliography{GSL_FT_QRF_refs}

@article{Witten:2018zxz,
    author = "Witten, Edward",
    title = "{APS Medal for Exceptional Achievement in Research: Invited article on entanglement properties of quantum field theory}",
    eprint = "1803.04993",
    archivePrefix = "arXiv",
    primaryClass = "hep-th",
    doi = "10.1103/RevModPhys.90.045003",
    journal = "Rev. Mod. Phys.",
    volume = "90",
    number = "4",
    pages = "045003",
    year = "2018"
}

@article{DeVuyst:2024fxc,
    author = "De Vuyst, Julian and Eccles, Stefan and Hoehn, Philipp A. and Kirklin, Josh",
    title = "{Crossed products and quantum reference frames: on the observer-dependence of gravitational entropy}",
    eprint = "2412.15502",
    archivePrefix = "arXiv",
    primaryClass = "hep-th",
    doi = "10.1007/JHEP07(2025)063",
    number = {2025},
    journal = "JHEP",
    volume = "07",
    pages = "063",
    year = "2025",
    note = "[Erratum: JHEP 10, 234 (2025)]"
}

@inproceedings{Liu:2025krl,
    author = "Liu, Hong",
    title = "{Lectures on entanglement, von Neumann algebras, and emergence of spacetime}",
    booktitle = "{Theoretical Advanced Study Institute in Elementary Particle Physics 2023}: {Aspects of Symmetry}",
    eprint = "2510.07017",
    archivePrefix = "arXiv",
    primaryClass = "hep-th",
    reportNumber = "MIT-CTP/5938",
    month = "10",
    year = "2025"
}

@article{Hollands:2012sf,
    author = "Hollands, Stefan and Wald, Robert M.",
    title = "{Stability of Black Holes and Black Branes}",
    eprint = "1201.0463",
    archivePrefix = "arXiv",
    primaryClass = "gr-qc",
    doi = "10.1007/s00220-012-1638-1",
    journal = "Commun. Math. Phys.",
    volume = "321",
    pages = "629--680",
    year = "2013"
}

@article{Hollands:2024vbe,
    author = "Hollands, Stefan and Wald, Robert M. and Zhang, Victor G.",
    title = "{Entropy of dynamical black holes}",
    eprint = "2402.00818",
    archivePrefix = "arXiv",
    primaryClass = "hep-th",
    doi = "10.1103/PhysRevD.110.024070",
    journal = "Phys. Rev. D",
    volume = "110",
    number = "2",
    pages = "024070",
    year = "2024"
}

@article{Wall:2011hj,
    author = "Wall, Aron C.",
    title = "{A proof of the generalized second law for rapidly changing fields and arbitrary horizon slices}",
    eprint = "1105.3445",
    archivePrefix = "arXiv",
    primaryClass = "gr-qc",
    doi = "10.1103/PhysRevD.85.104049",
    journal = "Phys. Rev. D",
    volume = "85",
    pages = "104049",
    year = "2012",
    note = "[Erratum: Phys.Rev.D 87, 069904 (2013)]"
}

@article{Talkner:2007sf,
    author = "Talkner, Peter and Lutz, Eric and H{\"a}nggi, Peter",
    title = "{Fluctuation theorems: Work is not an observable}",
    doi = "10.1103/PhysRevE.75.050102",
    journal = "Phys. Rev. E",
    volume = "75",
    pages = "050102",
    year = "2007"
}

@article{Bekenstein:1973ur,
    author = "Bekenstein, Jacob D.",
    title = "{Black holes and entropy}",
    doi = "10.1103/PhysRevD.7.2333",
    journal = "Phys. Rev. D",
    volume = "7",
    pages = "2333--2346",
    year = "1973"
}

@article{Hawking:1975vcx,
    author = "Hawking, Stephen W.",
    title = "{Particle creation by black holes}",
    doi = "10.1007/BF02345020",
    journal = "Commun. Math. Phys.",
    volume = "43",
    pages = "199--220",
    year = "1975",
    note = "[Erratum: Commun. Math. Phys. 46, 206 (1976)]"
}

@article{Araki:1976zv,
    author = "Araki, Huzihiro",
    title = "{Relative entropy of states of von Neumann algebras}",
    doi = "10.2977/prims/1195191148",
    journal = "Publ. Res. Inst. Math. Sci.",
    volume = "11",
    pages = "809--833",
    year = "1976"
}

@article{Bisognano:1975ih,
    author = "Bisognano, Joseph J. and Wichmann, Eyvind H.",
    title = "{On the duality condition for a Hermitian scalar field}",
    doi = "10.1063/1.522605",
    journal = "J. Math. Phys.",
    volume = "16",
    pages = "985--1007",
    year = "1975"
}

@book{Haag:1996,
    author = "Haag, Rudolf",
    title = "{Local Quantum Physics: Fields, Particles, Algebras}",
    edition = "2",
    publisher = "Springer",
    address = "Berlin",
    doi = "10.1007/978-3-642-61458-3",
    year = "1996"
}

@article{Takesaki:1973,
    author = "Takesaki, Masamichi",
    title = "{Duality for crossed products and the structure of von Neumann algebras of type III}",
    doi = "10.1007/BF02392041",
    journal = "Acta Math.",
    volume = "131",
    pages = "249--310",
    year = "1973"
}

@article{Witten:2022qdt,
    author = "Witten, Edward",
    title = "{Gravity and the crossed product}",
    eprint = "2112.12828",
    archivePrefix = "arXiv",
    primaryClass = "hep-th",
    doi = "10.1007/JHEP10(2022)008",
    number = {2022},
    journal = "JHEP",
    volume = "10",
    pages = "008",
    year = "2022"
}

@article{Chandrasekaran:2022eqq,
    author = "Chandrasekaran, Venkatesa and Penington, Geoff and Witten, Edward",
    title = "{Large N algebras and generalized entropy}",
    eprint = "2209.10454",
    archivePrefix = "arXiv",
    primaryClass = "hep-th",
    doi = "10.1007/JHEP04(2023)009",
    number = {2023},
    journal = "JHEP",
    volume = "04",
    pages = "009",
    year = "2023"
}

@article{Chandrasekaran:2022cip,
    author = "Chandrasekaran, Venkatesa and Longo, Roberto and Penington, Geoff and Witten, Edward",
    title = "{An algebra of observables for de Sitter space}",
    eprint = "2206.10780",
    archivePrefix = "arXiv",
    primaryClass = "hep-th",
    doi = "10.1007/JHEP02(2023)082",
    number = {2023},
    journal = "JHEP",
    volume = "02",
    pages = "082",
    year = "2023"
}

@article{KudlerFlam:2023qfl,
    author = "Kudler-Flam, Jonah and Leutheusser, Samuel and Satishchandran, Gautam",
    title = "{Generalized black hole entropy is von Neumann entropy}",
    eprint = "2309.15897",
    archivePrefix = "arXiv",
    primaryClass = "hep-th",
    doi = "10.1103/PhysRevD.111.025013",
    journal = "Phys. Rev. D",
    volume = "111",
    number = "2",
    pages = "025013",
    year = "2025"
}

@article{Jensen:2023yxy,
    author = "Jensen, Kristan and Sorce, Jonathan and Speranza, Antony J.",
    title = "{Generalized entropy for general subregions in quantum gravity}",
    eprint = "2306.01837",
    archivePrefix = "arXiv",
    primaryClass = "hep-th",
    doi = "10.1007/JHEP12(2023)020",
    number = {2023},
    journal = "JHEP",
    volume = "12",
    pages = "020",
    year = "2023"
}

@misc{Mondal:2026,
    author = "Mondal, Avinandan and Prabhu, Kartik",
    title = "{Perturbative semiclassical entropy of dynamical black holes}",
    eprint = "2603.02320",
    archivePrefix = "arXiv",
    primaryClass = "hep-th",
    year = "2026"
}

@article{IyerWald:1994,
    author = "Iyer, Vivek and Wald, Robert M.",
    title = "{Some properties of Noether charge and a proposal for dynamical black hole entropy}",
    eprint = "gr-qc/9403028",
    archivePrefix = "arXiv",
    doi = "10.1103/PhysRevD.50.846",
    journal = "Phys. Rev. D",
    volume = "50",
    pages = "846--864",
    year = "1994"
}

@article{Jarzynski:1997,
    author = "Jarzynski, Christopher",
    title = "{Nonequilibrium equality for free energy differences}",
    doi = "10.1103/PhysRevLett.78.2690",
    journal = "Phys. Rev. Lett.",
    volume = "78",
    pages = "2690--2693",
    year = "1997"
}

@article{Crooks:1999,
    author = "Crooks, Gavin E.",
    title = "{Entropy production fluctuation theorem and the nonequilibrium work relation for free energy differences}",
    doi = "10.1103/PhysRevE.60.2721",
    journal = "Phys. Rev. E",
    volume = "60",
    pages = "2721--2726",
    year = "1999"
}

@article{Campisi:2011,
    author = "Campisi, Michele and H{\"a}nggi, Peter and Talkner, Peter",
    title = "{Colloquium: Quantum fluctuation relations: Foundations and applications}",
    eprint = "1012.2268",
    archivePrefix = "arXiv",
    primaryClass = "cond-mat.stat-mech",
    doi = "10.1103/RevModPhys.83.771",
    journal = "Rev. Mod. Phys.",
    volume = "83",
    pages = "771--791",
    year = "2011"
}

@misc{Kurchan:2000,
    author = "Kurchan, Jorge",
    title = "{A quantum fluctuation theorem}",
    eprint = "cond-mat/0007360",
    archivePrefix = "arXiv",
    year = "2000"
}

@misc{Tasaki:2000,
    author = "Tasaki, Hal",
    title = "{Jarzynski relations for quantum systems and some applications}",
    eprint = "cond-mat/0009244",
    archivePrefix = "arXiv",
    year = "2000"
}

@article{Kwon:2019,
    author = "Kwon, Hyukjoon and Kim, M. S.",
    title = "{Fluctuation theorems for a quantum channel}",
    eprint = "1810.03150",
    archivePrefix = "arXiv",
    primaryClass = "quant-ph",
    doi = "10.1103/PhysRevX.9.031029",
    journal = "Phys. Rev. X",
    volume = "9",
    number = "3",
    pages = "031029",
    year = "2019"
}

@article{PageWootters:1983,
    author = "Page, Don N. and Wootters, William K.",
    title = "{Evolution without evolution: Dynamics described by stationary observables}",
    doi = "10.1103/PhysRevD.27.2885",
    journal = "Phys. Rev. D",
    volume = "27",
    pages = "2885--2892",
    year = "1983"
}

@article{Giacomini:2019,
    author = "Giacomini, Flaminia and Castro-Ruiz, Esteban and Brukner, {\v{C}}aslav",
    title = "{Quantum mechanics and the covariance of physical laws in quantum reference frames}",
    eprint = "1712.07207",
    archivePrefix = "arXiv",
    primaryClass = "quant-ph",
    doi = "10.1038/s41467-018-08155-0",
    journal = "Nature Commun.",
    volume = "10",
    pages = "494",
    year = "2019"
}

@article{Bardeen:1973gs,
    author = "Bardeen, James M. and Carter, Brandon and Hawking, Stephen W.",
    title = "{The four laws of black hole mechanics}",
    doi = "10.1007/BF01645742",
    journal = "Commun. Math. Phys.",
    volume = "31",
    pages = "161--170",
    year = "1973"
}

@article{LeeWald:1990,
    author = "Lee, Joohan and Wald, Robert M.",
    title = "{Local symmetries and constraints}",
    doi = "10.1063/1.528801",
    journal = "J. Math. Phys.",
    volume = "31",
    pages = "725--743",
    year = "1990"
}

@article{WaldZoupas:2000,
    author = "Wald, Robert M. and Zoupas, Andreas",
    title = "{A general definition of `conserved quantities' in general relativity and other theories of gravity}",
    eprint = "gr-qc/9911095",
    archivePrefix = "arXiv",
    doi = "10.1103/PhysRevD.61.084027",
    journal = "Phys. Rev. D",
    volume = "61",
    pages = "084027",
    year = "2000"
}

@article{Sewell:1982,
    author = "Sewell, Geoffrey L.",
    title = "{Quantum fields on manifolds: PCT and gravitationally induced thermal states}",
    doi = "10.1016/0003-4916(82)90285-8",
    journal = "Annals Phys.",
    volume = "141",
    pages = "201--224",
    year = "1982"
}

@article{KayWald:1991,
    author = "Kay, Bernard S. and Wald, Robert M.",
    title = "{Theorems on the uniqueness and thermal properties of stationary, nonsingular, quasifree states on space-times with a bifurcate Killing horizon}",
    doi = "10.1016/0370-1573(91)90015-E",
    journal = "Phys. Rept.",
    volume = "207",
    pages = "49--136",
    year = "1991"
}

@article{PuszWoronowicz:1978,
    author = "Pusz, Wieslaw and Woronowicz, Stanislaw L.",
    title = "{Passive states and KMS states for general quantum systems}",
    doi = "10.1007/BF01614224",
    journal = "Commun. Math. Phys.",
    volume = "58",
    pages = "273--290",
    year = "1978"
}

@article{GaoWald:2001,
    author = "Gao, Sijie and Wald, Robert M.",
    title = "{The `physical process' version of the first law and the generalized second law for charged and rotating black holes}",
    eprint = "gr-qc/0106071",
    archivePrefix = "arXiv",
    doi = "10.1103/PhysRevD.64.084020",
    journal = "Phys. Rev. D",
    volume = "64",
    pages = "084020",
    year = "2001"
}

@article{Connes:1973,
    author = "Connes, Alain",
    title = "{Une classification des facteurs de type III}",
    doi = "10.24033/asens.1247",
    journal = "Ann. Sci. Ecole Norm. Sup.",
    volume = "6",
    pages = "133--252",
    year = "1973"
}

@article{HollandsKovacsReall:2022,
 author={Hollands, Stefan and Kov{\'a}cs, {\'A}ron D. and Reall, Harvey S.},
 title={The second law of black hole mechanics in effective field theory},
 number = {2022},
    journal = {JHEP}, volume={08}, pages={258}, year={2022},
 eprint={2205.15341}, archivePrefix={arXiv}, primaryClass={hep-th}}

@article{FaulknerSperanza:2024,
 author={Faulkner, Thomas and Speranza, Antony J.},
 title={Gravitational algebras and the generalized second law},
 number = {2024},
    journal = {JHEP}, volume={11}, pages={099}, year={2024},
 eprint={2405.00847}, archivePrefix={arXiv}, primaryClass={hep-th}}

@article{Kirklin:2024,
 author={Kirklin, Josh}, title={Generalised second law beyond the semiclassical regime},
 number = {2025},
    journal = {JHEP}, volume={07}, pages={192}, year={2025},
 eprint={2412.01903}, archivePrefix={arXiv}, primaryClass={hep-th}}

@misc{Shajiee:2026,
 author={Shajiee, V. R. and Sheikh-Jabbari, M. M. and Taghiloo, V.},
 title={Who Writes the Gravitational Second Law of Thermodynamics?},
 year={2026}, eprint={2608.16997}, archivePrefix={arXiv}, primaryClass={hep-th}}

@article{AharonovSusskind:1967,
 author={Aharonov, Yakir and Susskind, Leonard},
 title={Charge and angular momentum as observables in quantum mechanics},
 journal={Phys. Rev.}, volume={155}, pages={1428--1431}, year={1967},
 doi={10.1103/PhysRev.155.1428}}

@article{AharonovKaufherr:1984,
 author={Aharonov, Yakir and Kaufherr, T.}, title={Quantum frames of reference},
 journal={Phys. Rev. D}, volume={30}, pages={368--385}, year={1984},
 doi={10.1103/PhysRevD.30.368}}

@article{Popescu:2018,
 author={Popescu, Sandu and Sainz, Ana Bel{\'e}n and Short, Anthony J. and Winter, Andreas},
 title={Quantum Reference Frames and Their Applications to Thermodynamics},
 year={2018}, eprint={1804.03730}, archivePrefix={arXiv}, primaryClass={quant-ph},
 journal={Phil. Trans. R. Soc. A}, volume={376}, pages={20180111},
 doi={10.1098/rsta.2018.0111}}

@book{Takesaki:1970,
 author={Takesaki, Masamichi},
 title={Tomita's Theory of Modular Hilbert Algebras and its Applications},
 series={Lecture Notes in Mathematics}, volume={128},
 publisher={Springer}, address={Berlin}, year={1970}}

@article{Sorce:2023,
 author={Sorce, Jonathan}, title={An intuitive construction of modular flow},
 number = {2023},
    journal = {JHEP}, volume={12}, pages={079}, year={2023},
 eprint={2309.16766}, archivePrefix={arXiv}, primaryClass={hep-th}}

@article{Cirafici:2024a,
 author={Cirafici, Michele},
 title={Fluctuation theorems, quantum channels and gravitational algebras},
 number = {2024},
    journal = {JHEP}, volume={11}, pages={089}, year={2024},
 eprint={2408.04219}, archivePrefix={arXiv}, primaryClass={hep-th}}

@article{Cirafici:2024b,
 author={Cirafici, Michele}, title={On the nonequilibrium dynamics of gravitational algebras},
 journal={Class. Quantum Grav.}, volume={41}, pages={235006}, year={2024},
 eprint={2402.03939}, archivePrefix={arXiv}, primaryClass={hep-th}}

@article{Cirafici:2025,
 author={Cirafici, Michele}, title={Gravitational Algebras and Applications to Nonequilibrium Physics},
 journal={Universe}, volume={11}, pages={24}, year={2025},
 eprint={2412.17674}, archivePrefix={arXiv}, primaryClass={hep-th}}

@article{RignonBret:2026a,
 author={Rignon-Bret, Antoine and Vilatte, Matthieu},
 title={Black hole thermodynamics at null infinity. I. Dual generalized second law},
 journal={Phys. Rev. D}, volume={114}, pages={065012}, year={2026},
 eprint={2601.03353}, archivePrefix={arXiv}, primaryClass={hep-th}}

@article{RignonBret:2026b,
 author={Rignon-Bret, Antoine and Vilatte, Matthieu},
 title={Black hole thermodynamics at null infinity. II. Open systems, Markovian dynamics, and work extraction from nonrotating black holes},
 journal={Phys. Rev. D}, volume={114}, pages={065011}, year={2026},
 eprint={2601.03356}, archivePrefix={arXiv}, primaryClass={hep-th}}

@article{Albash:2013,
 author={Albash, Tameem and Lidar, Daniel A. and Marvian, Milad and Zanardi, Paolo},
 title={Fluctuation theorems for quantum processes},
 journal={Phys. Rev. E}, volume={88}, pages={032146}, year={2013},
 eprint={1212.6589}, archivePrefix={arXiv}, primaryClass={quant-ph}}

@article{Kubo:1957,
 author={Kubo, Ryogo},
 title={Statistical-mechanical theory of irreversible processes. I. General theory and simple applications to magnetic and conduction problems},
 journal={J. Phys. Soc. Jpn.}, volume={12}, pages={570--586}, year={1957}}

@article{MartinSchwinger:1959,
 author={Martin, Paul C. and Schwinger, Julian}, title={Theory of many-particle systems. I},
 journal={Phys. Rev.}, volume={115}, pages={1342--1373}, year={1959}}

@article{HaagHugenholtzWinnink:1967,
 author={Haag, Rudolf and Hugenholtz, Nico M. and Winnink, Marinus},
 title={On the equilibrium states in quantum statistical mechanics},
 journal={Commun. Math. Phys.}, volume={5}, pages={215--236}, year={1967}}

@article{Klinger:2026,
 author={Klinger, Marc S. and Kudler-Flam, Jonah and Satishchandran, Gautam},
 title={Generalized black hole entropy is von Neumann entropy. II. The complete symmetry group and edge modes},
 journal={Phys. Rev. D}, volume={114}, pages={065010}, year={2026},
 eprint={2601.07910}, archivePrefix={arXiv}, primaryClass={hep-th}}

@misc{CovariantRegulator:2023,
 author={Kudler-Flam, Jonah and Leutheusser, Samuel and Rahman, Adel A. and Satishchandran, Gautam and Speranza, Antony J.},
 title={A covariant regulator for entanglement entropy: proofs of the Bekenstein bound and QNEC},
 year={2023}, eprint={2312.07646}, archivePrefix={arXiv}, primaryClass={hep-th}}

@article{Spohn:1978,
 author={Spohn, Herbert},
 title={Entropy production for quantum dynamical semigroups},
 journal={J. Math. Phys.}, volume={19}, number={5}, pages={1227--1230},
 year={1978}, doi={10.1063/1.523789}}

@article{ChetriteMallick:2012,
 author={Chetrite, Raphael and Mallick, Kirone},
 title={Quantum Fluctuation Relations for the Lindblad Master Equation},
 journal={J. Stat. Phys.}, volume={148}, pages={480--501}, year={2012},
 doi={10.1007/s10955-012-0557-z}, eprint={1112.1303}, archivePrefix={arXiv}, primaryClass={cond-mat.stat-mech}}

@article{DeChiaraImparato:2022,
 author={De Chiara, Gabriele and Imparato, Alberto},
 title={Quantum fluctuation theorem for dissipative processes},
 journal={Phys. Rev. Research}, volume={4}, pages={023230}, year={2022},
 doi={10.1103/PhysRevResearch.4.023230}, eprint={2108.05937}, archivePrefix={arXiv}, primaryClass={quant-ph}}

@article{Manzano:2018,
 author={Manzano, Gonzalo and Horowitz, Jordan M. and Parrondo, Juan M. R.},
 title={Quantum Fluctuation Theorems for Arbitrary Environments: Adiabatic and Nonadiabatic Entropy Production},
 journal={Phys. Rev. X}, volume={8}, pages={031037}, year={2018},
 doi={10.1103/PhysRevX.8.031037}, eprint={1710.00054}, archivePrefix={arXiv}, primaryClass={quant-ph}}

@article{Rastegin:2013,
 author={Rastegin, Alexey E.},
 title={Non-equilibrium equalities with unital quantum channels},
 journal={J. Stat. Mech.}, volume={2013}, pages={P06016}, year={2013},
 doi={10.1088/1742-5468/2013/06/P06016}, eprint={1301.0855}, archivePrefix={arXiv}, primaryClass={quant-ph}}

@article{RasteginZyczkowski:2014,
 author={Rastegin, Alexey E. and {\.{Z}}yczkowski, Karol},
 title={Jarzynski equality for quantum stochastic maps},
 journal={Phys. Rev. E}, volume={89}, pages={012127}, year={2014},
 doi={10.1103/PhysRevE.89.012127}, eprint={1307.5370}, archivePrefix={arXiv}, primaryClass={quant-ph}}

@article{Mishra:2018,
 author={Mishra, Akash and Chakraborty, Sumanta and Ghosh, Avirup and Sarkar, Sudipta},
 title={On the physical process first law for dynamical black holes},
 journal={JHEP}, number={2018}, volume={09}, pages={034}, year={2018},
 doi={10.1007/JHEP09(2018)034}, eprint={1709.08925}, archivePrefix={arXiv}, primaryClass={hep-th}}

@misc{Jaksic:2011entropic,
 author={Jak{\v{s}}i{\'c}, Vojkan and Ogata, Yoshiko and Pautrat, Yan and Pillet, Claude-Alain},
 title={Entropic Fluctuations in Quantum Statistical Mechanics. An Introduction},
 year={2011}, eprint={1106.3786}, archivePrefix={arXiv}, primaryClass={math-ph}}

@article{Benoist:2024two,
 author={Benoist, Tristan and Bruneau, Laurent and Jak{\v{s}}i{\'c}, Vojkan and Panati, Annalisa and Pillet, Claude-Alain},
 title={A note on two-times measurement entropy production and modular theory},
 journal={Lett. Math. Phys.}, volume={114}, pages={32}, year={2024},
 eprint={2310.10582}, archivePrefix={arXiv}, primaryClass={math-ph}}

@article{Upadhyaya:2024surprisal,
 author={Upadhyaya, Twesh and Braasch, Jr., William F. and Landi, Gabriel T. and Yunger Halpern, Nicole},
 title={Non-Abelian transport distinguishes three usually equivalent notions of entropy production},
 journal={PRX Quantum}, volume={5}, pages={030355}, year={2024},
 doi={10.1103/PRXQuantum.5.030355}, eprint={2305.15480}, archivePrefix={arXiv}, primaryClass={quant-ph}}

@article{DimockKay:1987,
  author  = {Dimock, Jonathan and Kay, Bernard S.},
  title   = {Classical and quantum scattering theory for linear scalar fields on the {Schwarzschild} metric {I}},
  journal = {Annals Phys.},
  volume  = {175},
  pages   = {366--426},
  year    = {1987},
  doi     = {10.1016/0003-4916(87)90214-4}
}

@article{Araki:1977relative,
    author = {Araki, Huzihiro},
    title = {Relative Entropy for States of von Neumann Algebras II},
    journal = {Publ. Res. Inst. Math. Sci.},
    volume = {13},
    number = {1},
    pages = {173--192},
    year = {1977},
    doi = {10.2977/PRIMS/1195190105}
}

@article{Rovelli:1991clock,
  author = {Rovelli, Carlo},
  title = {Time in quantum gravity: An hypothesis},
  journal = {Phys. Rev. D},
  volume = {43}, pages = {442--456}, year = {1991},
  doi = {10.1103/PhysRevD.43.442}
}

@article{Marolf:1995clock,
  author = {Marolf, Donald},
  title = {Quantum observables and recollapsing dynamics},
  journal = {Class. Quantum Grav.},
  volume = {12}, pages = {1199--1220}, year = {1995},
  doi = {10.1088/0264-9381/12/5/011},
  eprint = {gr-qc/9404053}, archivePrefix = {arXiv}
}

@article{Hoehn:2019trinity,
  author = {H{\"o}hn, Philipp A. and Smith, Alexander R. H. and Lock, Maximilian P. E.},
  title = {Trinity of relational quantum dynamics},
  journal = {Phys. Rev. D},
  volume = {104}, pages = {066001}, year = {2021},
  doi = {10.1103/PhysRevD.104.066001},
  eprint = {1912.00033}, archivePrefix = {arXiv}, primaryClass = {quant-ph}
}

@article{FrancoisRavera:2024,
  author = {Fran{\c c}ois, Jordan T. and Ravera, Lucrezia},
  title = {Geometric Relational Framework for General-Relativistic Gauge Field Theories},
  journal = {Fortschr. Phys.},
  volume = {73}, pages = {2400149}, year = {2025},
  doi = {10.1002/prop.202400149},
  eprint = {2407.04043}, archivePrefix = {arXiv}, primaryClass = {gr-qc}
}

@article{IsoOkazawaZhang:2010,
  author = {Iso, Satoshi and Okazawa, Susumu and Zhang, Sen},
  title = {Non-equilibrium fluctuations of black hole horizons and the generalized second law},
  journal = {Phys. Lett. B},
  volume = {705}, pages = {152--156}, year = {2011},
  doi = {10.1016/j.physletb.2011.09.114},
  eprint = {1008.1184}, archivePrefix = {arXiv}, primaryClass = {gr-qc}
}

@article{IsoOkazawa:2011,
  author = {Iso, Satoshi and Okazawa, Susumu},
  title = {Stochastic equations in black hole backgrounds and non-equilibrium fluctuation theorems},
  journal = {Nucl. Phys. B},
  volume = {851}, pages = {380--419}, year = {2011},
  doi = {10.1016/j.nuclphysb.2011.05.021},
  eprint = {1104.2461}, archivePrefix = {arXiv}, primaryClass = {hep-th}
}

@article{BrownKuchar:1995dust,
  author = {Brown, J. David and Kucha{\v r}, Karel V.},
  title = {Dust as a standard of space and time in canonical quantum gravity},
  journal = {Phys. Rev. D},
  volume = {51}, pages = {5600--5629}, year = {1995},
  doi = {10.1103/PhysRevD.51.5600},
  eprint = {gr-qc/9409001}, archivePrefix = {arXiv}
}

@article{Rovelli:2002partial,
  author = {Rovelli, Carlo},
  title = {Partial observables},
  journal = {Phys. Rev. D},
  volume = {65}, pages = {124013}, year = {2002},
  doi = {10.1103/PhysRevD.65.124013},
  eprint = {gr-qc/0110035}, archivePrefix = {arXiv}
}

@article{Dittrich:2007complete,
  author = {Dittrich, Bianca},
  title = {Partial and complete observables for {Hamiltonian} constrained systems},
  journal = {Gen. Relativ. Gravit.},
  volume = {39}, pages = {1891--1927}, year = {2007},
  doi = {10.1007/s10714-007-0495-2},
  eprint = {gr-qc/0411013}, archivePrefix = {arXiv}
}

\end{document}